\documentclass[10pt,tightenlines,eqsecnum,floats,aps,amsmath,
amssymb,nofootinbib,prd,noshowpacs,notitlepage,longbibliography,twocolumn,
showpacs]{revtex4-1}
\usepackage[utf8]{inputenc}
\usepackage{amsmath,amsthm,latexsym,amssymb,amsfonts,color,graphicx,tikz}
\usetikzlibrary{patterns}
\usepackage[colorlinks=true,citecolor=blue,urlcolor=blue]{hyperref}
\usepackage{longtable}

\begin{document}

\title{Coupled intertwiner dynamics:\\ A toy model for coupling matter to spin foam models}

\author{Sebastian Steinhaus}
\email{sebastian.steinhaus@desy.de}

\affiliation{
 II. Institute for Theoretical Physics,
 \small University of Hamburg,\\
 \small Luruper Chaussee 149,\\
 \small D-22761 Hamburg, Germany 
  }
  

\begin{abstract}
The universal coupling of matter and gravity is one of the most important features of general relativity. In quantum gravity, in particular spin foams, matter couplings have been defined in the past, yet the mutual dynamics, in particular if matter and gravity are strongly coupled, are hardly explored, which is related to the definition of both matter and gravitational degrees of freedom on the discretisation. However extracting this mutual dynamics is crucial in testing the viability of the spin foam approach and also establishing connections to other discrete approaches such as lattice gauge theories.

Therefore, we introduce a simple 2D toy model for Yang--Mills coupled to spin foams, namely an Ising model coupled to so--called intertwiner models defined for $\text{SU}(2)_k$. The two systems are coupled by choosing the Ising coupling constant to depend on spin labels of the background, as these are interpreted as the edge lengths of the discretisation. We coarse grain this toy model via tensor network renormalization and uncover an interesting dynamics: the Ising phase transition temperature turns out to be sensitive to the background configurations and conversely, the Ising model can induce phase transitions in the background. Moreover, we observe a strong coupling of both systems if close to both phase transitions.
\end{abstract}

\maketitle

\section{Introduction}

The universal coupling to all types of (known) matter is one of the most profound features of general relativity. This property has proven to be invaluable in testing the theory, make progress in cosmology and interpret phenomena such as black holes as originating from the gravitational collapse of matter. However, general relativity as a classical and thus deterministic theory of a dynamical spacetime is at odds with the fundamental quantum nature of matter, which is described by quantum field theories on a fixed, not necessarily flat, background spacetime.

It is expected that this inconsistency can be cured by quantising gravity \cite{quantising-gravity}. Inevitably this not only requires this theory of quantum gravity to possess a classical limit consistent with vacuum general relativity, but it must also be possible to couple matter to it and obtain a dynamics for the composite system. Most importantly, this should allow us to study systems, for which gravity and matter are strongly coupled, e.g. close to the Big Bang, where one can expect to find new physics. However, these regions are currently not accessible by experiments. On the other hand, the theory must not conflict already well--tested physics. Hence, it should also be possible to identify a regime, in which the effective dynamics (of quantum matter and gravity) behave like quantum field theory on an almost fixed background spacetime, e.g. consistent with data observed at accelerators. Moreover, the coupling of matter might reveal new properties and deepen our understanding of the theory of quantum gravity, as it has been demonstrated e.g. in \cite{eichhorn-matter} in the context of asymptotic safety \cite{asymptotic-safety1}.

The status of coupling matter to candidate theories of quantum gravity is as diverse as the theories themselves. For approaches relying heavily on quantum field theory methods, e.g. asymptotic safety \cite{asymptotic-safety1}, in particular choosing (albeit not necessarily fixing) a background spacetime, coupling matter to gravity is (at least conceptionally) straightforward, whereas theories relying on (either auxiliary or fundamental) discretisations, such as spin foam models \cite{alexreview,spinfoams-zako}, causal dynamical triangulations \cite{cdt-review} or causal sets \cite{causal-set}, additionally face the issue of defining gravity and matter on the same discrete structure.

A candidate theory of quantum gravity defined on a discretisation, so--called spin foam models \cite{alexreview,spinfoams-zako}, motivate this paper. Spin foams are a path integral approach related to loop quantum gravity \cite{cbook,thomasbook} and are defined on a 2--complex, i.e. a collection of vertices, edges and faces, which is frequently chosen to be dual to a triangulation. These elements of the foam carry representation theoretic objects of the underlying symmetry group; for Euclidean signature this is $\text{SO}(4)$, whereas $\text{SL}(2,\mathbb{C})$ is chosen for Lorentzian signature. Irreducible representations can be found on the faces, whereas on the edges projectors onto an invariant subspace (in the tensor product of vectorspaces of representations meeting at this edge) are located. The spin foam model assigns an amplitude to a configuration of the foam, i.e. a colouring with representations and intertwiners, where one obtains the full amplitude after summing over all configurations. If the 2--complex has a boundary, the associated boundary Hilbert space is that of loop quantum gravity on the boundary graph\footnote{As most spin foam models are defined for triangulations, this does not contain all boundary graphs allowed in loop quantum gravity. See \cite{kkl} for a spin foam model resolving this issue and also \cite{zipfel} on the challenge of linking loop quantum gravity and spin foam models. On the other hand, in \cite{bf-vacuum,bf-vacuum2} a new representation of loop quantum gravity based on graphs dual to triangulations has been defined, which could be closer to spin foam models.}; thus a spin foam assigns an amplitude to one possible transition of loop quantum gravity states.

Readers familiar with lattice gauge theories may recognize that the data carried by a spin foam is very similar to the strong coupling expansion of lattice gauge theories. Indeed, spin foam models can be understood as generalized lattice gauge theories, since they carry projectors onto a smaller invariant subspace instead of the full Haar projectors. In the usual holonomy representation, this requires the introduction of additional group elements assigned to the faces of the discretisation \cite{holonomy-sf,holonomy-sf2}. In fact, these similarities can already be noticed for the classical continuum action, where Yang--Mills and gravity can be written as topological BF theory \cite{bf-theory} plus constraints, which break the too many symmetries of the topological theory. The crucial difference is however that the simplicity constraints\footnote{Implementing these constraints in spin foam models results in the projectors onto smaller invariant subspaces, which is however not uniquely defined.} in the Plebanski formulation of general relativity \cite{plebanski} are manifestly background independent, whereas for BF--Yang--Mills \cite{bf-yangmills} the constraints explicitly depend on a background metric.

The before mentioned similar mathematical structure makes pure lattice gauge theories a very natural choice for (tentatively) examining matter couplings to spin foams, where one can focus on the conceptual issues of the coupling of matter and gravity, which is another motivation for the present paper. In fact several different ways to couple Yang--Mills to spin foams have already been defined in the literature: In \cite{crane,crane1,mikovic} the system of gravity and matter is described by a topological quantum field theory, in which propagating degrees of freedom arise either by symmetry breaking in a low energy phase or matter is represented by defects in the geometry. Conversely, \cite{oriti-pfeiffer,mikovic1} define lattice gauge theory on a triangulation (dual to the foam) by modifying the Wilson action to depend on the geometric data provided by the foam, which can be interpreted as a geometry dependent coupling constant -- an idea we will revive and test in this work for a toy model. In contrast to this, \cite{simone-3dYM} derived a discrete action for (BF--) Yang--Mills \cite{bf-yangmills} coupled to topological 3D gravity from the classical action and pointed out that in order to consistently make contact with boundary Hilbert spaces of the canonical theory, both theories ought to be discretized on the same discrete structure, here the 2--complex. In yet a different approach \cite{smolin-ym}, Yang--Mills and gravity are unified in a larger symmetry group, which possesses $\text{SO}(4)$ (or $\text{SL}(2,\mathbb{C})$) as a subgroup and the remainder of the group is associated to the Yang--Mills degrees of freedom. In the context of coupling fermions to spin foams \cite{carlo-muxin}, the introduction of $\text{U}(1)$ Yang--Mills has been discussed in \cite{sf-fermion-marseille}. However, despite this multitude of interesting ideas, it has not been possible to thoroughly study the dynamics of the coupled system or even establish a link to usual lattice gauge theories defined on a flat background lattice.

Similar to the pure gravity case, this issue is rooted in the introduction of the discretisation: Both for gravity and matter, the theory is not uniquely defined, e.g. for Yang--Mills one can choose the Wilson action or the heat kernel action \cite{kogut-review}, but moreover the crucial diffeomorphism symmetry of general relativity is broken \cite{dittrich-diffeo,bahrdittrich-broken,bahrdittrich-improve,perfect-disc,discrete-review} and the theory depends on the chosen discretisation. In fact this also includes the chosen coupling (mechanism) of the two theories. However, physical predictions such as expectation values of observables, should not depend on the chosen regulator, which inevitably raises the question how to remove it. In spin foams alone this is a topic of debate \cite{matteo-summing}, where two main options have been advertised: the first considers a sum over all discretisations (and even topologies as put forward by group field theories \cite{oriti-gft}), which (at least formally) removes the cut--off, whereas the second one proposes a refinement-- / coarse--graining procedure that relates theories, boundary states and observables defined on different discretisations and can be understood as a (generalized) renormalization procedure \cite{fotini-coarse,oeckl-coarse,zapata-coarse,bianca-solve,time-evo,philipp-quant1,hoehn1,hoehn2,hoehn3,philipp-quant2,bahr-ren}, since a notion of `scale' is absent\footnote{In contrast to standard lattice gauge theories, where the lattice spacing $a$ is inherited from the background spacetime, geometric information has to be inferred from the geometric variables of the spin foam. One already faces a similar issue if one considers irregular lattices \cite{oeckl-book}.}. In this work we follow the latter idea, which eventually requires us to renormalize the coupled system, e.g. the EPRL--model \cite{eprl,eprl2} coupled to $\text{U}(1)$-- or $\text{SU}(3)$--Yang--Mills theory, which clearly is beyond the scope of this paper and beyond the possibilities of current renormalization techniques, despite recent progress in coarse graining algorithms for lattice gauge theories \cite{decorated-tensor} and analogue spin foam models \cite{abelian_tnw,abelian_tnw2,s3_tnw,q_spinnet}.

Instead, the goal of this paper is to introduce a toy model, in fact similar in spirit to the work \cite{cdt-ising1,cdt-ising2} for 2D causal dynamical triangulations, that shares several crucial properties with spin foams and lattice gauge theories, but is considerably simpler such that we can study its renormalization group flow, here via tensor network renormalization \cite{levin,guwen}. As the physical system we are interested in, our toy model consists of two parts: a `spin--foam inspired' background, given by `intertwiner models' \cite{invariant_int}, which are based on the quantum group $\text{SU}(2)_k$ \cite{biedenharn,yellowbook}, and a $\mathbb{Z}_2$ Ising model chosen for the matter part. Both systems can be understood as a dimensional reduction of their 4D counterparts, in which the local symmetry is replaced by a global symmetry. Thus both systems pertain a crucial ingredient, namely intertwiners, which are located on the vertices of the lattice\footnote{As an Abelian theory the Ising model only has trivial intertwiner spaces, an extension to non--Abelian (finite) groups is possible, see e.g. \cite{sf-finite}. Note also that the intertwiner models \cite{invariant_int} are also simpler than spin net models \cite{abelian_tnw,s3_tnw,q_spinnet}, which are also called analogue spin foam models.}. We choose this lattice to be a regular square lattice. Then, these two systems are coupled by choosing the Ising coupling constant $\beta$ to be a function of background spin labels $j$\footnote{This possibility has been pointed out to us by Bianca Dittrich, who also raised the idea to examine a coupling $\beta(j_e) \sim \frac{1}{j_e}$.}, similar to \cite{oriti-pfeiffer,mikovic1}, based on the interpretation that the background spins give the distance between two Ising spins. This idea actually resonates with the idea of local coupling constants for lattice gauge theories defined on irregular lattices in \cite{oeckl-book}, where the coupling constants are assigned to parts of the discretisation, e.g. edges or faces. Therefore we study the dynamics of {\it coupled intertwiner degrees of freedom} and are particularly interested in how the two systems affect each other. To do so, we study the Ising model on different backgrounds (including superpositions) for two different ways of coupling the two systems and extract phase diagrams of the model.

The paper is organized as follows: in section \ref{sec:toy-model} we introduce both the background and the Ising model and two different ways to couple the two theories. Section \ref{sec:renormalization} describes the coarse graining algorithm. We split the results in two parts: section \ref{sec:top-background} concerns the Ising model (for different temperatures) on a topological background, i.e. one fixed point intertwiner for the background, in section \ref{sec:superposition} we superpose several fixed point intertwiners. We close with a summary and discussion in section \ref{sec:discussion}. In appendix \ref{app:quantum-group} we have included some additional information on the quantum group $\text{SU}(2)_k$ and the graphical calculus developed in \cite{invariant_int,q_spinnet} helpful to understand some calculations in this article. Appendix \ref{app:tensor} explains tensor network renormalization in more detail and gives a more thorough derivation of the formulae.

\section{The toy model -- Ising spins on a dynamical background} \label{sec:toy-model}


In this section we introduce the toy model of interest: in section \ref{sec:grav-part} we first present the `gravitational' part followed by the (uncoupled) Ising model in section \ref{sec:matter-part}. In section \ref{sec:coupled-system} we discuss the coupling of the two systems in more detail, including assumptions and a brief discussion on choosing the dual discretisation related to \cite{oriti-pfeiffer} and \cite{simone-3dYM}. More importantly, we introduce two different couplings in sections \ref{sec:length-coupling} and \ref{sec:area-coupling}, for which we can already deduce a very different qualitative behaviour.

\subsection{The `gravitational' part -- Intertwiner models} \label{sec:grav-part}

On the `gravity' side, we pick the so--called intertwiner models defined in \cite{invariant_int} by Dittrich and Kami{\'n}ski as our dynamical background. These models are defined on a graph with 3--valent vertices with an underlying symmetry governed by the quantum group $\text{SU}(2)_k$ at root of unity, i.e. the quantum deformation of the universal enveloping algebra $\mathcal{U}_q(\mathfrak{su}(2))$ with deformation parameter $q = \exp{\frac{i \pi}{k+2}}$, see e.g. \cite{biedenharn,yellowbook} and appendix \ref{app:quantum-group} for more details. $k \in \mathbb{N}$ is also called the level of the quantum group and defines a natural upper cut--off on the spins $j$ with $j_{\text{max}}=\frac{k}{2}$. Each 3--valent vertex is dual to a triangle, such that each of its links pierces exactly one edge of a triangle. Each of the links, and hence also the dual edges, carries a spin $j \, \in \, \frac{\mathbb{N}}{2}$ and a magnetic index $m \in \frac{\mathbb{N}}{2}$ with $-j \, \leq \, m \, \leq j$. To the vertex itself, we assign an amplitude: The dependence on the magnetic indices and triangle inequalities for the spins $\{j_e\}$ are encoded in the Clebsch--Gordan coefficients ${}_q \mathcal{C}^{j_1 \, j_2 \, j_3}_{m_1 \, m_2 \, m_3}$  of the quantum group $\text{SU}(2)_k$. Additionally, we assign a factor $a(j_1,j_2,j_3)$ to the vertex, which only depends on the representation labels. Thanks to the imposed triangle inequalities, we will interpret the spin labels $j_e$ as the length of the edges of the triangulation.

To sum up, the basic building blocks of our model -- expressed in a graphical notation\footnote{The peculiarities of the quantum group require us to specify a preferred direction, here the edges of the graphs come with an orientation pointing from bottom up. This is not crucial for most part of this work, thus we have included more details in appendix \ref{app:quantum-group}. See also \cite{q_spinnet} and \cite{invariant_int} for more thorough discussions of this graphical notation.} -- are:
\begin{align} \label{eq:cg1}
& a(j_1,j_2,j_3) \, {}_q \mathcal{C}^{j_1 \, j_2 \, j_3}_{m_1 \, m_2 \, m_3} \; = \; \nonumber \\
=\; & a(j_1,j_2,j_3)
\begin{tikzpicture}[baseline,scale=0.75]
\draw (0,-0.5) -- (0.5,0.) -- (0.5,0.6);
\draw (0.5,0.) -- (1,-0.5);
\node at (0.5,0.) {};
\node [below] at (0,-0.5) {$j_1$};
\node [above] at (0.5,0.5) {$j_3$};
\node [below] at (1,-0.5) {$j_2$};
\end{tikzpicture} =
\begin{tikzpicture}[baseline,scale=0.75]
\draw (0,-0.5) -- (0.5,0.) -- (0.5,0.6);
\draw (0.5,0.) -- (1,-0.5);
\node at (0.5,0.) {\textbullet};
\node [below] at (0,-0.5) {$j_1$};
\node [above] at (0.5,0.5) {$j_3$};
\node [below] at (1,-0.5) {$j_2$};
\end{tikzpicture} \quad ,
\end{align}

\begin{align} \label{eq:cg2}
& a'(j_1,j_2,j_3) \, {}_{\bar{q}} \mathcal{C}^{j_1 \, j_2 \, j_3}_{m_1 \, m_2 \, m_3} \; = \; \nonumber \\
= \; & a'(j_1,j_2,j_3)
\begin{tikzpicture}[baseline,scale=0.75]
\draw (0,-0.6) -- (0,0) -- (-0.5,0.5);
\draw (0,0) -- (0.5,0.5);
\node [below] at (0,-0.6) {$j_3$};
\node [above] at (0.5,0.5) {$j_1$};
\node [above] at (-0.5,0.5) {$j_2$};
\end{tikzpicture} =
\begin{tikzpicture}[baseline,scale=0.75]
\draw (0,-0.6) -- (0,0) -- (-0.5,0.5);
\draw (0,0) -- (0.5,0.5);
\node at (0,0) {\textbullet};
\node [below] at (0,-0.6) {$j_3$};
\node [above] at (0.5,0.5) {$j_1$};
\node [above] at (-0.5,0.5) {$j_2$};
\end{tikzpicture} \quad ,
\end{align}
where $\bar{q}$ denotes the complex conjugate of the deformation parameter $q$.

In general, the factors $a(\{j_e\})$ and $a'(\{j_e\})$ can be chosen freely, but we are rather interested in the particular case of triangulation independent, i.e. topological, models. For a model to be triangulation independent, the predictions of the theory, e.g. expectation values of observables and the partition function, must be independent on the chosen triangulation of the manifold. As shown by Pachner \cite{pachner,pachner1}, two triangulations of the same manifold are related to one another by a consecutive application of certain local changes of the triangulation, known as Pachner moves.

There exist three Pachner moves in 2D: the 2--2 Pachner move, the 3--1 Pachner move and its inverse, the 1--3 Pachner move. These moves are illustrated e.g. in equations \eqref{eq:2-2move} and \eqref{eq:3-1move}. The 2--2 move essentially exchanges the edge shared between two triangles by an edge connecting the `tips' of the original triangles. This gives two new triangles, such that the total number of 2--simplices in the triangulation is unchanged. For the 3--1 move on the other hand, one considers three triangles, where all three triangles share a common vertex. These three triangles are combined into one triangle by removing this (interior) vertex resulting in a coarser triangle; the inverse 1--3 move reverts the process by placing a vertex in the center of the triangle and connecting it to the vertices of the coarse triangle.

In the systems investigated in this work, invariance under Pachner moves directly translates into conditions on the amplitudes associated to the building blocks / triangles, in particular $a(\{j_e\})$. Using the notation introduced above, we can write these conditions pictorially (see \cite{invariant_int} for more details):
\begin{widetext}
\begin{itemize}
\item {\bf `Tilting' condition}:

\begin{equation}
\begin{tikzpicture}[baseline,scale=0.75]
\draw (-0.,-0.5) -- (0,0) -- (-0.6,0.3);
\draw (0,0) -- (0.6,0.3) -- (0.6,0.8);
\draw (0.6,0.3) -- (1.2,0);
\node at (0,0) {\textbullet};
\node at (0.6,0.3) {\textbullet};
\node at (-0.8,0.45) {$j_2$};
\node at (0.6,1.00) {$j_1$};
\node at (-0.,-0.7) {$j_3$};
\node at (1.35,-0.15) {$j_4$};
\node at (0.25,0.4) {$j_6$};
\end{tikzpicture} 
\quad = \quad
\begin{tikzpicture}[baseline,scale=0.75]
\draw (0.0,0.3) -- (0.0,0.8);
\draw (-0.5,0) -- (0.0,0.3) -- (0.6,0) -- (0.6,-0.5);
\draw (0.6,0) -- (1.1,0.3);
\node at (0,0.3) {\textbullet};
\node at (0.6,0.) {\textbullet};
\node at (-0.,1.0) {$j_2$};
\node at (-0.6,-0.1) {$j_3$};
\node at (0.6,-0.7) {$j_4$};
\node at (1.3,0.5) {$j_1$};
\node at (0.4,0.4) {$j_6$};
\end{tikzpicture}
\quad =: \quad
\begin{tikzpicture}[baseline,scale=0.75]
\draw (-0.2,-0.6) -- (0,0) -- (-0.2,0.6);
\draw (0,0) -- (1.2,0.);
\draw (1.4,-0.6) -- (1.2,0.) -- (1.4,0.6);
\node at (0,0) {\textbullet};
\node at (1.2,0) {\textbullet};
\node at (-0.35,-0.8) {$j_3$};
\node at (-0.35,0.8) {$j_2$};
\node at (1.55,-0.8) {$j_4$};
\node at (1.55,0.8) {$j_1$};
\node at (0.65,0.25) {$j_6$};
\end{tikzpicture}  \quad , 
\end{equation} 
which leads to the following relation for the factors $a(\{j\})$:
\begin{equation}
a'_{623} \, a_{641} = a_{362} \, a'_{164} \quad . 
\end{equation}

\item {\bf 2--2 Pachner move}:
\begin{equation} \label{eq:2-2move}
\sum_{j_6}
\begin{tikzpicture}[baseline,scale=0.75]
\draw (-0.2,-0.6) -- (0,0) -- (-0.2,0.6);
\draw (0,0) -- (1.2,0.);
\draw (1.4,-0.6) -- (1.2,0.) -- (1.4,0.6);
\node at (0,0) {\textbullet};
\node at (1.2,0) {\textbullet};
\node at (-0.35,-0.8) {$j_3$};
\node at (-0.35,0.8) {$j_2$};
\node at (1.55,-0.8) {$j_4$};
\node at (1.55,0.8) {$j_1$};
\node at (0.65,0.25) {$j_6$};
\end{tikzpicture} 
\quad = \quad \sum_{j_5}
\begin{tikzpicture}[baseline,scale=0.75]
\draw (-0.5,0.9) -- (0,0.5) -- (0,-0.5) -- (-0.5,-0.9);
\draw (0.5,0.9) -- (0,0.5) -- (0,-0.5) -- (0.5,-0.9);
\node at (0,0.5) {\textbullet};
\node at (0,-0.5) {\textbullet};
\node at (-0.7,1.1) {$j_2$};
\node at (0.7,1.1) {$j_1$};
\node at (-0.7,-1.1) {$j_3$};
\node at (0.7,-1.1) {$j_4$};
\node at (0.3,0) {$j_5$};
\end{tikzpicture}
\end{equation} 
Together with the following identity on the Clebsch--Gordan coefficients \cite{invariant_int}
\begin{equation}
\begin{tikzpicture}[baseline,scale=0.75]
\draw (-0.2,-0.6) -- (0,0) -- (-0.2,0.6);
\draw (0,0) -- (1.2,0.);
\draw (1.4,-0.6) -- (1.2,0.) -- (1.4,0.6);
\node at (-0.35,-0.8) {$j_3$};
\node at (-0.35,0.8) {$j_2$};
\node at (1.55,-0.8) {$j_4$};
\node at (1.55,0.8) {$j_1$};
\node at (0.65,0.25) {$j_6$};
\end{tikzpicture} 
\quad = \quad \sum_{j_5}
\left[
\begin{matrix}
j_1 & j_2 & j_5 \\
j_3 & j_4 & j_6
\end{matrix}
\right]
\sqrt{\frac{d_{j_5}}{d_{j_6}}}
\begin{tikzpicture}[baseline,scale=0.75]
\draw (-0.5,0.9) -- (0,0.5) -- (0,-0.5) -- (-0.5,-0.9);
\draw (0.5,0.9) -- (0,0.5) -- (0,-0.5) -- (0.5,-0.9);
\node at (-0.7,1.1) {$j_2$};
\node at (0.7,1.1) {$j_1$};
\node at (-0.7,-1.1) {$j_3$};
\node at (0.7,-1.1) {$j_4$};
\node at (0.3,0) {$j_5$};
\end{tikzpicture} \quad ,
\end{equation} 
where the symbol in the square brackets is a modified $\{6j\}$ symbol and $d_j = [2j+1]$ denotes the quantum dimension of spin $j$, which is the quantum number of the classical dimension. Again, see appendix \ref{app:quantum-group} for more details. One obtains the following condition on the factors $a(\{j\})$.
\begin{equation}
\sqrt{\frac{1}{d_{j_6}}}\; a'_{623} \, a_{641} = \sum_{j_5} 
\left[
\begin{matrix}
j_1 & j_2 & j_5 \\
j_3 & j_4 & j_6
\end{matrix}
\right] \sqrt{\frac{1}{d_{j_5}}} \; a'_{125} \, a_{345} \quad .
\end{equation}

\item {\bf 3--1 Pachner move}:
\begin{equation} \label{eq:3-1move}
\sum_{j_4,j_5,j_6}
\begin{tikzpicture}[baseline,scale=0.75]
\draw (0.6,-1.1) -- (0.6,-0.45) -- (0.25,0.15) -- (-0.05,0.65);
\draw (0.25,0.15) -- (0.95,0.15);
\draw (0.6,-0.45) -- (0.95,0.15) -- (1.25,0.65);
\node at (0.25,0.15) {\textbullet};
\node at (0.95,0.15) {\textbullet};
\node at (0.6,-0.45) {\textbullet};
\node at (0.6,-1.35) {$j_3$};
\node at (-0.3,0.9) {$j_2$};
\node at (1.5,0.9) {$j_1$};
\node at (0.2,-0.25) {$j_4$};
\node at (1.0,-0.25) {$j_5$};
\node at (0.6,0.4) {$j_6$};
\end{tikzpicture} 
\quad = \quad c
\begin{tikzpicture}[baseline,scale=0.75]
\draw (0.6,-1.1) -- (0.6,-0.15) -- (-0.05,0.65);
\draw (0.6,-0.15) -- (1.25,0.65);
\node at (0.6,-0.15) {\textbullet};
\node at (0.6,-1.35) {$j_3$};
\node at (-0.3,0.9) {$j_2$};
\node at (1.5,0.9) {$j_1$};
\end{tikzpicture} \quad ,
\end{equation}
which gives the following equation for the factors $a(\{j\})$:
\begin{equation}
c \, a'_{123} = \sum_{j_4,j_5,j_6}
\left[
\begin{matrix}
j_1 & j_2 & j_3 \\
j_4 & j_5 & j_6
\end{matrix}
\right]
(-1)^{j_4 + j_5 - j_3} \frac{1}{\sqrt{d_{j_3} d_{j_4}}} \, a_{651} \, a'_{624} \, a'_{543} \quad ,
\end{equation}
and
\begin{equation}
\sum_{j_4,j_5,j_6}
\begin{tikzpicture}[baseline,scale=0.75]
\draw (0.6,1.1) -- (0.6,0.5) -- (0.25,-0.1) -- (-0.05,-0.65);
\draw (0.25,-0.1) -- (0.95,-0.1);
\draw (0.6,0.5) -- (0.95,-0.1) -- (1.25,-0.65);
\node at (0.25,-0.1) {\textbullet};
\node at (0.95,-0.1) {\textbullet};
\node at (0.6,0.5) {\textbullet};
\node at (0.6,-0.4) {$j_6$};
\node at (0.2,0.35) {$j_4$};
\node at (1.05,0.45) {$j_5$};
\node at (0.6,1.35) {$j_3$};
\node at (-0.3,-0.9) {$j_1$};
\node at (1.5,-0.9) {$j_2$};
\end{tikzpicture} 
\quad = \quad c
\begin{tikzpicture}[baseline,scale=0.75]
\draw (0.6,1.1) -- (0.6,0.2) -- (-0.05,-0.65);
\draw (0.6,0.2) -- (1.25,-0.65);
\node at (0.6,0.2) {\textbullet};
\node at (0.6,1.35) {$j_3$};
\node at (-0.3,-0.9) {$j_1$};
\node at (1.5,-0.9) {$j_2$};
\end{tikzpicture} \quad ,
\end{equation} 
which gives the following equation for the factors $a(\{j\})$:
\begin{equation}
c \, a_{123} = \sum_{j_4,j_5,j_6} 
\left[
\begin{matrix}
j_1 & j_2 & j_3 \\
j_4 & j_5 & j_6
\end{matrix}
\right]
(-1)^{j_4 + j_5 - j_3} \frac{1}{\sqrt{d_{j_3} d_{j_4}}} \, a'_{641} \, a_{625} \, a_{453} \quad .
\end{equation}

\end{itemize}
\end{widetext}

Even if the factors $a(\{j_e\})$ (and $a'(\{j_e\})$) satisfy these conditions, i.e. the theory is topological, these factors are not uniquely determined. Further input is required, which usually is a restriction on the allowed spins. In \cite{invariant_int} a wide variety of these models has been identified, here we briefly review the ones most interesting in this work. Note that from now on, we restrict ourselves to integer spins, $j \in \mathbb{N}$, i.e. representations of $\text{SO}(3)_k$. Moreover, for all these topological theories, $a(\{j_e\}) = a'(\{j_e\})$, such that we can simplify the notation:

\begin{itemize}
\item {\bf Maximal spin $J \leq j_{\text{max}}$:} \\
This is the class of topological models we will be mostly interested in in this work. They are labelled by a spin $J \in \mathbb{N}$, which indicates that all spins $j\leq J$ are allowed and excited, that is all couplings, which are allowed by the quantum group, involving these spins are allowed. $J$ can be smaller or equal to the maximal spin of the quantum group, $j_{\text{max}} = \frac{k}{2}$. The corresponding factor $a^J(\{j_e\})$ is given by

\begin{align}
& a^J(j_1,j_2,j_3) = \sqrt{(-1)^{J-j_1}} \sqrt{(-1)^{J-j_2}} \sqrt{(-1)^{J-j_3}} \; \times \nonumber \\
 \times & \; (-1)^{2J-j_1-j_2} \sqrt{\frac{[2j_1+1][2j_2+1]}{[J+1]}}
\left[
\begin{matrix}
j_1 & j_2 & j_3 \\
\frac{J}{2} & \frac{J}{2} & \frac{J}{2}
\end{matrix}
\right]
\; .
\end{align}

Hence, by choosing $J$, we have a direct handle on the maximal spin allowed on the edges of the lattice. Moreover, it is straightforward to consider linear combinations of these theories as it has also been studied in \cite{q_spinnet}, albeit for more complex spin net models.

\item {\bf Only $j=0$ and $j=j_{\text{max}}$ excited:}

For even levels $k$ of the quantum group, the maximal spin $j_{\text{max}}=\frac{k}{2}$ is an integer and is of quantum dimension $d_{j_{\text{max}}} = 1$. Hence $j_{\text{max}}$ can only couple with itself to the trivial representation $j=0$; $j_{\text{max}} \otimes j_{\text{max}} = 0$. One can construct a topological theory where only these two representations are allowed by choosing $a(\{j\})$ as follows \cite{invariant_int}:
\begin{align} \label{eq:just0andmax}
a(0,0,0) = & a'(0,0,0) = 1 \nonumber \\
a(j_{\text{max}},j_{\text{max}},0) = & a'(j_{\text{max}},j_{\text{max}},0) = 1 \, .
\end{align}

\item {\bf Only $j=0$, $j=\frac{j_{\text{max}}}{2}$ and $j=j_{\text{max}}$ excited:}

For this model, we consider $k$ to be a multiple of four, $k=4l$. Then $j_{\text{max}}=2l$ and $\frac{j_{\text{max}}}{2} = l$. As in the previous example, $j_{\text{max}}=2l$ is of quantum dimension one, thus again all factors $a(\{j\})$ which just contain $j=0$ and $j_{\text{max}}$ are equal to one (unless the coupling is forbidden), see \eqref{eq:just0andmax}. Thus the only remaining factors are those containing $j=l$: Those factors $a(\{j\})$ containing either $j=0$ or $j=2l$ are proportional to the quantum dimension of $j=l$:
\begin{align}
a(0,l,l) = a(l,0,l) = a(l,l,0) & = \sqrt{d_l} \nonumber \\
a(2l,l,l) = a(l,2l,l) = a(l,l,2l) & = \sqrt{d_l} \quad .
\end{align}
Thus the only non--trivial amplitude remaining is $a(l,l,l)$, which is given by:
\begin{equation}
a(l,l,l) = a'(l,l,l) = \sqrt{d_l (d_l -2)} \quad .
\end{equation}
Note that these types of fixed points cannot be constructed for all $k=4l$, since in a few cases the 2--2 move cannot be solved, e.g. for $k=12$, $k=24$ or $k=32$. See \cite{invariant_int} for more details. Therefore we restrict ourselves to the working cases, namely $k=4$, which is identical (up to signs) to the model $J=2$ discussed previously and $k=8$.
\end{itemize}

\subsubsection{Intertwiner models for 4--valent vertices:}

The discussion so far only concerned 3--valent intertwiners. In order to define the background for a square lattice, we have to define 4--valent intertwiners.

In general, one can construct $N$--valent intertwiners from 3--valent ones by choosing a 3--valent graph with $N$ outer edges and specifying additional $N-3$ internal representations\footnote{The graph is also called a recoupling scheme. Additionally we assume multiplicity free representation categories, i.e. in a tensor product of two irreps, each irrep appears at most once.}. In our case, the 3--valent vertices come with additional restrictions on the allowed irreps, which we have encoded in the factors $a(\{j\})$. These restrictions translate into restrictions on the $N$--valent intertwiners, if only the allowed representations for 3--valent intertwiners appear in the graph. Thus the set of allowed representations for 3--valent vertices determine those for $N$--valent ones; this is known as Reisenberger's construction principle \cite{reisenberger}, see also \cite{q_spinnet}. Note that different recoupling schemes generically result in different $N$--valent intertwiners.

In the 4--valent case, there exist three different recoupling schemes:
\begin{equation}
A=
\begin{tikzpicture}[baseline,scale=0.75]
\draw (-0.5,-0.75) -- (0,-0.25) -- (0,0.25) -- (-0.5,0.75)
      (0.5,-0.75) -- (0,-0.25)
      (0,0.25) -- (0.5,0.75)
      (-0.5,-1) node {$j_3$}
      (0.5,-1) node {$j_4$}
      (0.5,1.25) node {$j_1$}
      (-0.5,1.25) node {$j_2$}
      (0.3,0) node {$j_5$};
      \node at (0,-0.25) {\textbullet};
      \node at (0,0.25) {\textbullet};
\end{tikzpicture}\; ;\quad
B=
\begin{tikzpicture}[baseline,scale=0.75]
\draw (-1,0.5) -- (-0.5,0) -- (0,0) -- (0.5,0.5)
      (-0.5,0) -- (-1,-0.5)
      (0,0) -- (0.5,-0.5)
      (-1,-0.75) node {$j_3$}
      (-1,0.75) node {$j_2$}
      (0.5,0.75) node {$j_1$}
      (0.5,-0.75) node {$j_4$}
      (-0.25,0.25) node {$j_5$};
      \node at (-0.5,0) {\textbullet};
      \node at (0,0) {\textbullet};
\end{tikzpicture}\; ;\quad
C=\begin{tikzpicture}[baseline,scale=0.75]
\draw  (-0.75,0.75) -- (-0.25,0.25) -- (0.25,0.25) -- (0.75,0.75)
	(0.25,0.25) -- (-0.5,-0.75)
	 (-0.25,0.25) -- (0,0)
	 (0.1,-0.1) -- (0.7,-0.75)
      (-0.5,-1) node {$j_3$}
      (-0.75,1) node {$j_2$}
      (0.7,1) node {$j_1$}
      (0.7,-1) node {$j_4$}
      (0,0.5) node {$j_5$};
      \node at (-0.25,0.25) {\textbullet};
      \node at (0.25,0.25) {\textbullet};
\end{tikzpicture}\quad .
\end{equation}
Due to our choice of factors $a(\{j\})$, we see that the recoupling schemes $A$ and $B$ are related to one another by a 2--2 Pachner move and thus give the same set of allowed representations and intertwiners. The case $C$, as it is also discussed in \cite{q_spinnet}, is not satisfied for all models, in particular not the ones labelled by $J$, due to braiding in the quantum group, but still gives the same set of allowed representations as cases $A$ and $B$. Nevertheless, in this work we will restrict ourselves to the planar case, i.e. cases $A$ and $B$.

Crucially, any choice of factors $a(\{j\})$ giving a triangulation independent model, specifies a fixed point of the renormalization scheme (if only the background is considered) we will use in this work, and thus a special family of intertwiners, $\left| \iota^4_a \right \rangle$ in the 4--valent case. Given this, we define the projector on the 4--valent vertices, $\mathcal{P}^4_v$ as:
\begin{equation}
\mathcal{P}^4_v = \left| \iota^4_a \right \rangle \left \langle \iota^4_a \right | \quad ,
\end{equation}
which satisfy the projector condition $\mathcal{P} \cdot \mathcal{P} = \mathcal{P}$ by definition. We can also define superpositions of intertwiners, which will however generically not satisfy this projector condition any more:
\begin{equation}
\mathcal{P}^4_v = \sum_i \alpha_i \left| \iota^4_{a_i} \right \rangle \left \langle \iota^4_{a_i} \right | \quad ,
\end{equation}
where $\{\alpha_i\}$ are the coefficients of the linear superposition. To put this in context to the models under discussion here, let us express $\mathcal{P}$ for any choice of triangulation independent factors $a(\{j\})$:
\begin{align}\label{eq:4-valent-int}
\mathcal{P}^4_v(\{j\},\{m\})& = \sum_{j_5}\frac1{d_{j_5}}
\begin{tikzpicture}[baseline,scale=0.75]
\draw (-0.5,-1) -- (0,-0.5) -- (0,0.5) -- (-0.5,1)
      (0.5,-1) -- (0,-0.5)
      (0,0.5) -- (0.5,1)
      (-0.5,-1.25) node {$j_3$}
      (0.5,-1.25) node {$j_4$}
      (0.5,1.25) node {$j_1$}
      (-0.5,1.25) node {$j_2$}
      (0.25,0) node {$j_5$};
      \node at (0,-0.5) {\textbullet};
      \node at((0,0.5) {\textbullet};
\end{tikzpicture}
            \nonumber\\
           & = \sum_{j_6}\frac1{d_{j_6}}
\begin{tikzpicture}[baseline,scale=0.75]
\draw (-1,0.5) -- (-0.5,0) -- (0,0) -- (0.5,0.5)
      (-0.5,0) -- (-1,-0.5)
      (0,0) -- (0.5,-0.5)
      (-1,-0.75) node {$j_3$}
      (-1,0.75) node {$j_2$}
      (0.5,0.75) node {$j_1$}
      (0.5,-0.75) node {$j_4$}
      (-0.25,-0.3) node {$j_6$};
      \node at (-0.5,0) {\textbullet};
      \node at (0,0) {\textbullet};
\end{tikzpicture}
\;
\quad .
\end{align}
To check the projector property for \eqref{eq:4-valent-int}, the projectors have to be concatenated vertically (due to the special direction of the quantum group), see also \cite{invariant_int,time-evo} for more details. For the rest of the work, we drop the superscript, since we always work on a square lattice. Eventually, we have the following partition function for the quantum gravity inspired background:
\begin{equation}
Z = \sum_{\{j_e\},\{m_e\}} \mathcal{P}_v(\{j_e\}_{e \supset v},\{m_e\}_{e \supset v}) \quad .
\end{equation}

\subsection{The `matter' part -- Ising spins} \label{sec:matter-part}

The goal of this work is to couple a matter system to this gravitationally inspired, possibly dynamical, background. The simplest, non--trivial option `living' on a lattice is the Ising model, a $\mathbb{Z}_2$ spin system. On a given lattice, here with equidistant vertices, it is defined as follows:

On the vertices $v$ of the lattice, one places group elements $g_v \in \mathbb{Z}_2$; their interactions, restricted to nearest neighbours, are described by edge weights assigned to the edges $e$:

\begin{equation}
\omega_e(g_{s(e)} g_{t(e)}^{-1}, \beta ) = \exp \{\beta \; g_{s(e)} g_{t(e)}^{-1} \} \quad ,
\end{equation}
where the inverse `temperature' $\beta \sim \frac{1}{k T}$ is the coupling constant of the theory. $g_v \in \{-1,1\}$ and the group multiplication is simply given by scalar multiplication. Each edge $e$ comes with an orientation, where $s(e)$ and $t(e)$ denote the source and target of the edge $e$ respectively. Inverting the orientation of an edge results in inverting the argument of the corresponding edge weight, $\omega_{e^{-1}}(g) = \omega_e (g^{-1})$. Since $g=-1$ is its own inverse, changes of orientation do not affect the Ising model. The introduction of an orientation is rather a choice of notational convenience and necessary for the quantum groups.

Given these ingredients, the Ising model is defined via its partition function
\begin{equation} \label{eq:ising-part}
Z = \sum_{\{g_v\}} \prod_e \exp\{\beta \; g_{s(e)} g_{t(e)}^{-1} \} \quad ,
\end{equation}
where the sum is over all configurations $\{g_v\}$ of group elements assigned to the vertices. This sum of configurations is weighted with the edge weights, which assign a higher weight to aligned spins, i.e. $g_{s(e)} = g_{t(e)}$. \eqref{eq:ising-part} also has a global symmetry: if all spins are simultaneously flipped, i.e. $g_v = \pm 1 \rightarrow g_v = \mp 1 \forall v$, the partition function is unchanged.

As it is well--known from Onsager's solution \cite{onsager}, this model exhibits two different phases, separated by a second order phase transition at the critical inverse temperature $\beta_{\text{crit}} \approx 0.4406 \ldots$, in fact two different effects are dominating in the two phases. For $\beta > \beta_{\text{crit}}$, i.e. low temperature, the system is in the ordered phase, in which neighbouring spins are more likely to be aligned to minimize the energy of the configuration. For $\beta < \beta_{\text{crit}}$, i.e. high temperature, the spins essentially decouple and are not aligned any more. Still energy minimization and hence spin alignment is preferable, yet the higher the temperature, the smaller the energy gap becomes. Since there are many more configurations for which the spins are not aligned, they dominate the partition function.

For the purpose of this work, it will be useful to briefly discuss the (group) Fourier transform of the Ising model, see also \cite{sf-finite,abelian_tnw}. Any function $\omega(g)$ on $\mathbb{Z}_2$ can be expanded in characters $\chi$\footnote{For non--Abelian groups this only applies to class functions, which have the property $\omega(g) = \omega(h\, g \, h^{-1})$ for all $h \in G$. Clearly, all functions on Abelian groups are class functions.}:

\begin{align}
\omega(g) & = \sum_k \; \tilde{\omega}_k \; \chi_k(g) \quad , \\
\tilde{\omega}_k & = \frac{1}{2} \sum_g \; \omega(g) \; \overline{\chi_k(g)} \quad .
\end{align}
$k \in \{0,1\}$ denote the irreducible representations of $\mathbb{Z}_2$ and $\chi_k(g)$ is the character of the group element $g$ in the representation $k$. $\overline{\chi_k}$ denotes the complex conjugate character. This character is given by:
\begin{align}
\chi_0(g) & = 1 \; \forall g \in \mathbb{Z}_2 \quad , \\
\chi_1(g) & = \left\{ 
\begin{matrix}
1 & \text{if } g=1 \quad , \\
-1 & \text{if } g=-1 \quad .
\end{matrix}
\right.
\end{align}
The Fourier transformed edge weights $\tilde{\omega}_k$ are then given by:
\begin{equation}
\tilde{\omega}_k = \left\{
\begin{matrix}
\cosh(\beta) & \text{for } k=0 \quad , \\
\sinh(\beta) & \text{for } k=1 \quad .
\end{matrix}
\right.
\end{equation}
Eventually, we can expand the partition function into Fourier components:
\begin{align}
Z = & \sum_{\{g_v\}} \prod_e \exp\{\beta \, g_{s(e)} g_{t(e)}^{-1} \} \\
  = & \frac{1}{2^{E}} \sum_{\{g_v\}} \sum_{\{k_e\}} \prod_e \tilde{\omega}_{k_e} \; \chi_{k_e}(g_{s(e)} g_{t(e)}^{-1}) \\
  = & \frac{1}{2^{E}} \sum_{\{k_e\}} \prod_e \tilde{\omega}_{k_e} \sum_{\{g_v\}} \prod_v \prod_{e \supset v} \chi_{k_e}(g_v^{o(e,v)}) \\
  = & \sum_{\{k_e\}} \prod_e \tilde{\omega}_{k_e} \prod_v \delta^{(2)}\left(\sum_{e \supset v} (-1)^{o(e,v)} k_e\right) \quad ,
\end{align}
where $E$ denotes the number of edges in the lattice. We have used the fact that $\chi_k(g \cdot h) = \chi_k(g) \; \chi_k(h)$, such that the partition factorizes for all $g_v$ and the sum can be performed. As a result one obtains the delta function on $\mathbb{Z}_2$, denoted $\delta^{(2)}(k)$, on the vertices of the lattice, which require that the representations $k_e$ on the edges $e$ meeting at the vertex $v$ sum to zero (modulo 2).

Following this transformation one can define the dual theory to the Ising model \cite{savit-review}, by introducing new variables on the vertices of the dual lattice, which depend on the $\{k_e\}$ in such a way that all constraints on the original vertices are satisfied. Remarkably, the dual variables are again Ising spins (elements of $\mathbb{Z}_2$), which like the originial Ising model only interact with their nearest neighbours. The new coupling constant depends on the original $\beta$, yet is large if $\beta$ is small and vice versa. Hence, the Ising model is also called a self--dual theory, which relates the low / high temperature regime of the Ising model to the high / low temperature regime of another Ising model respectively.

This particular property of the Ising model will simplify the discussion of the coupling to the gravitational background in the following section.

\subsection{The coupled system -- some basic observations} \label{sec:coupled-system}

\begin{figure}
\includegraphics[width=0.3\textwidth]{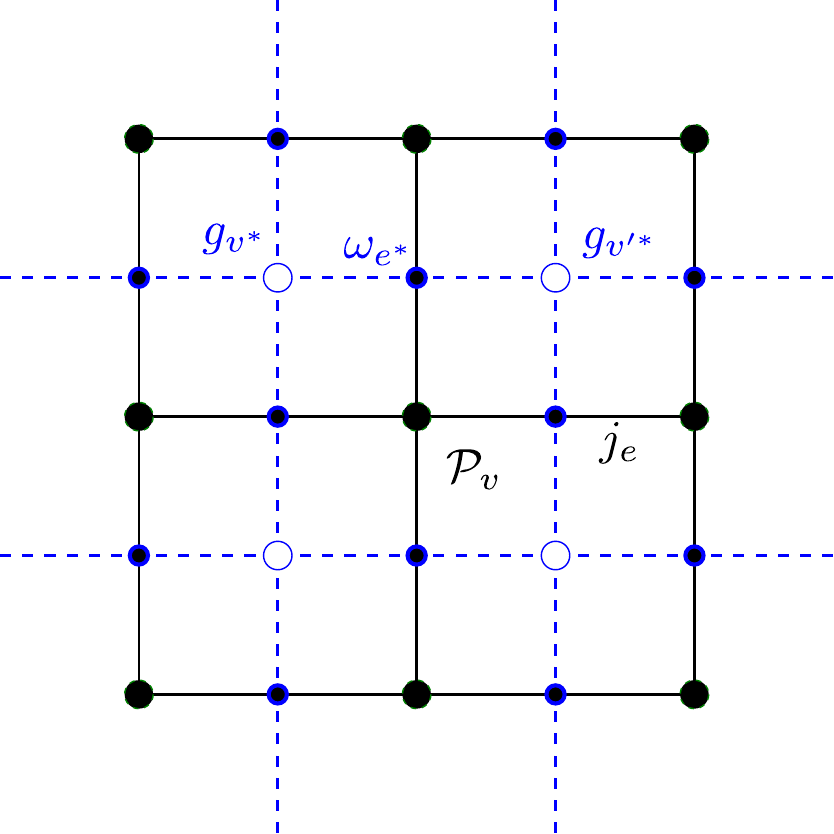}
\caption{The coupled model, with Ising spins on the dual lattice. The gravitational model is solely placed on the black lattice, with $\text{SU}(2)_k$ projectors / intertwiners $\mathcal{P}_v(\{j_e\}_{e \supset v})$ on the vertices $v$ and representations $j_e$ on the edges. The blue dual lattice carries the Ising model, with Ising spins $g_{v^*}$ on the dual vertices $v^*$ and edge weights $\omega_{e^*}$ on the dual edges $e^*$. Note that every edge $e$ pierces its dual edge $e^*$ only once, such that we can interpret the spin $j_e$ as the length or distance between the two Ising spins on $e^*$. There we will implement the interaction between the Ising and background degrees of freedom, indicated by the blue--black dots.
\label{fig:lattice-vs-dual}}
\end{figure}

In the previous sections \ref{sec:grav-part} and \ref{sec:matter-part} we have introduced two discrete systems, a quantum gravity inspired one and a simple matter model. The goal of this work is to sensibly couple these models and study their collective dynamics. To be more precise, the gravitational part is intended to be the background on which the Ising model `lives', providing a dynamical notion of lengths which should influence the Ising spin interactions. To do so, we work with the following assumptions:

\begin{itemize}
\item Both theories live on the `same' lattice\footnote{This can in principle mean that one theory lives on the dual lattice, a possibility we do not wish to exclude a priori.}, in the sense that we do not build in a different scaling behaviour by hand, e.g. say that the gravitational degrees of freedom should be `finer' than the Ising degrees of freedom. This choice is motivated by interpreting the lattice as a regulator for both theories, where additionally the lattice itself does not carry a notion of length -- this notion is supposed to be provided by the gravitational theory. 
\item The two theories will be coupled by modifying the matter part, concretely by implementing a dependence on the gravitational labels into the edge weights of the Ising model. One can interpret this as a coupling constant $\beta$ depending on the spins of the gravitational model, for which we will introduce two schemes below. This choice is motivated from the coupling of matter to general relativity: While gravity is not affected by the matter part in the action, the matter part is sensitive to gravity via the dynamical metric $g_{\mu \nu}$ that enters in the volume element, the contraction of spacetime indices and covariant derivatives.
\item For the rest of this paper we will consider only regular (with respect to the combinatorics) square lattices, i.e. we consider 4--valent vertices where every pair of vertices share at most one edge. The main reason is that coarse graining via tensor network renormalization \cite{levin,guwen} can be straightforwardly implemented and will allow us to study the phase structure of the model. Moreover the dual lattice is also simply a shifted square lattice and we can compare the coupled system to the pure Ising model on an equilateral square lattice, e.g. observe changes in the phase transition temperature. 
\end{itemize}
As we have emphasized before, we will interpret the gravity inspired theory as a dynamical background on which we place the Ising model. This can in principle be done in two ways:

One possibility is to place the Ising spins on the dual of the original lattice, which is demonstrated in fig. \ref{fig:lattice-vs-dual}. Indeed, a spin foam model coupled to a pure (Yang--Mills) lattice gauge theory has been defined on a triangulation in \cite{oriti-pfeiffer}. While the 4D Barrett--Crane model \cite{Barrett-Crane} `lived' on the dual 2--complex, the lattice gauge theory has been placed on the triangulation, where the spin foam model provided the geometric data, i.e. areas of triangles and volumes of 4--simplices. These two theories have been coupled by implementing the dependence on the geometric data into the face weights of the lattice gauge theory: via scaling arguments to obtain the proper continuum limit, the Wilson action has to scale like the volume of the building block, here the 4--simplex, however the plaquette variables scale like the area squared of the plaquette. This mismatch\footnote{In ordinary 4D lattice gauge theory the underlying lattice is hypercubic and equilateral, such that the area squared of any plaquette scales as the volume of a 4D hypercube.} has to be cured by a proper normalisation, which could be interpreted as a locally modified coupling constant of the lattice gauge theory, dependent on the geometry arising from the spin foam.

\begin{figure}
\includegraphics[width=0.3\textwidth]{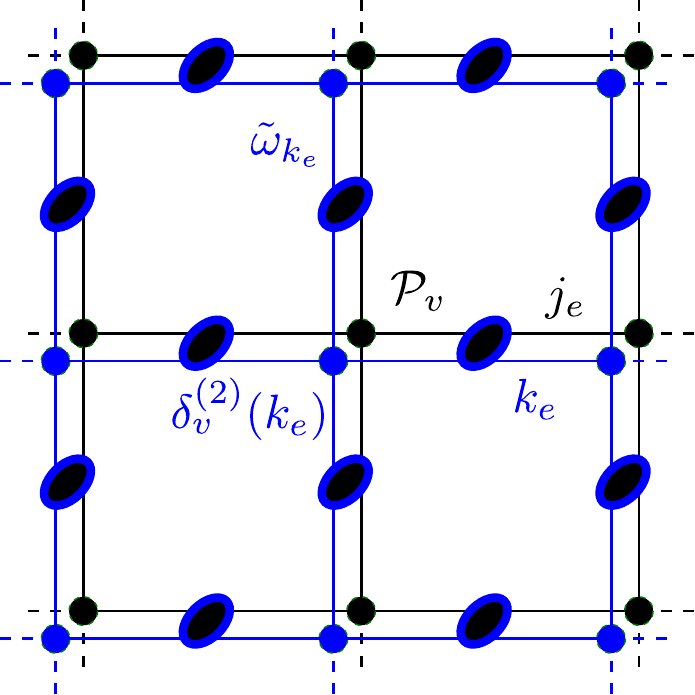}
\caption{The coupled model, with Ising spins on the same lattice, after the Fourier transform to irreducible representations $k_e$ has been performed. The gravitational model is represented by the black lattice, with $\text{SU}(2)_k$ projectors / intertwiners $\mathcal{P}_v(\{j_e\}_{e \supset v})$ on the vertices $v$ and representations $j_e$ on the edges. The blue lattice carries the Ising model, with irreducible representation $k_e$ and (transformed) edge weights $\tilde{\omega}_{k_e}$ on the edges $e$ and a `Gauss' constraint $\delta^{(2)}_v(k_e):=\delta^{(2)}(\sum_{e \supset v} (-1)^{o(e,v)} k_e)$ on the vertices. Again, we interpret the $\text{SU}(2)_k$ spin $j_e$ as the distance between two Ising spins on the vertices, which enters into the edge weights $\tilde{\omega}_{k_e}$, such that the interaction of the two sets of degrees of freedom is only via the edges. To underline this, we have drawn both lattices distinctly, with blue--black dots indicating the interaction on the edges.
\label{fig:lattice-vs-dual2}}
\end{figure}

However, the definition of the two coupled theories on dual lattices / complexes has a drawback that has been pointed out in \cite{simone-3dYM}: Since both theories live on related but essentially different lattices, the states and Hilbert spaces one defines in the canonical theory, which are essentially product states \cite{cbook} are not boundary states of the spin foam model, such that this model cannot be used to compute transition amplitudes between states of the canonical theory. To avoid this, one should rather define both theories on the same discretisation. In the context of our toy model, this version is illustrated in fig. \ref{fig:lattice-vs-dual2}.

In our concrete case, the chosen discretisation of the Ising model (with respect to the background) is less important, since the Ising model is self--dual; essentially we explore the same system with inverted temperature\footnote{This is also true for the Ising model coupled to the background via a $j$ dependent coupling constant, as introduced below. }. Nevertheless, these two choices are still worth exploring since the choice will impact the coarse graining procedure explained in section \ref{sec:renormalization}, for which the partition function is written as a contraction of (local) tensors assigned to the vertices of the lattice. To achieve this in the two different discretisation schemes requires a different choice of (equivalent) variables for the Ising spins. In the dual case, we stick to the original Ising spins on the dual vertices, whereas in the ordinary discretisation it is preferable to use the Fourier transformed ones.


After this extended discussion on the discretisation schemes, we have to focus now on the coupling of the background and the Ising model, for which a priory many choices are possible. As discussed above, we implement this coupling by solely modifying the Ising model, i.e. by implementing a dependence on the background variables in the edge weights $\omega_e$ (or conversely their Fourier transforms $\tilde{\omega}_k$), concretely by choosing $\beta$ to be a function of the underlying geometry, here of the spins $j_e$; hence $\beta(j_e)$.

The reader might wonder, why this is supposed to be a reasonable choice. Again the reasons stem from analogies. First of all, the toy model at hand is supposed to give insights into quantum gravity coupled to lattice gauge theory, since 2D spin systems share statistical similarities to 4D lattice gauge theories \cite{kogut-review}. If we consider again the classical continuum coupling of gravity to Yang--Mills theory, we observe that the metric only couples to the curvature of the Yang--Mills connections, not directly to the connection. Thus, we rather avoid coupling the background degrees of freedom directly to the Ising spins, but rather their edge--wise product. In such a way we also avoid breaking the global $\mathbb{Z}_2$ symmetry of the Ising model, such that it can still be Fourier transformed and remains self--dual\footnote{Note that we are not implying that a direct coupling breaks this symmetry necessarily, but a well--defined definition might not be obvious. E.g. a direct product of Yang--Mills and gravitational holonomies might not exist for certain choices of symmetry groups or just work in a common representation. A direct coupling seems to be rather suited if a larger symmetry group is used to describe the coupled degrees of freedom.}. Moreover, a background dependent coupling constant is consistent with our intuition of the Ising model. In both discretisation schemes, we can interpret the spin $j_e$ as governing the distance between two spins on this (dual) edge. If $j_e$ is increased, the distance between the spins should also increase and result in a weaker interaction, hence a smaller effective $\beta$\footnote{Of course, the interpretation of this model is significantly different from the standard idea of the Ising model, where the spins sit in a rigid lattice (or rather crystal).}. Finally, the idea of local coupling constants is a notion discussed for lattice gauge theories on irregular lattices \cite{oeckl-book}. Here the irregularity of the lattice arises from irregular colourings of background spins $j_e$, which can serve as a guideline for the choice of $\beta(j_e)$.

In the sections \ref{sec:length-coupling} and \ref{sec:area-coupling} we will present two different scenarios on how to model the dependence of $\beta$ on $j_e$.

\subsubsection{The `length' coupling} \label{sec:length-coupling}

The first approach is a rather simple one, since it only takes the distance between two spins into account, thus called `length coupling'. The coupling looks as follows:

\begin{equation}
\beta(j_e) := \frac{\beta_0}{j_e + \frac{1}{2}} \quad ,
\end{equation}
where $\beta_0$ is the standard parameter of the Ising model. $j_e + \frac{1}{2}$ is chosen to avoid an ill behaviour for $j_e=0$. Furthermore it is the edge length appearing in the asymptotics of the Ponzano--Regge model \cite{PR,pr-model}, a model for 3D discrete Riemannian quantum gravity. This Ising coupling $\beta(j)$ is large for small spins, actually largest for spin $j=0$. In order to have a direct comparison to the Ising model on an equilateral lattice, we rescale $\beta_0$, such that $\beta'_0 = \frac{1}{2} \beta_0$. Then one obtains the standard Ising coupling constant if all $j=0$.

From the definition of this coupling constant, we observe already a few basic properties. For $\beta_0 \ll 1$, i.e. high temperature, the Ising model is very insensitive to the size of the edge lengths, since all configurations come with (almost) the same strength. Hence, for $\beta_0 = 0$ we recover the background without Ising model on it. However for growing $\beta_0$, we notice that the couplings fall off with roughly $\frac{1}{j_e}$, such that we can expect two things: First, if $j_e > 0$ are excited / allowed, the phase transition of the Ising model might be pushed towards larger $\beta_0$ in comparison to the standard case, since the Ising spins effectively see a weaker coupling. Secondly, for $\beta_0 \gg 1$, $\beta(0) \gg \beta(j) \, \forall j \neq 0$, such that we expect the Ising model to greatly favour $j=0$ configurations and thus might induce a phase transition in the background from `$j>0$ excited' to `only $j=0$ excited'. 

\subsubsection{The `area--length' coupling} \label{sec:area-coupling}

The second option we will investigate is far less obvious, but also interesting. Instead of just assigning weaker and weaker coupling constants $\beta$ to larger spins, we rather want to modify the coupling constant by a dimensionless quantity. Therefore we have to compare the edge length of the edge to another geometric quantity; the only other local option is the (squareroot of the) area the edge is part of. This idea is quite similar to the modification done in 4D \cite{oriti-pfeiffer,mikovic1}, where one considers the ratio of the area squared of a triangle and the 4D volume of a 4--simplex. In 2D, independent of the chosen discretization (or valency of the vertices), every edge is shared by two faces, such that we define:

\begin{equation}
\beta(\{j_e\}):= \beta_0 \frac{\frac{1}{2} \sum_{f\supset e} \sqrt{\text{Ar}(f)}}{j_e + \frac{1}{2}} \quad ,
\end{equation}
where $\text{Ar}(f)$ denotes the area of the face $f$.

First of all we note that, if we restrict our discussion to square lattices, in which all edges have the same length, we recover the standard Ising coupling constant. In the case of a dynamical background, we are required to define the area of the face $f$ as a function of the edge lengths of the face. Instead of constructing a suitable area operator, we will go with a poor man's version: we simply multiply the edge lengths of two adjacent edges, i.e. spanning two orthogonal directions, as if we had rectangular angles at each corner of the tetragon. Of course, these will give very different answers depending on which pair of edge lengths we pick, but this can be improved e.g. by averaging over these four choices.

Clearly, such an area coupling is more non--local than the length coupling previously discussed, actually to a varying degree. E.g. the area averaging renders the edge weight dependent on seven edge labels, which are also subject to additional constraints on the bounding vertices of the squares, in particular (generalized) triangle inequalities. Therefore, to define this coupling information of six vertices of the (background) lattice is required, which is obviously at odds with the local coarse graining scheme\footnote{To disentangle this dependence and write down an amplitude for each vertex of the lattice is non--obvious.}. Instead we simplify this interaction by introducing a more local coupling constant, defined for each pair of edge and vertex:

\begin{equation}
\beta(\{j_e\}) = \frac{\frac{1}{2} \left(\sqrt{j_{\text{left}} + \frac{1}{2}} +\sqrt{j_{\text{right}} + \frac{1}{2}} \right)}{\sqrt{j_e + \frac{1}{2}}} \quad ,
\end{equation}
where $j_{\text{left}}$ and $j_{\text{right}}$ denote the edge labels `left' and `right' of the edge $e$ as seen from the vertex $v$. In short, we simply define the area of the squares locally from the edge lengths of the local vertex. Again, if all edge lengths are equal, we recover the standard Ising coupling constant. Moreover, note that this coupling constant is invariant under a global rescaling of all edge lengths.

In comparison to the length coupling, general statements on the possible behaviour of the composite model are not as obvious. Of course, for $\beta_0 \ll 1$, we recover the uncoupled background, since the Ising model essentially assigns constant weights to all background configurations. However for larger $\beta_0$ no immediate consequences can be read off, since the coupling on the edge is not only determined by the local $j_e$, but also its neighbouring ones. E.g. if the neighbouring spins are larger than $j_e$ the coupling constant will be larger than $\beta_0$ (and thus larger than the configuration for all $j_e=0$). But that statement is not sufficient to determine whether larger spins get suppressed or not, since it only concerns one edge: A configuration that increases the coupling constant for one edge generically assigns a lower coupling constant to the neighbouring ones. Hence the question of possible phases depends on the weights and number of certain configurations, yet a general mechanism for suppressing spins larger than $j=0$ for $\beta_0 \gg 1$ does not seem to exist.

To make this statement more precise, let us consider a configuration, which is allowed by the coupling rules and deviates from the regular square case. For the time being, we ignore the amplitude of the background and just consider the influence of the Ising edge weights. A typical configuration is e.g. twice spin $j=0$ and twice spin $j=j'>0$ on the four edges -- all possible permutations of these spins are allowed by the coupling rules of $\text{SU}(2)_k$. If we then consider the product of all four edge weights, e.g. for alternating spins and all $k_e=0$ normalized by the equilateral case, we obtain:
\begin{align} \label{eq:alternating-weight}
& \frac{\prod_{i=1}^4 \tilde{\omega}_0(j_i,j_{i+1},j_{i-1})}{(\tilde{\omega}_0(0,0,0))^4} = \nonumber \\
& = \frac{\cosh^2\left(\beta_0 \sqrt{\frac{\frac{1}{2}}{j' + \frac{1}{2}}}\right) \cosh^2\left(\beta_0 \sqrt{\frac{j' + \frac{1}{2}}{\frac{1}{2}}}\right) }{\cosh^4(\beta_0)} \quad ,
\end{align}
since the spins $j'$ are `surrounded' by spins $j=0$ and vice versa. Thus we have a hyperbolic cosine squared with an effectively larger coupling and one with an effectively smaller one. Similarly this also occurs for more general cases with non--equal spins $\{j_i\}$ on the four edges of the vertex.

Remarkably, it turns out that the product of four edge weights (for the same $k_e$ configuration) for configurations deviating from the all spins $j$ equal case is larger in general than for equilateral configurations. Thus, the Ising spins effectively see a larger coupling if the lattice spacings are not all equal. The explanation is the exponential growth of the hyperbolic cosine (or sine for $k_e=1$), where the shift to larger $\beta_0$ dominates over the shift to smaller (still positive) $\beta_0$. Therefore, this weight is particularly large if the four spins $j$ around a vertex are very different: The maximal effective coupling occurs if the spins on the edges of the vertex alternate between smallest ($j=0$) and largest spin of the quantum group ($j=j_{\text{max}}$) -- the most `non--equilateral' configuration, see \eqref{eq:alternating-weight} for $j'=j_{\text{max}}$. If the deviation from the equilateral case is very small, e.g. only spins $j$ and $j+1$ are excited with $j \gg 1$, then the effective coupling quickly converges to the equilateral one.

Thus we can expect a very different behaviour from the `area'--coupling than the `length'--coupling: A mechanism to suppress configurations with larger spins $j$, appears to be absent in the `area'--coupling. Preferred configurations are rather characterized non--locally by their (ir)regularity, where the lowest weight is assigned to the equilateral configurations (independent of the spin $j$) and the highest weights are assigned to the most irregular configurations alternating between smallest and largest spins. One implication is that configurations with spins $j \neq 0$ will occur that are equipped with a larger weight than the configuration with all $j=0$. Thus one can expect a shift of the Ising phase transition towards smaller $\beta_0$ if more spins $j$ are allowed, since the coupling is effectively larger. A second implication concerns the background, which might be affected by the preference of irregular configurations\footnote{For $\beta_0 \rightarrow \infty$, this weight for irregular configurations diverges with respect to all $j=0$. Since the background always couples to the trivial representation and is normalized with respect to it, the background probably will not stay unaffected.}.

In order to answer these questions and properly examine the differences between the two couplings of the Ising model to the background, it is imperative to study the dynamics of the system beyond just a single vertex. To do so, we will introduce a coarse graining algorithm, known as tensor network renormalization, in the following section, which will allow us to extract the phase structure of the coupled system.

\section{Coarse graining algorithm -- tensor network renormalization} \label{sec:renormalization}

In section \ref{sec:toy-model} we have thoroughly motivated and introduced a system of Ising spins coupled to a dynamical background as a toy model for lattice gauge theory coupled to discrete quantum gravity, where both theories are regularized by the same (or the dual) lattice. The goal of the remainder of this work is to study the composite dynamics of this novel system. By this we particularly mean the composite phases of the system, i.e. the regions in the parameter space of the theory, in which the system shows the same qualitative behaviour. An example would be the Ising model in the (dis)ordered phase on a specific background, e.g. all spins $j \leq J$ are allowed. Moreover, we intend to investigate the quantitative changes each of the models experiences, e.g. whether the phase transition temperature of the Ising model shifts due to the deviations of the background away from equilateral lattices and whether the background is affected by the Ising model.

In this context, the discretisation plays a crucial role. Unless we discuss topological theories, the results of the theory will depend on the chosen discretisation. In the standard context, this discretisation inherits a length scale from the fixed background and one describes how the dynamics of the system changes as one considers a different scale, which is the standard idea of renormalization. Since a background scale is absent in our situation, we have to relate theories defined on different discretisations instead, e.g. the system on a finer discretisation to a system on a coarser discretisation, via a transformation, here coarse graining, which should not change the partition function and expectation values of observables. We can understand this transformation as a map from finer to coarser degrees of freedom. 

A coarse graining algorithm well suited for studying systems without a direct reference to a length scale is tensor network renormalization \cite{levin,guwen}, originally developed in condensed matter physics. Note that Monte--Carlo simulations are not applicable, since the intertwiner models \cite{invariant_int} are inherently complex; the same also holds for spin foam models. The basic idea of this algorithm is to rewrite the partition function as a contraction of multidimensional arrays, the tensors, which encode all the dynamical information of the system\footnote{For lattice gauge theories, tensor network representations exist as well, see e.g. \cite{abelian_tnw,blocking-lgt}, but are very costly due to storing redundant (gauge) information. In \cite{decorated-tensor} an algorithm for 3D lattice gauge theory has been implemented and tested, in which not all information is stored in the tensor network. The data encoding gauge symmetries are rather used to `decorate' the tensor network, with the gauge conditions solved.}. Frequently the tensors have the same number of indices as the original lattice and are pictorially represented by a vertex with as many legs as indices. An index contracted with another tensor is represented by connecting the two legs of the tensors. Hence the partition function is represented by many tensors connected to one another according to the combinatorics of the lattice --  a network.

The range of the index is frequently referred to as the bond dimension $\chi$. In our case, as we deal with finite groups and quantum groups that come with a natural cut--off on the representation labels, the bond dimension of the (initial) tensor is finite. However a straightforward definition for systems with (compact) Lie groups, e.g. $\text{U}(1)$ or $\text{SU}(2)$, as their underlying symmetry group is not obvious, unless one can introduce a approximative cut--off as in the strong coupling expansion or use a different scheme to write the system with a finite bond dimension \cite{tagz,phi4} or incorporate Monte--Carlo methods for the contraction of indices \cite{vidalMC}.

\begin{figure}
\includegraphics[width=0.4\textwidth]{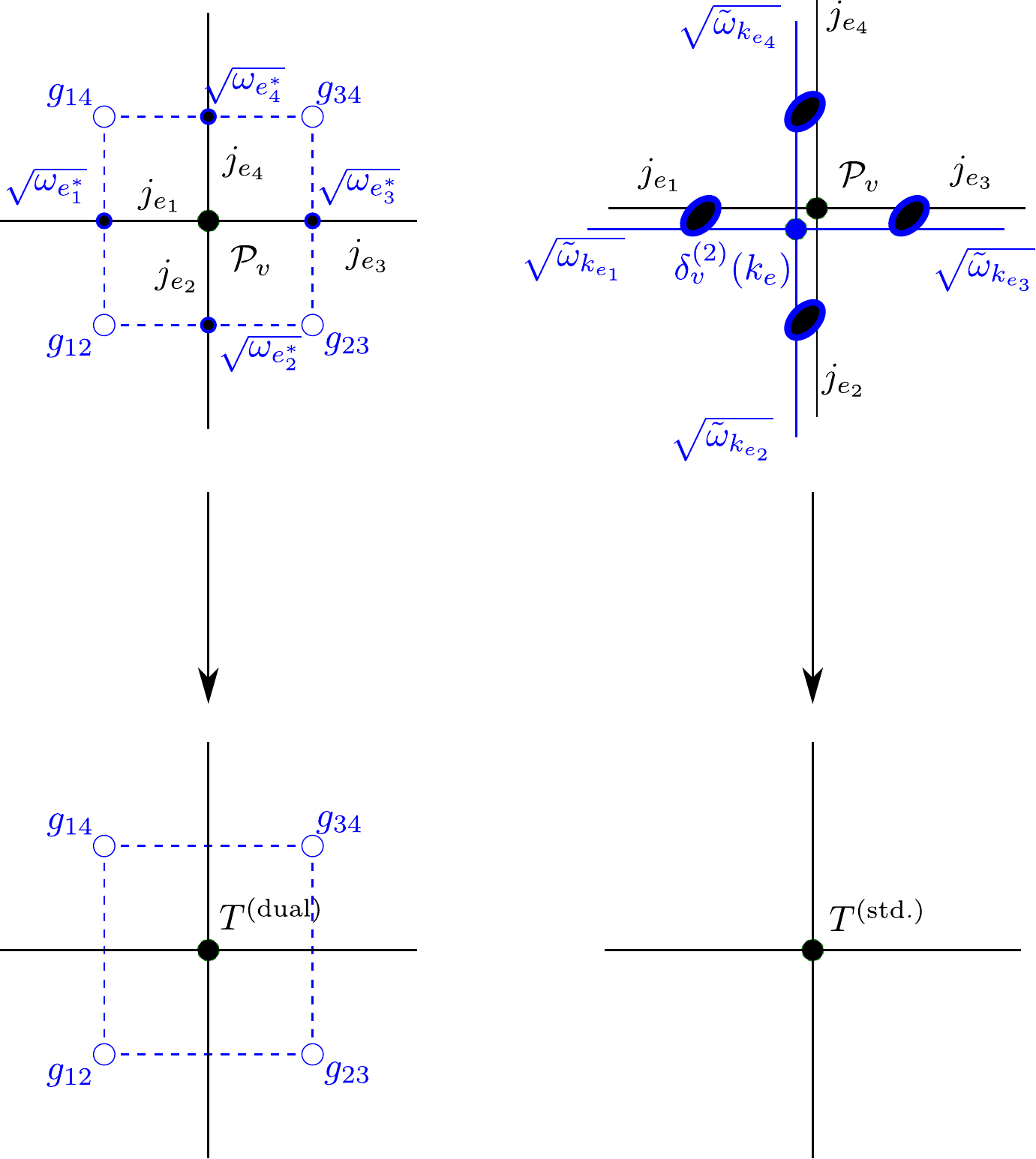}
\caption{The initial amplitudes for the dual and the standard discretisation. The partition function is chopped into local amplitudes, tensors, along the edges. Since the same variables are shared between the separated tensors, only the edge weights, $\omega_{e^*}$ and $\tilde{\omega}_{k_e}$ respectively, are split. Then the amplitudes are combined into tensors, a `decorated' tensor for the dual theory, and a `standard' tensor for the normal one.
\label{fig:local-amplitude}}
\end{figure}

In our concrete example, we chop the partition function into local amplitudes, which we can associate to the vertices $v$ of the lattice. Schematically, we show this in fig. \ref{fig:local-amplitude}. These local amplitudes are summarized as a tensor $T$ on the vertex $v$ of the lattice:
\begin{align}
& T^{(\text{dual})} (\{j_e\},\{g_{v^*}\}) =  \; \nonumber \\
= \;& \mathcal{P}_v (\{j_e\}_{e \supset v}) \; \prod_{e^* \text{dual to } e \supset v} \sqrt{\omega (g_{s(e^*)} g_{t(e^*)},\{j_e\})} \; , \\
& T^{(\text{std.})} (\{j_e\},\{k_e\}) = \; \nonumber \\
= \; & \mathcal{P}_v (\{j_e\}_{e \supset v}) \;  \delta^{(2)}\left(\sum_{e \supset v} (-1)^{o(e,v)} k_e\right) \prod_{e \supset v} \sqrt{\tilde{\omega}_{k_e} (\{j_e\})} \; .
\end{align}
Note that there is a slight difference in the definition of the tensor if we have chosen either the lattice or its dual for the Ising spins. In the former case, the standard construction goes through, since the variables solely live on the edges $e$ and their mutual dependence is encoded on the vertices $v$. In the latter case, the dual, this only applies to the background, however the Ising spins live on the dual vertices $v^*$ with edge weights on the dual $e^*$. In fact, splitting of the edge weights is straightforward, but one has to keep track of the Ising spins, a `decoration' to the tensor network. How to coarse grain such `decorated' tensor networks has been developed in \cite{decorated-tensor} and turns out to be as straightforward as the standard algorithm \cite{levin,guwen}. Due to the self--duality of the Ising model, we nevertheless drop this option and focus on the model, in which both systems live on the same lattice, see also fig. \ref{fig:lattice-vs-dual2}. To simplify the notation, we will from now on drop the superscript.  

To obtain the whole partition function again, the tensors are connected, i.e. the shared indices are summed over, and the partition function is written as the tensor trace of the tensor network:
\begin{equation}
Z = \text{Ttr} \prod_v (T \, T \, \ldots) \quad .
\end{equation}
So far, we have obviously only rewritten the problem into a different form, which is more local than the original one\footnote{Here local means that each tensor is only connected to a nearest neighbour tensor.}. The idea of the algorithm is then to locally transform / coarse grain the tensor network, such that original partition function is approximated by a coarser, still local tensor network. In short, the tensors themselves, and as such the dynamical ingredients of the system, get renormalized.

There exist a plethora of different tensor network algorithms \cite{levin,guwen,corner-matrix,HOSVD,vidal-evenbly}, which more or less only differ in the scheme of how several fine tensors are transformed into a coarse one. Such a transformation can be straightforward, e.g. on a square lattice, one can simply combine four tensors on the corners of a square into one by contracting the inner legs. One straightforwardly arrives at a new tensor, however with a squared index size, pictorially represented by double edges. Continuing with this endeavour, also with respect to inevitable numerical simulations, is fruitless and requires a truncation for two reasons: a practical one, since no computer possesses infinite memory and infinite computational time to contract the indices, and an interpretative one, in order to compare the coarse to the fine tensor. Hence one is required to define new coarse degrees of freedom arising from the fine ones, such that one preserves an interpretation and can simultaneously truncate the number of degrees of freedom with a good control on the error being made.

At this stage, we will not repeat the whole introduction and derivation of tensor network renormalization and instead briefly discuss the triangular algorithm, originally introduced in \cite{decorated-tensor}. The derivation of the formulae is straightforward from the related 4--valent algorithm \cite{levin,guwen,q_spinnet}, which we explain in more detail in appendix \ref{app:tensor}.

\subsection{The triangular algorithm}

%

As the name suggests, the triangular algorithm shifts the perspective from 4--valent tensors to 3--valent ones: Instead of starting with one tensor $T$, we start with four tensors $S_i$. These tensors can be obtained e.g. from the first step of the 4--valent algorithm, in which one splits the 4--valent tensor across both diagonals into two pairs of 3--valent ones, see appendix \ref{app:tensor} for a more detailed explanation.

\begin{figure}
\includegraphics[width=0.45\textwidth]{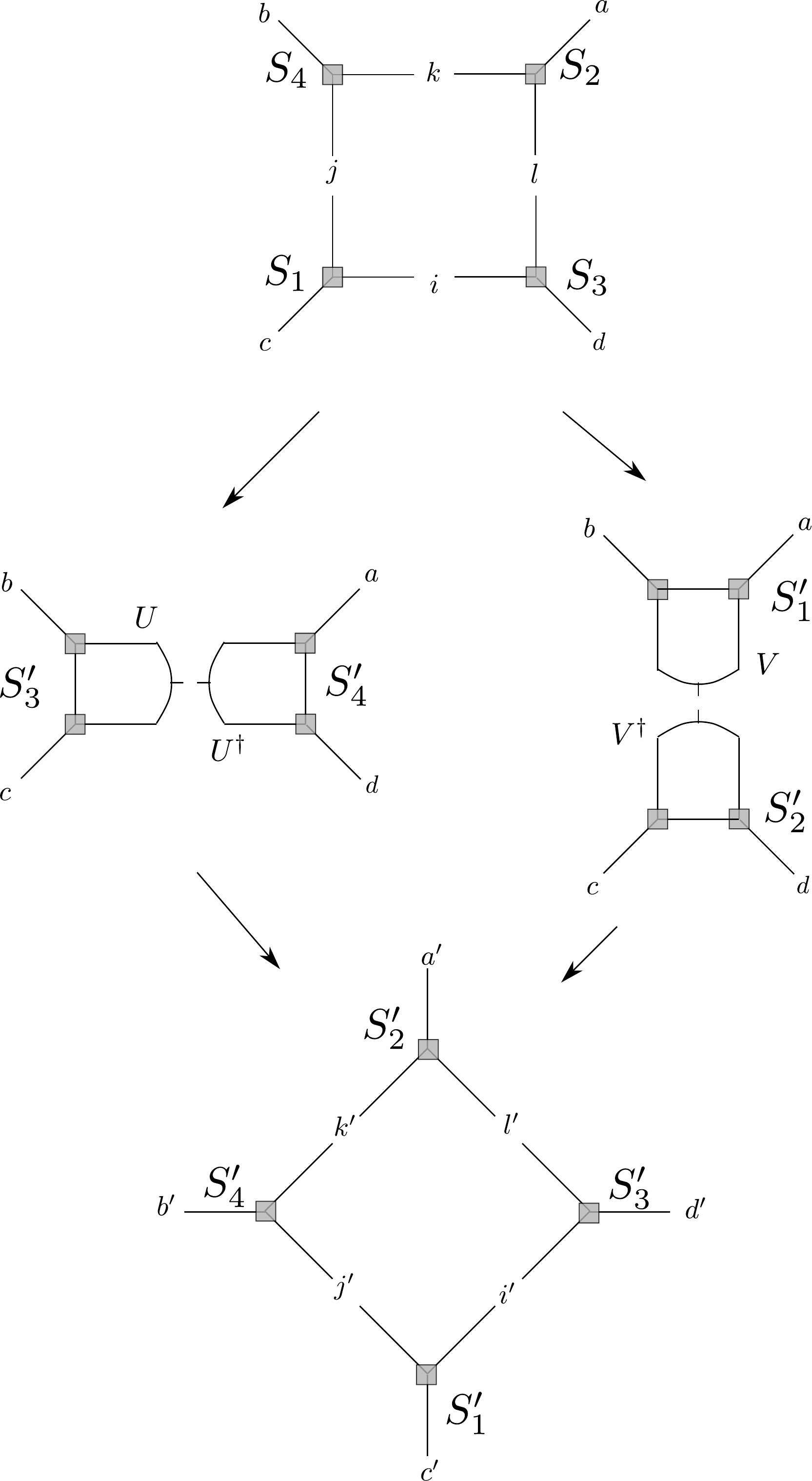}
\caption{The triangular algorithm. Instead of one 4--valent tensor, forming a rectangular tensor network, we work with four 3--valent tensors $S_i$. We glue them pairwise into intermediate 4--valent tensors, compute an embedding map from this tensor, here called $U$ and $V$, which maps two fine edges into one effective coarse edge without affecting the partition function since $U U^\dagger = V V^\dagger = 1$. Thus we obtain four new 3--valent tensors $S'_i$ and iterate the procedure.
\label{fig:tria-algo}}
\end{figure}

The triangular algorithm itself is shown in fig. \ref{fig:tria-algo}: To compute a new 3--valent tensor, two of the $S_i$ are combined pairwise, where the shared edge gets contracted, resulting in a new 4--valent tensor. To arrive again at a 3--valent vertex, the two `parallel' edges in fig. \ref{fig:tria-algo} have to be combined into one, since they represent the `finest' degrees of freedom of the system; for this we have to introduce an embedding map, which is computed via a singular value decomposition (SVD).

Let us discuss this for a more concrete example: Suppose we want to construct the $S'_1$ in fig. \ref{fig:tria-algo} from $S_2$ and $S_4$. We define the intermediate tensor $\tilde{S}_1$ as:
\begin{equation}
(\tilde{S}_1)_{a b; j l} = \sum_k (S_2)_{a; k l} \, (S_4)_{b; j k} \quad .
\end{equation}
We have to define an embedding map for the indices $(j l)$ into a new effective index $c'$. Therefore we perform a SVD on the following matrix:
\begin{equation}
(\tilde{S}_1)_{(a b); (j l)} = \sum_i U_{(a b), i} \, \lambda_i \, V^\dagger_{(j l),i} \quad .
\end{equation}
$U_{(a b),i}$ and $V_{(j l),i}$ are the singular vectors, $\lambda^{(l)}_i$ the singular values of the matrix $(\tilde{S}_1)_{(a b);(j l)}$, where the indices $(a d)$ and $(j l)$ simply denote product indices. $U$ and $V$ are unitary matrices, i.e. the singular vectors are orthonormal, while the singular values are non--negative and ordered in size, $\lambda_1 \geq \lambda_2 \geq ...\geq \lambda_N \geq 0$.

$U$ and $V$ can be interpreted as mapping the indices $(ab)$ and $(jl)$ to $i$ respectively. The (relative) value of $\lambda_i$, e.g. with respect to the largest singular value, determines how important the index $i$ is in reconstructing $\tilde{S}_1$. Thus we use $V$ both as an embedding map, mapping the indices $(j l)$ into a new effective index $i$, and as a truncation, cutting off the index range of the new edge. To do so, we have to multiply it from the right to the tensor $\tilde{S}_1$ and sum over the indices $(j l)$. Then we obtain the new $S'_1$ as:
\begin{equation}
(S'_1)_{c'; a b} = U_{(a b),c'} \lambda_{c'} \quad ,
\end{equation}
where we have used the fact that $V$ is a unitary matrix. However, in order to leave the partition function unchanged, it is necessary to contract the new $S'_1$ on the other side with $V^\dagger$ obtained from $\tilde{S}_1$, as it is also implied in fig. \ref{fig:tria-algo}. Hence the formula for $S'_2$ contains the explicit contraction\footnote{In fact, it is necessary to compute the embedding maps from both $\tilde{S}_1$ and $\tilde{S}_2$ and check, which one of them leads to a smaller overall error and thus a better approximation, similar to the higher order SVD used in \cite{HOSVD}.}. Also note that this algorithm is not as `symmetric' as the 4--valent one: Every 3--valent tensor has a special edge, namely the one obtained after applying the latest embedding map.

Of course, similar to the 4--valent algorithms employed in \cite{abelian_tnw,abelian_tnw2,s3_tnw,q_spinnet}, it is beneficial to explicitly preserve the underlying symmetries of the tensors under coarse graining because of saving of computational resources and keeping an interpretation of the new coarse degrees of freedom in terms of the original variables. Thanks to representation theory, we know that some couplings are forbidden, e.g. for the Ising model if the Gauss constraints are not satisfied. To exploit this as much as possible, one writes the matrices that are to be split in a block diagonal form; we call this the recoupling basis. In the non--Abelian or quantum group case, this amounts to a precontraction of magnetic indices, as the dependence on those is not changed under coarse graining. Each block is then labelled by a set of representation labels, e.g. $(j,k)$ for our toy model, called the `intertwiner channel'. Then instead of performing the SVD on the entire matrix at once, one performs one SVD per block matrix, such that one endows the block label onto the new effective edge.

For the triangular algorithm, this symmetry protecting algorithm can be straightforwardly derived from its 4--valent counterpart. Therefore, we here only present the formulae, the actual derivation of the 4--valent version is put in appendix \ref{app:tensor}. 

Let us start with the block diagonal form of the 3--valent tensor, which is similar to equations \eqref{eq:S1} to \eqref{eq:S4} of the 4--valent algorithm:
\begin{widetext}
\begin{align} \label{eq:S1-3}
(S_1)^{j_5,k_5}_{m_5}(\{I\}_{\{1,2\}},\{m\}_{\{1,2\}})= &(\hat{S}_1)^{(j_5,k_5)}(\{I\}_{\{1,2\}}) \, \delta^{(2)}(k_1 + k_2 - k_5)
\begin{tikzpicture}[baseline,scale=0.75]
\draw  (0,-0.75) -- (0,0) -- (-0.5,0.5)
      (0,0) -- (0.5,0.5)
         (0.5,0.75) node {$j_1$}
      (-0.5,0.75) node {$j_2$}
      (0.25,-0.5) node {$j_5$};
\end{tikzpicture}
\quad ,\\ \label{eq:S2-3}
(S_2)^{j_5,k_5}_{m_5}(\{I\}_{\{3,4\}},\{m\}_{\{3,4\}})= & (\hat{S}_2)^{(j_5,k_5)}(\{I\}_{\{3,4\}}) \, \delta^{(2)}(k_3 + k_4 - k_5)
\begin{tikzpicture}[baseline,scale=0.75]
\draw (-0.5,-0.5) -- (0,0) -- (0,0.75)
      (0.5,-0.5) -- (0,0)
      (-0.5,-0.75) node {$j_3$}
      (0.5,-0.75) node {$j_4$}
      (0.25,0.5) node {$j_5$};
\end{tikzpicture}
\quad ,\\ \label{eq:S3-3}
(S_3)^{j_6,k_6}_{m_6}(\{I\}_{\{2,3\}},\{m\}_{\{2,3\}})= & (\hat{S}_3)^{(j_6,k_6)}(\{I\}_{\{2,3\}}) \, \delta^{(2)}(k_2 + k_3 - k_6)
\begin{tikzpicture}[baseline,scale=0.75]
\draw (-1.25,0.5) -- (-0.75,0) --(0,-0.3)
      (-0.75,0) -- (-1.25,-0.5)
      (-1,-0.75) node {$j_3$}
      (-1,0.75) node {$j_2$}
      (-0.25,0.25) node {$j_6$};
\end{tikzpicture} 
\quad ,\\ \label{eq:S4-3}
(S_4)^{j_6,k_6}(\{I\}_{\{4,1\}},\{m\}_{\{4,1\}})= &(\hat{S}_4)^{(j_6,k_6)}(\{I\}_{\{4,1\}})\; \delta^{(2)}(k_4 + k_1 - k_6)
\begin{tikzpicture}[baseline,scale=0.75]
\draw (-0.75,0.25) -- (0,0) -- (0.5,0.5)
      (0,0) -- (0.5,-0.5)
      (0.5,0.75) node {$j_1$}
      (0.5,-0.75) node {$j_4$}
      (-0.5,-0.1) node {$j_6$};
\end{tikzpicture}
\quad ,
\end{align}
\end{widetext}
where we have summarized the representation labels $j_e$, $k_e$ as $\{I_e\}$. The diagrams encode the dependence on the magnetic indices, which remains unchanged under coarse graining due to preserving the symmetries.

Analogue to the 4--valent algorithm, the next step is the computation of the intermediate 4--valent tensor in its block--diagonal form to which the SVD will be applied. Remarkably, this expression can almost immediately be read off equations \eqref{eq:new-effective-tensor} and \eqref{eq:new-effective-tensor2}, together with the fact that identity \eqref{eq:9j-symbol} splits into two $6j$ symbols. This allows us to essentially cut equations \eqref{eq:new-effective-tensor} and \eqref{eq:new-effective-tensor2} in half to obtain the equations for $\hat{\tilde{S}}_1$ and $\hat{\tilde{S}}_2$:
\begin{widetext}
\begin{align} \label{eq:intermediate-S1}
\hat{\tilde{S}}_1^{(j_5,k_5)}(I_1,I_2;I_c,I_a)&=\sum_{b}\sum_{\{m\}}
\frac{\sqrt{(-1)^{j_c + j_a + j_5}}} {\sqrt{d_{j_5}}\sqrt{d_{j_b}}}
\sqrt{ d_{j_1}d_{j_2}} \; \left[
\begin{matrix}
\, j_c \, & \, j_a \,  & \, j_5 \, \\
\, j_1 \, & \, j_2 \, & \, j_b \,
\end{matrix}
\right]  \delta^{(2)}(k_1 + k_2 - k_5)
 \nonumber \\
& \times
(\hat{S}_2)^{(j_1,k_1)}(\{I\}_{\{b,a\}})
(\hat{S}_4)^{(j_2,k_2)}(\{I\}_{\{c,b\}})
\quad , \\ \label{eq:intermediate-S2}
\hat{\tilde{S}}_2^{(j_5,k_5)}(I_3,I_4;I_a,I_c)&=\sum_{d}\sum_{\{m\}}
\frac{\sqrt{(-1)^{j_c + j_a + j_5}}} {\sqrt{d_{j_5}}\sqrt{d_{j_d}}}
\sqrt{d_{j_3}d_{j_4}} \; \left[
\begin{matrix}
\, j_c \, & \, j_a \,  & \, j_5 \, \\
\, j_4 \, & \, j_3 \, & \, j_d \,
\end{matrix}
\right]\delta^{(2)}(k_3 + k_4 - k_5)\nonumber\\
&\times
(\hat{S}_1)^{(j_3,k_3)}(\{I\}_{\{d,c\}})
(\hat{S}_3)^{(j_4,k_4)}(\{I\}_{\{a,d\}})
\quad .
\end{align}
Analogously, one obtains the equations for $\hat{\tilde{S}}_3$ and $\hat{\tilde{S}}_4$ from the equation of $\hat{T}_2$, but essentially it is merely a suitable permutation of labels in the equation. Eventually, we can compute the embedding maps by performing a SVD on $\hat{\tilde{S}}_i$, e.g. on $\hat{\tilde{S}}_1$:
\begin{equation}
\left(\hat{\tilde{S}}_1^{(j_5,k_5)}\right)_{(I_1,I_2);(I_c,I_a)} = \sum_{i_1} U^{(j_5,k_5)}_{(I_1,I_2);i_1} \; \lambda^{(j_5,k_5)}_{i_1} \; (V^\dagger)^{(j_5,k_5)}_{(I_c,I_a),i_1} \quad , 
\end{equation}
where $i_1$ denotes the multiplicity of the channel $(j_5,k_5)$ in this particular example. If this is the proper embedding map, i.e. the one with least error approximating both for $\tilde{\hat{S}}_1$ and $\tilde{\hat{S}}_2$, then we eventually obtain for the new $\hat{S'}_1$\footnote{We assume that the embedding maps obtained from $\tilde{\hat{S}}_1$ is the one giving the smallest error, therefore we also suppress the additional superscript $(1)$.}:
\begin{equation} \label{eq:simple-contraction}
(\hat{S'}_1)^{(j_5,k_5,i_1)}(\{I\}_{\{1,2\}}) = U^{(j_5,k_5)}_{(I_1,I_2);i_1} \lambda^{(j_5,k_5)}_{i_1} \quad ,
\end{equation}
whereas we find for $\hat{S}_2$:
\begin{equation}
(\hat{S'}_2)^{(j_5,k_5,i_1)}(\{I\}_{\{3,4\}}) = \sum_{a,c} \hat{\tilde{S}}_2^{(j_5,k_5)}(I_3,I_4;I_a,I_c) (V^\dagger)^{(j_5,k_5)}_{(I_c,I_a),i_1} \quad.
\end{equation}
\end{widetext}
The derivation of $\hat{S'}_3$ and $\hat{S'}_4$ works analogously.

Fortunately, in the models under discussion here, all the $\hat{S}_i$ turn out to be identical, also under coarse graining because the same recurrence relation holds for all of them\footnote{This is due to the choice of splitting of the signs in \eqref{eq:new-effective-tensor2} symmetrically in \eqref{eq:intermediate-S1} and \eqref{eq:intermediate-S2}. In principle, a different splitting is possible, e.g. assigning the sign just to one of them. However, we have found that our choice is numerically more stable and consistent with results of the 4--valent algorithm.}. As a result, it will be sufficient from now on to just work with one 3--valent tensor instead of four, which simplifies equation \eqref{eq:intermediate-S1}, but more importantly always allows us to use \eqref{eq:simple-contraction}. Thus we greatly reduce the amount of data necessary to be stored during the numerical simulations and additionally avoid the explicit contraction of (roughly) $2\times 2 \times j_{\text{max}}$ indices. Here this might appear to be insignificant, but as soon as more complicated systems are examined, for example $\text{SU}(2)_k \times \text{SU}(2)_k$ spin nets, which roughly require quartic amount of resources in comparison to the background discussed here, these gains are crucial.

\subsubsection{Truncation scheme and identification of phases} \label{sec:truncation}

For our study of the presented toy model, we will use a very simple truncation scheme for the singular values. Instead of comparing the singular values obtained from all blocks and keeping the $\chi$ largest of these, we take only one singular value for each block $(j,k)$. This certainly is a drastic simplification, but it seems sufficient to capture many important features of the model: Taking one singular value for each block is close to the initial definition of the models, in particular we can capture the fixed points of both the background and the Ising model and thus can qualitatively study the phase diagram of the model. Certainly, we are aware that this approximation breaks down in certain regimes of the model, e.g. close to the phase transition of the Ising model, where more singular values have to be taken into account, which results in shifts of the position of the phase transitions. However, numerical precision is a rather secondary concern, since we are dealing with a toy model tailored to demonstrate the potentially rich dynamics of matter coupled to spin foams. Moreover, as many other tensor network renormalization methods, our scheme requires an infinite bond dimension to study (second order) phase transitions. A recently developed algorithm \cite{vidal-evenbly} allows the study of phase transitions at relatively low bond dimension --  we leave an adaptation of our current algorithm for future research.

Fortunately, this truncation scheme allows for a straightforward identification of the phases of the model via the singular values in the intertwiner channels $(j,k)$, $j$ for the background, $k$ for the Ising model. These can then be summarized in a $(j_{\text{max}}+1) \times 2$ matrix. E.g. the topological background given by $J$ (see again section \ref{sec:grav-part}) is given if all intertwiner channels $j \leq J$ have one non--vanishing singular value equal to one and all channels $j>J$ have only vanishing singular values. Hence the singular values directly indicate which spins $j$ are allowed. Similarly for the Ising model, the disordered phase is given if only the channel $k=0$ has one non--vanishing singular value equal to one, the ordered phase if both $k=0$ and $k=1$ have one non--vanishing singular value equal to one. Let us conclude with an example for the combined model: If only the channels $(j,k)= (0,0), \, (1,0), \, (2,0), \, (0,1), \, (1,1), \, (2,1)$ possess one singular value equal to one while all others vanish, this indicates the background model $J=2$ and the ordered Ising phase.

In the sections \ref{sec:top-background} and \ref{sec:superposition} we will discuss the results obtained from coarse graining the Ising model coupled to the dynamical background. In the next section \ref{sec:top-background} we first focus on the influence of the background on the Ising model by choosing a topological fixed point for the background.

\section{The Ising model coupled to a topological background} \label{sec:top-background}

In the remainder of this paper, we will present the results of applying the coarse graining algorithm introduced in section \ref{sec:renormalization} to the model of Ising spins coupled to a dynamical background discussed in section \ref{sec:coupled-system}. As we have explained in section \ref{sec:grav-part}, there exist many different choices for the background, encoded in the factors $a(\{j\})$ modifying the Clebsch--Gordan coefficients of the quantum group. Recall that these factors are chosen such that the local amplitudes are invariant under all 2D Pachner moves, such that they describe topological systems. Additionally, one can also consider linear combinations of the factors $a(\{j\})$ to go beyond topological backgrounds.

To order the results pedagogically, we start by discussing the cases in which the background is given by a topological theory, that is only one $a(\{j\})$ is chosen instead of a superposition, which will be addressed in section \ref{sec:superposition}. Still, this ansatz allows for an interesting interpretation. Since the background by itself is topological, it is also a fixed point of the coarse graining algorithm: usually for small deviations around this fixed point, the system flows back to this fixed point. Here, the Ising model at $\beta_0 > 0$ serves as a deviation, such that the composite system is not a fixed point and actually flows under renormalization. Hence, it is foremost interesting to discuss how this new background geometry affects the Ising model, in particular the change of its phase transition temperature, e.g. if larger edge lengths are allowed. For larger $\beta_0$ we can also expect changes in the background.

Let us introduce the initial tensors for the start of the coarse graining procedure. The 4--valent tensor $T$ is of the general form (see also appendix \ref{app:tensor}, e.g. \eqref{eq:form-of-tensor}):
\begin{widetext}
\begin{equation}
T (\{j_i\},\{m_i\},\{k_i\}) = \sum_{j_5} {\hat{T}_1}^{(j_5,k_5)}(\{j_i\},\{k_i\}) \delta^{(2)}(k_1 + k_2 - k_5) \delta^{(2)}(k_3 + k_4 - k_5)
\begin{tikzpicture}[baseline,scale=0.75]
\draw (-0.5,-1) -- (0,-0.5) -- (0,0.5) -- (-0.5,1)
      (0.5,-1) -- (0,-0.5)
      (0,0.5) -- (0.5,1)
      (-0.5,-1.25) node {$j_3$}
      (0.5,-1.25) node {$j_4$}
      (0.5,1.25) node {$j_1$}
      (-0.5,1.25) node {$j_2$}
      (0.25,0) node {$j_5$};
\end{tikzpicture}
 \quad .
\end{equation}
The block diagonal part $\hat{T}_1$ is given as follows:
\begin{equation} \label{eq:intial-amplitude}
\hat{T}^{(j_5,k_5}_1(\{j_i\},\{k_i\}) = \frac{a(j_1,j_2,j_5)a(j_3,j_4,j_5)}{d_{j_5}} \sqrt{\tilde{\omega}_{k_1}(j_1;\{j_e\}) \tilde{\omega}_{k_2}(j_2;\{j_e\}) \tilde{\omega}_{k_3}(j_3;\{j_e\}) \tilde{\omega}_{k_4}(j_4;\{j_e\})} \quad ,
\end{equation}
where the argument of $\tilde{\omega}$ indicates that the Ising edge weights can also depend on neighbouring spins $j_e$, e.g. in the `area coupling'. To obtain from \eqref{eq:intial-amplitude} the amplitudes for the triangular algorithm is straightforward. In case the $\tilde{\omega}_{k_e}$ is only of function of $j_e$, \eqref{eq:intial-amplitude} factorizes and can be trivially split; if it is a more non--local function, one has to apply a SVD on \eqref{eq:intial-amplitude} completely analogous to the 4--valent algorithm described in appendix \ref{app:tensor}. The 3--valent tensors obtained from that give the initial amplitudes.
\end{widetext}

In the following, we will summarize the results first for the `length coupling', followed by the `area coupling'. We discuss the choices of topological fixed points for the background presented in section \ref{sec:grav-part} for different levels $k$ of the quantum group $\text{SU}(2)_k$.

\subsection{`Length coupling' -- Results}

Let us briefly recap the length coupling introduced in section \ref{sec:length-coupling}. A dependence of the Ising edge weight $\tilde{\omega}_{k_e}$ is implemented by modifying the coupling constant $\beta \sim \frac{\beta_0}{j_e}$, that is the larger the distance between two spins, the weaker their interaction.

From this coupling mechanism, we already conjectured that for $\beta_0 > 1$ spins $j \geq 0$ will get suppressed by the Ising edge weights and there should exist a phase for which the Ising model is in its ordered phase (low temperature) on a background with all $j=0$. Conversely, in the limit $\beta_0 \rightarrow 0$, i.e. high temperature, we reobtain the pure background, such that we conjecture that there exists a phase with the disordered Ising model on the initially chosen topological background.

{\bf All spins $J \leq j_{\text{max}}$ allowed:}
The largest class of topological fixed points we are considering for the background are those which allow all spins $j \leq J$, where $J \in \mathbb{N}$ and $J \leq j_{\text{max}} = \frac{k}{2}$. That is for larger level $k$, we can consider more models. The results are summarized in tables \ref{tab:length1} and \ref{tab:length2}, since we find two phase transitions for almost all backgrounds:

\begin{table*}
\begin{center}
\begin{tabular}{| c || c | c | c | c | c | c | c |}
\hline
Transition in $\beta$--interval & $k=4$ & $k=5$ & $k=6$ & $k=7$ & $k=8$ \\
 \hline \hline
$J=0$ & $[0.388,0.389]$ & $[0.388,0.389]$ & $[0.388,0.389]$ & $[0.388,0.389]$ & $[0.388,0.389]$ \\
\hline
$J=1$ & $[0.694,0.695]$ & $[0.716,0.717]$ & $[0.729,0.73]$  & $[0.738,0.739]$ & $[0.744,0.745]$ \\
\hline
$J=2$ & $[0.821,0.822]$ & $[0.913,0.914]$ & $[0.969,0.97]$  & $[1.006,1.007]$  & $[1.032,1.033]$ \\
\hline
$J=3$ & n.a.            & n.a.            & $[1.061,1.062]$ & $[1.158,1.159]$ & $[1.224,1.225]$ \\
\hline
$J=4$ & n.a.            & n.a.            & n.a.            & n.a.            & $[1.295,1.296]$ \\
\hline
\end{tabular}
\caption{ Interval for parameter $\beta$ marking the phase transition of the Ising model from the disordered to the ordered phase in the `length' coupling. The background model, labelled by $J$, does not change.
\label{tab:length1}}
\end{center}
\end{table*}

\begin{table*}
\begin{center}
\begin{tabular}{| c || c | c | c | c | c | c | c |}
\hline
Transition in $\beta$--interval & $k=4$ & $k=5$ & $k=6$ & $k=7$ & $k=8$\\
 \hline \hline
$J=1$ & $[1.658,1.659]$ & $[1.693,1.694]$ & $[1.715,1.716]$ & $[1.731,1.732]$ & $[1.742,1.743]$ \\
\hline
$J=2$ & $[1.69,1.691]$  & $[1.761,1.762]$ & $[1.81,1.811]$  & $[1.845,1.846]$ & $[1.87,1.871]$  \\
\hline
$J=3$ & n.a.            & n.a.            & $[1.838,1.839]$ & $[1.898,1.899]$ & $[1.943,1.944]$  \\
\hline
$J=4$ & n.a.            & n.a.            & n.a.            & n.a.            & $[1.965,1.966]$ \\
\hline
\end{tabular}
\caption{ Interval for parameter $\beta$ marking the phase transition of the background from $j_e \leq J$ to $j_e=0$ in the `length' coupling. The Ising model stays in the ordered phase.
\label{tab:length2}}
\end{center}
\end{table*}
First of all, we can verify both previously made conjectures: For $\beta_0 \ll 1$ there exists an extended phase, in which the geometry is given by the chosen topological background (all $j \leq J$ allowed) and the Ising model is found in its disordered / high temperature phase. For $\beta \gg 1$ we find another extended phase, in which the background is restricted to $j=0$ and the Ising model is in its ordered phase. However, these two phases are not directly separated by a phase transition (of geometric and Ising degrees of freedom). In between there exists a phase parametrized by the ordered Ising model on the initially chosen topological background. Hence we observe two phase transitions, the first (for rising $\beta_0$) is a transition purely of the Ising model on the topological background, whereas the second is a transition purely of the geometry, triggered by the particular coupling (via edge lengths alone) of the Ising model to the background.

Besides the appearance of three different phases of the composite system, it is interesting to study the influence of the background system on the position of the phase transitions. The case $J=0$, essentially the Ising model on a regular square lattice, is our reference point. Due to the approximation discussed in section \ref{sec:truncation} of the used algorithm, we find the critical temperature at $0.388 < \beta_0 < 0.389$ instead of $\beta_{\text{crit}} \approx 0.4406...$. This also implies that the other phase transition locations are subject to change if the accuracy is increased. Hence we consider a very precise determination of the phase transition temperatures (in this approximation) to be a fruitless endeavour and rather interpret the findings more qualitatively.

Let us consider the phase transition of the Ising model (see table \ref{tab:length1}) first: We can study this data from two perspectives. Either we consider a particular background $J$ and study the change of the phase transition if we increase the level $k$ of the quantum group (a row) or we examine how the temperature changes for the same level $k$, while increasing the maximum allowed spin $J$ (a column).

In the former case, we observe that for a given $J$, the phase transition parameter increases for growing $k$, yet this increase is not uniform for increasing $k$. The increase is generically largest between $J=j_{\text{max}}$ and the next largest level $k$, and appears to decrease under increasing $k$ further, such that it might eventually converge. The interpretation of this behaviour is that, for the same underlying geometry, the Ising model effectively has a weaker coupling $\beta$, such that its phase transition occurs for larger $\beta_0$, even though the edge weights are not changed in different quantum groups. Hence the origin of this behaviour must lie in the background geometry: Take e.g. the model $J=2$ for $k=4$, i.e. $J= j_{\text{max}} =2$. From the coupling rules / Clebsch--Gordan coefficients, one can determine that there exist only three possible ways how two spins $j_1, j_2$ can couple to a third spin $j=2$, namely $(0,2)$, $(1,1)$ and $(2,0)$ -- $j=2$ here is of quantum dimension one. If we go to $k=5$, there are then five ways, namely also $(1,2)$ and $(2,1)$ in addition to the previous ones. As a result, there exist now more configurations in the partition function with larger edge lengths and hence an effectively lower temperature $\beta$ of the Ising model, pushing the phase transition towards higher $\beta_0$. If the quantum group level $k$ is increased further, even more couplings are allowed, however these involve spins $j>J$, which are disallowed by the geometry (or only excited briefly under coarse graining), such that the phase transition temperature is less affected.

On the other hand, if we keep $k$ fixed and increase $J$, we also observe an increase in the phase transition parameter $\beta_0$, where the gap between $J$ and $J+1$ decreases as one approaches $J = j_{\text{max}}$ of the quantum group, the jump from $J=0$ to $J=1$ being the most significant. This is a very much expected result considering that $\beta(j)$ falls off with $1/j$ in the edge weights of the Ising model and the fact that increasing $J$ allows not only larger spins $j$, but also more configurations of the geometry with non--vanishing $j$ on it. The latter effect can be seen e.g. for $k=8$: even though only for $J=4$ the largest edge lengths get excited, the increase in $\beta_{\text{crit.}}$ is rather small. Since $j=4$ is of quantum dimension one, only few new couplings become allowed with respect to the $J=3$ model. Thus we see that the weakening of the Ising coupling is not solely determined by the (larger) edge lengths but also the amount of configurations, determined by the background, that allow these edge lengths to occur.

For the second phase transition (see table \ref{tab:length2}), that is a geometric phase transition from a topological fixed point to a geometry with only $j=0$ allowed (with the Ising model in its ordered phase in both cases), we perform a similar analysis: If we keep $J$ of the background fixed and increase the level $k$ of the quantum group, we again observe an increase in the transition parameter $\beta_0$. The reason is similar to the previous case: If the level of the quantum group is increased, more couplings (of non--vanishing spins $j$) become allowed increasing the number of configurations with non--vanishing spins. Thus $\beta_0$ has to be increased further for larger $k$ to sufficiently suppress spins $j > 0$ such that the system flows to a geometry with all $j=0$. A similar argument is valid in case of keeping $k$ fixed and increasing $J$, where we also increase the number of configurations with non--vanishing $j$.

This second phase transition is of course a very peculiar example of how the matter system influences the background. If the coupling $\beta_0$ of Ising spins becomes too strong, it forces the background to only allow its shortest possible distance between two vertices. This is a feature one would rather like to avoid for 4D gravity plus matter, such that one can see this geometric transition as an indication that $\beta_0$ should be restricted to smaller values or that the coupling $\beta(j)$ should be modified to favour spins $j>0$. 

{\bf Only $j=0$ and $j=j_{\text{max}}$ allowed:}
The next class of models we tested, is the background in which only representations with quantum dimension $d_j=1$ are allowed, that is only $j=0$ and $j=j_{\text{max}}$ for even levels $k$ of the quantum group. Again, we observe two phase transitions; their locations for various $k$ is given in table \ref{tab:0-max-length}.

\begin{table*}
\begin{center}
\begin{tabular}{| c || c | c | c | c |}
\hline
Transition in $\beta$--interval & $k=4$ & $k=6$ & $k=8$ 
 \\
 \hline \hline
Ising `disordered' $\rightarrow$ `ordered' & $[0.639,0.64]$ & $[0.666,0.667]$ & $[0.681,0.682]$
 \\
\hline \hline
$(j=0;j=j_{\text{max}})$ $\rightarrow$ $j=0$ & $[1.25,1.251]$ & $[1.175,1.176]$ & $[1.139,1.14]$ 
\\
\hline
\end{tabular}
\caption{Results for the background with only $j=0$ and $j=j_{\text{max}}$ allowed for even $k$ together with the Ising model in the `length' coupling. The first row is the first transition we find, a pure transition in the Ising model form `disordered' to `ordered' as $\beta_0$ is increased. The second transition is geometric, where for large $\beta_0$ just $j=0$ is allowed.
\label{tab:0-max-length}}
\end{center}
\end{table*}

The phase transitions are identical in type as for the previous background. On the one hand we find one transition at lower $\beta_0$, which is a pure transition of the Ising model on the background, from the disordered to the ordered phase. On the other hand we observe again the background transition induced by the length coupling, which forces the background to take its smallest edge length possible at larger $\beta_0$.

Concerning the Ising transition, the actual position of the phase transition is quite different from the $j \leq J$ backgrounds. In general, the phase transition can be found at lower $\beta_0$ (unless only $j=0$ is allowed) and increases only slowly as $k$ is increased. This is due to the peculiarity of this model: for even $k$, both $j=0$ and $j=j_{\text{max}}$ have quantum dimension $d_j=1$; the only allowed non--trivial coupling rule is $(j_{\text{max}},j_{\text{max}}) \rightarrow 0$. Thus, even though the Ising model is added in a non--trivial way, the background cannot flow away from this general configuration, i.e. no other spins $j$ can get excited during the renormalization procedure. Therefore, there are much fewer configurations of the background with non--vanishing spins, even in contrast to the lowest $j\leq J$ models. Moreover, this remains unchanged as $k$ is increased, yet the interaction between neighbouring Ising spins on an edge carrying $j_{\text{max}}$ is weaker for larger $k$, pushing the transition to larger $\beta_0$.

For the geometric transition to a background where only $j=0$ is allowed, we find a different situation with respect to the $j \leq J$ models. Instead of increasing $\beta_0$ for growing $J$ or $k$, we observe a decrease. Again, this is due to this peculiar model, in which no other spins except $j=0$ and $j=j_{\text{max}}$ are (and stay) excited. Since $j_{\text{max}}$ grows as $k$ gets increased, the edge weights for $j_{\text{max}}$ fall off faster with respect to $j=0$ for larger $k$, resulting in a transition to `just $j=0$' at lower $\beta_0$.

{\bf Only $j=0$, $j=\frac{j_{\text{max}}}{2}$ and $j=j_{\text{max}}$ allowed:}
The last background model to investigate together with the length coupling is the background with $j=0$, $j= \frac{j_{\text{max}}}{2}$ and $j=j_{\text{max}}$ excited. Since such a background can only exist if the level $k$ of the quantum group is a multiple of four (this is necessary, but not sufficient, see also section \ref{sec:grav-part}), we study it for only three cases. The results are summarized in table \ref{tab:0-half-max-length}:

\begin{table*}
\begin{center}
\begin{tabular}{| c || c | c | c |}
\hline
Transition in $\beta$--interval & $k=4$ & $k=8$ 
\\
 \hline \hline
Ising `disordered' $\rightarrow$ `ordered' & $[0.821,0.822]$ & $[1.093,1.094]$ 
 \\
\hline \hline
$(j=0;j=\frac{j_{\text{max}}}{2};j=j_{\text{max}})$ $\rightarrow$ $j=0$ & $[1.69,1.691]$ & $[1.576,1.577]$ 
 \\
\hline
\end{tabular}
\caption{Results for the background with only $j=0$, $j=\frac{j_{\text{max}}}{2}$ and $j=j_{\text{max}}$ allowed for $k=4l$, $l \in \mathbb{N}$, together with the Ising model in the `length' coupling. The first row is the first transition we find, a pure transition in the Ising model from `disordered' to `ordered' as $\beta_0$ is increased. The second transition is geometric, where for large $\beta_0$ just $j=0$ is allowed.
\label{tab:0-half-max-length}}
\end{center}
\end{table*}

As we have also observed for the other backgrounds, there are two phase transitions in the length coupling: the first one of the Ising model from `disordered' to `ordered', the second one in the geometry from the background to the background with only $j=0$. Across the different quantum groups, we observe that the Ising transition occurs at larger $\beta_0$, since the spins $j=\frac{j_{\text{max}}}{2}$ and $j=j_{\text{max}}$ increase for larger $k$, while the couplings among the spins $j$ (and thus the background configurations) remain unchanged. This results again in an effectively weaker interaction between the Ising spins as $k$ is increased. For the second transition, we see a decrease in the critical $\beta_0$ from $k=4 \rightarrow k=8$, similar to the $(0,j_{\text{max}})$ background.

More interesting is the direct comparison across the backgrounds (for the same quantum group) because we can directly study the changes that more excited spins have on the positions of phase transitions. Note that we only compare the results of $k=4$ and $k=8$.

Let us start with the pure Ising transition, while comparing the models $(0,j_{\text{max}})$ and $(0,\frac{j_{\text{max}}}{2},j_{\text{max}})$. For the latter, we generically observe the transition at larger $\beta_0$, which is clear, since more configurations with weaker Ising spin interactions (with respect to $j=0$) are allowed. The relation to the $j\leq J$ models is less obvious: For $k=4$ and $J=1$, $j_{\text{max}}$ is not excited, such that edges with the weakest Ising spin interactions are forbidden, yet the transition still occurs for a larger $\beta_0$ with respect to the $(0,j_{\text{max}})$ model. This is the case because the $J=1$ model permits more configurations with non--vanishing spins, giving less weight to the `only $j=0$' case. The $J=2$ model is (up to signs) identical to the $(0,\frac{j_{\text{max}}}{2},j_{\text{max}})$ model, both show consistent results. Since they allow all configurations of the background allowed by the quantum group, they show the largest phase transition parameters for $k=4$. For $k=8$ we observe a similar behaviour across the different backgrounds: despite the small weight assigned to edges carrying the maximal spin $j_{\text{max}}$, the transition occurs first for the model $(0,j_{\text{max}})$ followed by the $j\leq J$ models for $J=1$ and $J=2$. Remarkably, the latter is already very close to the transition in the $(0,\frac{j_{\text{max}}}{2},j_{\text{max}})$ model, $\beta_0$ gets increased further for $J>2$. Again we conclude that the amount of background configurations with non--vanishing spins $j$ quickly outweighs the weakening of Ising spin interactions by large spins $j$.

The observation is even more striking for the second transition towards background geometries, where only the shortest edge length $j=0$ is allowed: Both for $k=4$ and $k=8$ this transition occurs always first for the $(0,j_{\text{max}})$ model followed by the $(0,\frac{j_{\text{max}}}{2},j_{\text{max}})$ model, which coincides for $k=4$ with the $J=2$ model. For $k=8$ even all $J>0$ models show transitions at larger $\beta_0$ than the other two models.

Before discussing the area coupling, let us briefly summarize the results for the length coupling thus far: For all topological backgrounds, we observe two phase transitions, one purely for the Ising model, the other in the geometry towards configurations, which only allow for $j=0$, induced by the suppression of spins $j > 0$ by this particular coupling of the Ising model to the background. The actual position of these phase transitions is primarily determined by the number of configurations with non--vanishing spins, not the background that allow for the largest spins (due to the coupling rules the quantum group). However, unless $\beta_0$ is large enough such that spins $j>0$ are strongly suppressed, we observe that the system quickly flows back to the topological fixed point and the actual flow of the tensor only involves the Ising degrees of freedom. In that sense, the Ising model is sensitive to the background, yet both systems are not strongly coupled.

\subsection{`Area coupling' -- Results}

In contrast to the `length' coupling, the `area' coupling is more non--local in nature, as we have explained in section \ref{sec:area-coupling}, as it rather prescribes a coupling with respect to the geometry. Still it includes the idea that the interaction between two spins are weakened if their distance is increased, but it relates this distance to the areas of the squares this edge is shared by. Roughly speaking $\beta \sim \beta_0 \frac{\sum_{f \supset e}\sqrt{Ar_f}}{j_e}$, where we simply give the area of a square by multiplying its edge lengths.

As we have already stressed in section \ref{sec:area-coupling}, the mechanism to relate the edge length to the area in the definition of the modified coupling constant could prevent the suppression of spins $j>0$ for large $\beta_0$, however, due to the non--local nature of this coupling, a statement for an edge alone is not conclusive. Indeed, this coupling shifts the perspective rather towards the `regularity' of the configuration as it assigns generically {\it larger} weights to vertices for which the four spins $j_e$ do not agree -- the largest weight is assigned to spins alternating between $j=0$ and $j=j_{\text{max}}$, the smallest to equilateral configurations for any $j$.

{\bf All spins $J \leq j_{\text{max}}$ allowed:}
We start the discussion of the results again for the background models labelled by a spin $J$, i.e. those models that allow all spins $j\leq J$. In table \ref{tab:area} we have summarized the results:

\begin{table*}
\begin{center}
\begin{tabular}{| c || c | c | c | c | c | c | c |}
\hline
Transition in $\beta$--interval & $k=4$ & $k=5$ & $k=6$ & $k=7$ & $k=8$ \\
 \hline \hline
$J=0$ & $[0.388,0.389]$ & $[0.388,0.389]$ & $[0.388,0.389]$  & $[0.388,0.389]$  & $[0.388,0.389]$\\
\hline
$J=1$ & $[0.377,0.378]$ & $[0.377,0.378]$ & $[0.378,0.379]$  & $[0.378,0.379]$  & $[0.378,0.379]$\\
\hline
$J=2$ & $[0.373,0.374]$ & $[0.374,0.375]$ & $[0.375,0.376]$  & $[0.375,0.376]$  & $[0.376,0.377]$ \\
\hline
$J=3$ & n.a.            & n.a.            & $[0.373,0.374]$  & $[0.374,0.375]$  & $[0.374,0.375]$ \\
\hline
$J=4$ & n.a.            & n.a.            & n.a.             & n.a.             & $[0.373,0.374]$ \\
\hline
\end{tabular}
\caption{ Interval for parameter $\beta$ marking the phase transition from the `disordered' to the `ordered' phase (on the geometry with all spins $j_e \leq J$ allowed) in the `area' coupling.
\label{tab:area}}
\end{center}
\end{table*}
The results are qualitatively very different to the length coupling case (compare to tables \ref{tab:length1} and \ref{tab:length2}). Most notably, we only find the phase transition for the Ising model from the disordered phase to the ordered one, the background model does not change.

There exist geometric transitions for very larger $\beta_0$, e.g. $\beta_0 >3$ for the $J=j_{\text{max}}$ models or even $\beta_0 > 10$ for the $J=1$ models. This is due to the divergence of the edge weights as $\beta_0 \rightarrow \infty$ (if compared to all $j=0$), greatly shifting the model away from the topological background fixed point. One can identify some patterns, however the interpretation is less clear than in the length coupling and rather indicates that a modification of the coupling for large $\beta_0$ is necessary or $\beta_0$ should be restricted to smaller values.

The second significant change to the length coupling is the {\it decrease} of the phase transition temperature as the parameter $J$ is increased. Recall that increasing $J$ allows for more background configurations, in particular larger spins / edge lengths get excited. Among the added configurations are a few equilateral ones, but these are clearly outnumbered by irregular configurations, i.e. configurations in which not all edge lengths are equal. E.g. consider the case $k=4$ from $J=1$ to $J=2$: the latter adds three allowed couplings of two spins to the spin $j=2$, namely $(2,0)$, $(1,1)$ and $(0,2)$. On the level of one 4--valent tensor, this adds 3 non--equilateral colouring ($(2,2,0,0)$, $(2,2,1,1)$ and $(2,0,1,1)$) plus perturbations in contrast to just one new equilateral configurations\footnote{Note that the Ising weights are not equal for all perturbations.}. Since these irregular configurations come with a larger weight than the equilateral one, the Ising spins effectively see a stronger coupling such that the transition occurs at smaller $\beta_0$. Yet this shift in the phase transition parameter is much smaller than in the length coupling, the most significant jump is the one from $J=0$ to $J=1$. 

Also, if we keep $J$ fixed and instead increase $k$, the phase transition parameter changes only slightly, this is due to two effects. One effect applies to the models $J=j'$, if $j'=j_{\text{max}} = \frac{k}{2}$ as $k$ is increased by 1. As $k$ is increased $j'$ is no longer of quantum dimension one and (a few) new couplings are allowed, e.g. from $k=4 \rightarrow k=5$ the couplings $(1,2),(2,1) \rightarrow 2$ become possible. Thus again more background configurations are allowed, however the added ones are less `irregular' and thus average out the most irregular ones further, such that the Ising coupling is effectively slightly weaker. In the other cases, this mechanism is not present, however note that the amplitudes of the background model also change as $k$ gets increased.

{\bf Only $j=0$ and $j=j_{\text{max}}$ allowed:}

We continue our presentation of the results for the background model, in which only $j=0$ and $j_{\text{max}}$ are excited for even $k$, i.e. only representations with quantum dimension one. This is particularly interesting for the area coupling, since this background only supports the smallest and the largest coupling, such that we can expect to see an emphasis on very irregular background configurations. The results for the phase transition of the Ising model are summarized in table \ref{tab:0-max-area}:

\begin{table*}
\begin{center}
\begin{tabular}{| c || c | c | c | c |}
\hline
Transition in $\beta$--interval & $k=4$ & $k=6$ & $k=8$ \\
 \hline \hline
Ising `disordered' $\rightarrow$ `ordered' & $[0.36,0.361]$ & $[0.348,0.349]$ & $[0.338,0.339]$ \\
\hline
\end{tabular}
\caption{Results for the background with only $j=0$ and $j=j_{\text{max}}$ allowed for even $k$ together with the Ising model in the `area' coupling. We find only one transition, purely for the Ising model from `disordered' to `ordered' as $\beta_0$ is increased.
\label{tab:0-max-area}}
\end{center}
\end{table*}
As expected, only allowing configurations with $j=0$ and $j=j_{\text{max}}$ affects the phase transition parameter greatly: The parameter always occurs for a smaller parameter than for any of the $j \leq J$ models and decreases significantly as $k$, and thus $j_{\text{max}}$ are increased. This is straightforward to understand, since the coupling rules and thus the allowed geometric configurations do not change, but due to the growth of $j_{\text{max}}$ these configurations are more irregular such that the Ising spins experience a stronger coupling. Also note that for this model no geometric transitions occur, since the excited representations are of quantum dimension $d_j=1$ and cannot couple among themselves to any other representation. Moreover, in contrast to the length coupling, the area coupling does not suppress large spins $j$ as $\beta_0$ is increased.

{\bf Only $j=0$, $j=\frac{j_{\text{max}}}{2}$ and $j=j_{\text{max}}$ allowed:}
Eventually let us study the last background model, defined only if $k$ is a multiple of four, for which only $j=0$, $j=\frac{j_{\text{max}}}{2} = \frac{k}{4}$ and $j=j_{\text{max}}$ are excited. We can consider this model to lie `in between' the extreme case of only allowing $j=0$ and $j_{\text{max}}$ and the $j \leq J$ models. Again note that the model $(0, \frac{j_{\text{max}}}{2}, j_{\text{max}})$ agrees up to signs with the model $J=2$ for $k=4$. The results are summarized in table \ref{tab:0-half-max-area}: 

\begin{table*}
\begin{center}
\begin{tabular}{| c || c | c | c |}
\hline
Transition in $\beta$--interval & $k=4$ & $k=8$ \\
 \hline \hline
Ising `disordered' $\rightarrow$ `ordered' & $[0.373,0.374]$ & $[0.367,0.368]$ \\
\hline
\end{tabular}
\caption{Results for the background with only $j=0$, $j=\frac{j_{\text{max}}}{2}$ and $j=j_{\text{max}}$ allowed for $k=4l$, $l \in \mathbb{N}$, together with the Ising model in the `area' coupling. We find only one transition, just for the Ising model from `disordered' to `ordered' as $\beta_0$ is increased.
\label{tab:0-half-max-area}}
\end{center}
\end{table*}
The results agree with our expectations: for $k=4$ the results are consistent / identical to the $J=2$ model, for $k=8$ we find the transition at lower $\beta_0$ than for any $j \leq J$ model, but clearly higher than for the $(0,j_{\text{max}})$ model. We can interpret this as follows: The addition of one intermediate spin, here $\frac{j_{\text{max}}}{2}$, with respect to the $(0,j_{\text{max}})$ model adds many configurations that are in between the equilateral and the most irregular configurations (alternating between $j=0$ and $j=j_{\text{max}}$). Yet there are only very few of these extreme configurations, such that they get averaged out by the majority of intermediate configurations. This effect is enhanced once more representations are allowed, as in the $j \leq J$ models. Again, no geometric transitions occur: even though the excitation of other spins is allowed, the model always flows back to the ($j=0$, $j=\frac{j_{\text{max}}}{2}$, $j=j_{\text{max}}$) background.

Before we continue with the discussion of the Ising model coupled to a superposition of the background in section \ref{sec:superposition}, let us summarize the results for the area coupling: As anticipated from our discussion in section \ref{sec:area-coupling}, this coupling shifts the focus away from just the length of edges to the (ir)regularity of the lattice, i.e. the relative size of neighbouring edges meeting at the same vertex, where the Ising spins effectively see a stronger coupling for irregular configurations than for equilateral ones. Considering the dynamics, the most irregular configurations push the phase transition of the Ising model significantly towards smaller $\beta_0$, as we see from a direct comparison of the $(0,j_{\text{max}})$ model to the standard Ising model ($J=0$). However, both the equilateral and the most irregular configurations are only a small subset of configurations as soon as more representations $j_e$ are allowed. Therefore, the averaging over less irregular backgrounds, like in the ($j=0$, $j=\frac{j_{\text{max}}}{2}$, $j=j_{\text{max}}$) model and more profoundly in all $j \leq J$ models (for $J>0$), shifts the phase transition closer to the equilateral case and appears to be rather stable as the quantum group level $k$ is increased. Again, as for the length coupling, the background is very stable, i.e. the system quickly flows back to the topological fixed point; the system mainly flows in the Ising degrees of freedom. Thus, while the Ising model is sensitive to the different background configurations, both systems are not strongly coupled. This will change in section \ref{sec:superposition}, where we also superimpose background fixed point intertwiners, allowing these degrees of freedom to flow under renormalization as well.

\section{The Ising model coupled to a dynamical background} \label{sec:superposition}

After the examination of the Ising model coupled to a topological background, we generalize the background model by superposing (two) intertwiner models, which just amounts of superposing two different choices of parameters $a(\{j_e\})$ discussed in section \ref{sec:grav-part}. As a result, the background model alone is no longer a topological theory and will flow under renormalization. In this section, we therefore address the following points: First, we are interested to see, how the presence of the Ising model affects the flow of the background model, e.g. on the position of the phase transition between the two topological fixed points as $\beta_0$ is changed or on the appearance of background model different from both initial ones. Second, in section \ref{sec:top-background} we have observed significant effects on the phase transition of the Ising model due to the presence of the background, such that one may wonder how it reacts to a superposition of backgrounds. As a last point, the region in parameter space in which both the Ising model and the background model are close to a transition is interesting to investigate, yet it is very likely that an algorithm keeping more singular values than the one prescribed in section \ref{sec:truncation} has to be used.

Let us briefly outline the changes to the initial tensor due to the superposition of intertwiner models: The initial tensor $\hat{T}$ written in the recoupling basis, see eq. \eqref{eq:intial-amplitude}, is simply modified by summing over the factors $a(\{j_e\})$ associated to the topological background fixed point\footnote{All fixed point intertwiners have the same dependence on magnetic indices. Thus the transformation to the recoupling basis is unaffected by the superposition.}, to differentiate them we added a superscript: 
\begin{widetext}
\begin{align} \label{eq:intial-amplitude-sup}
\hat{T}^{(j_5,k_5}_1(\{j_i\},\{k_i\}) = & \frac{1}{d_{j_5}} \, \left( \alpha \, a^{(1)}(j_1,j_2,j_5)a^{(1)}(j_3,j_4,j_5)+ (1-\alpha) \, a^{(2)}(j_1,j_2,j_5)a^{(2)}(j_3,j_4,j_5) \right) \, \times \nonumber \\
& \sqrt{\tilde{\omega}_{k_1}(j_1;\{j_e\}) \tilde{\omega}_{k_2}(j_2;\{j_e\}) \tilde{\omega}_{k_3}(j_3;\{j_e\}) \tilde{\omega}_{k_4}(j_4;\{j_e\})} \quad ,
\end{align}
\end{widetext}
where $\alpha \in [0,1]$ is the superposition parameter. In principle one can also discuss superpositions of more models, we restrict ourselves to two models to present the results in a phase diagram for $\alpha$ and $\beta_0$. Note that in order to obtain the initial 3--valent tensor, one has to perform a singular value decomposition on \eqref{eq:intial-amplitude-sup}; the sum over factors $a(\{j_e\})$ does not factorize over 3--valent vertices.

Recalling the background models introduced in section \ref{sec:grav-part}, there are many possibilities to choose from; too many to cover in this article. Hence we will restrict the discussion to the superposition of two models from the $j\leq J$ class, namely $J=1$ and $J=3$ for $k=4$. This is an interesting choice, in particular since the model $J=3$ allows the spins $j=2$ and $j=3$ in contrast to the $J=1$ model. One can expect a possible transition to the model $J=2$ between them, even without the Ising model. Moreover, the models $J > 0$ are preferable to e.g. the $(0,j_{\text{max}})$ or the $J=0$ model, since the latter have very restrictive coupling rules, strongly suppressing the excitation of other spins. Of course, one can also enhance the study by going to larger levels $k$ of the quantum group, however this significantly increases the numerical effort.

\begin{figure}
\includegraphics[width=0.45\textwidth]{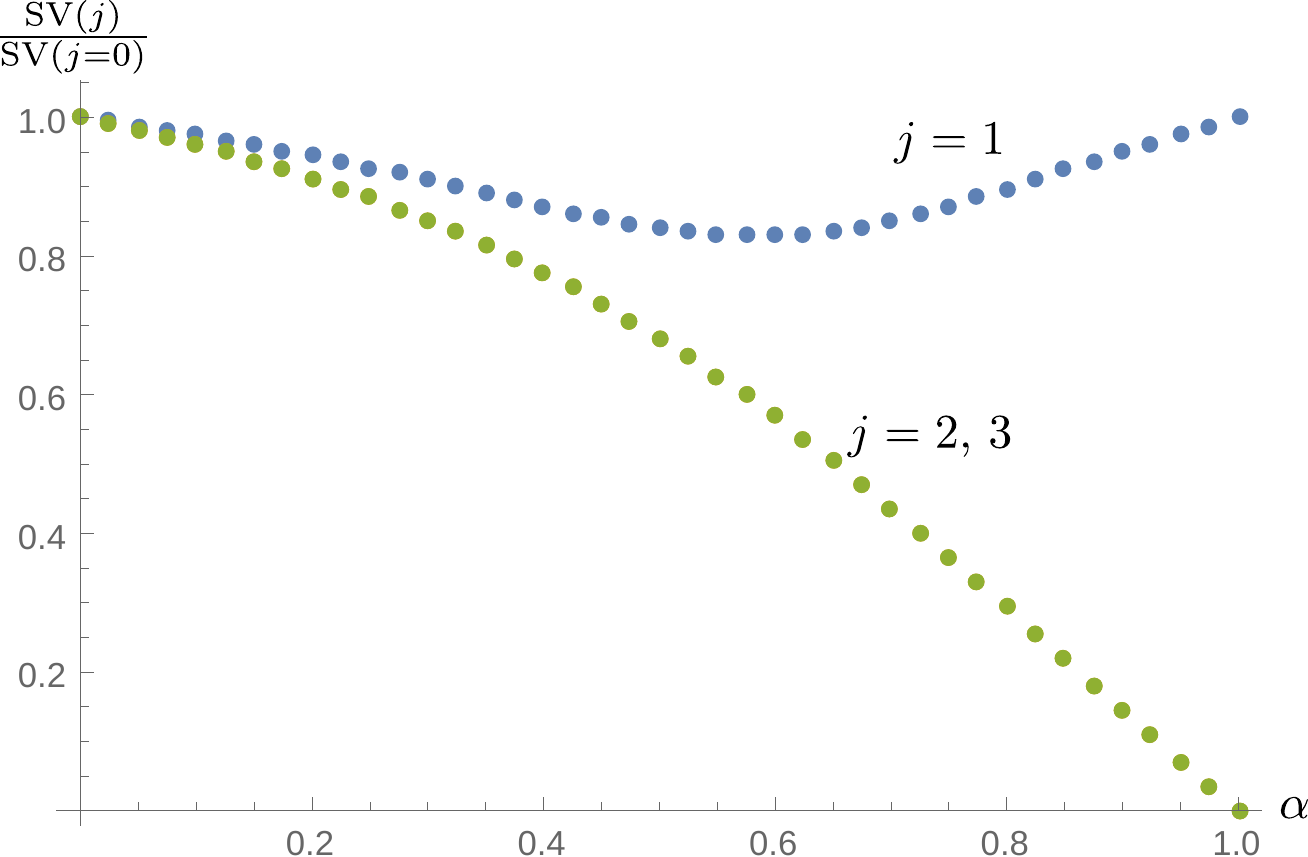}
\caption{Singular values for $j=1,\,2,\,3$ in the first splitting of the 4--valent tensors, normalized with respect to the one for $j=0$, for the superposition of the models $J=1$ ($\alpha=1$) and $J=3$ ($\alpha=0$). The values for $j=2$ and $j=3$, which have identical singular values, fall off quickly as $\alpha$ is increased, $j=1$ has a minimum around $\alpha=0.6$, which is actually close to the phase transition(s) from $J=3$ to $J=2$ to $J=1$ (for growing $\alpha$).
\label{fig:p=16-singular}}
\end{figure}

Before discussing the results for the two different couplings, let us briefly focus on the impact of the superposition of the $J=1$ and the $J=3$ model at $\beta_0=0$, i.e. without the Ising model, to get an idea of the geometries preferred by these geometries. To do so, let us study the singular values associated to the spins $j=1,\,2,\,3$ for varying $\alpha$ obtained from the first singular value decomposition splitting the 4--valent tensors (see eq. \eqref{eq:intial-amplitude-sup}) into 3--valent ones for the triangular algorithm in fig. \ref{fig:p=16-singular}.
As expected, we observe the quick decline of the singular values associated to the spins $j=2$ and $j=3$ as the system moves away from the $J=3$ to the $J=1$ model. The fate of the spin $j=1$ is more interesting: even though it is excited in both models, it goes through a minimum close to $\alpha=0.6$, which actually is very close to the phase transition between these two models. In fact this transition is not direct but rather interrupted by a tiny phase of the $J=2$ model spanning the interval $\alpha \in [0.60567,0.60568]$. The interpretation of this data is rather straightforward: As $\alpha$ grows, geometries containing spins $j=2$ and $j=3$ become less probable and eventually vanish as $\alpha=1$. Configurations containing $j=1$, which is excited in both superimposed models, become slightly disfavoured for intermediate values of $\alpha$, which conversely means an emphasis on the only other remaining spin $j=0$. Thus close to the phase transition, we can expect that configurations containing $j=0$ are favoured over configurations without $j=0$, however this effect is too weak to cause the model to flow to the phase, in which only $j=0$ is allowed. We will see that the Ising model, more precisely its phase transition parameter, is sensitive to these changes in the background in both coupling schemes. 

In sections \ref{sec:sup-length} and \ref{sec:sup-area} we present and qualitatively discuss the phase diagrams for the length and the area coupling respectively.

\subsection{Length coupling} \label{sec:sup-length}

\begin{figure}
\includegraphics[width=0.45\textwidth]{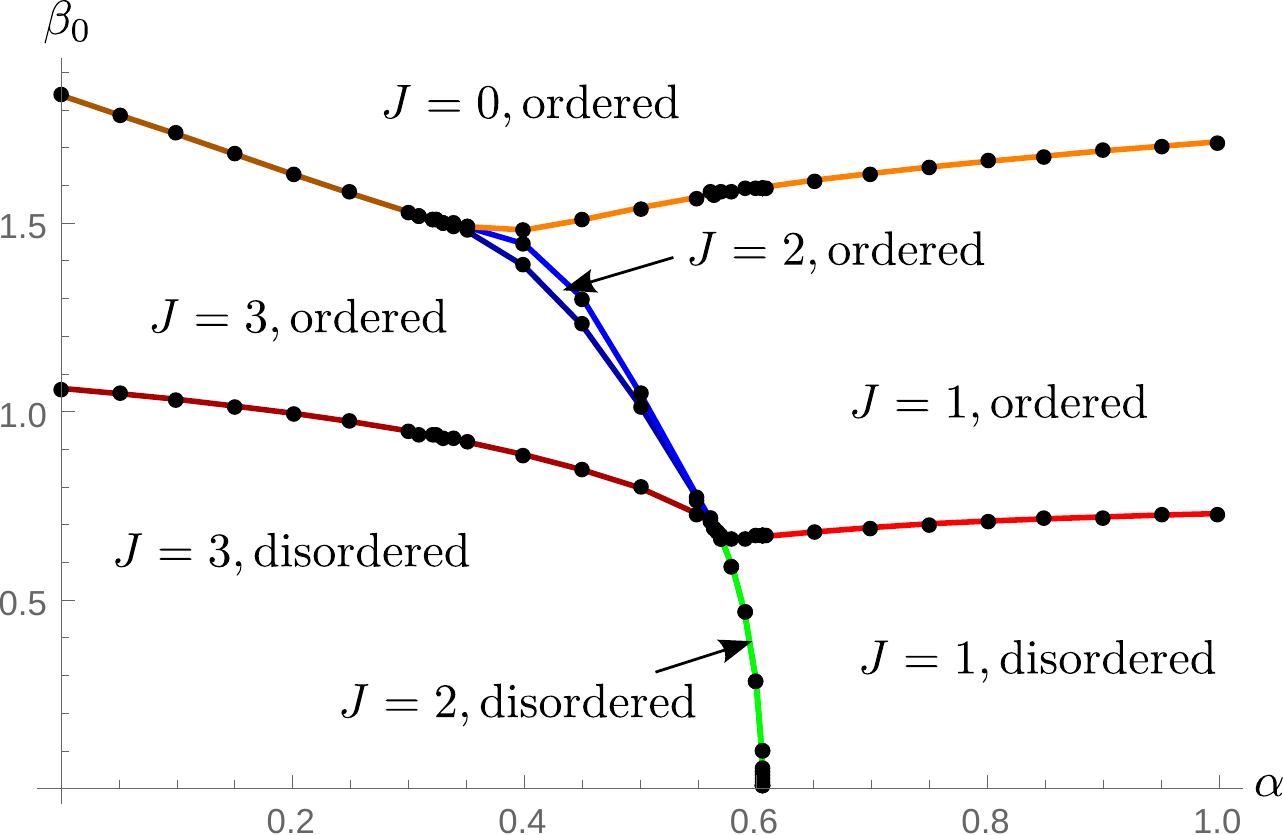}
\caption{Phase diagram for $k=6$, superimposing the models $J=1$ ($\alpha=1$) and $J=3$ ($\alpha=0$) in the length coupling. The black dots show the actual position of the phase transition; the connecting lines only indicate to which phase transition they belong. The white areas between the transitions belong to the same phase and are named accordingly. The $J=2$ phases are located between the green / blue lines, which is indicated by the arrows. The colours are chosen to according to the transition: red indicates the transition of the Ising model, blue the geometric transitions in the ordered Ising phase, green in the disordered one. Orange indicates the transition to the `just $j=0$' phase. 
\label{fig:p=16-length}}
\end{figure}

The phase diagram for the length coupling can be found in fig. \ref{fig:p=16-length}.
Let us first start with some basic observations: The vertical lines $\alpha=0$ and $\alpha=1$ are the pure intertwiner models for $J=3$ and $J=1$ respectively, the results are consistent with section \ref{sec:top-background}. The horizontal line $\beta_0 = 0$ represents the pure background model, since the Ising model assigns constant amplitudes to all background configurations. There we also observe the two geometric phase transitions mentioned above (which are difficult to resolve).

In general we observe four `main' phase transitions: the two horizontal lines indicate the transitions induced via the Ising model (coupling), the lower one for the Ising model transition from disordered to ordered, the upper one marks the transition to geometries, for which only $j=0$ is allowed, induced by the suppression of $j > 0$ in the matter coupling. The two vertical lines, which are mostly indistinguishable, mark the two geometric transitions via the $J=2$ model. Let us discuss these transitions separately:

Considering the Ising transition from the disordered to the ordered phase, we generically observe a drop in the phase transition parameter as we approach the geometric transition, where the $J=3$ line appears to merge with the geometric transitions before meeting the line from $J=1$. This drop in phase transition temperature implies that the Ising spins effectively see a stronger coupling as one moves away from the topological fixed points, where this is far more striking compared to the $J=3$ perspective. Considering the properties of the length coupling, the increase in the effective coupling for the Ising spins implies an emphasis on smaller spins in the background configurations. As we have concluded above from fig. \ref{fig:p=16-singular}, this is exactly what happens: Seen from $\alpha=0$, i.e. the $J=3$ model, the spins $j=2$ and $j=3$, which weaken the interaction between neighbouring Ising spins, are significantly suppressed as $\alpha$ is increased explaining the substantial drop in the phase transition parameter. Moreover, even the spin $j=1$, which is excited in both superimposed models, gets disfavoured close to the phase transition, such that also the effective coupling is weaker than in the $J=1$ model.



The story is similar for the second transition induced by the Ising model, namely the one to `just $j=0$' geometries. Again we observe a drop in the phase transition parameter from both sides close to the (significantly shifted) geometric transition. Recalling that this transition occurs due to the suppression of all spins $j>0$ by the Ising edge weights, it is clear that this suppression occurs for smaller $\beta_0$ if configurations allowing just $j=0$ become emphasized by the geometric superposition.

Eventually, let us discuss the geometric transition: as mentioned above, there are actually two geometric transitions, enclosing the emergent $J=2$ configuration. This pair of transitions starts out vertically from the x-axis, however quickly bends toward smaller $\alpha$, i.e. $J=3$, where it also becomes broader after crossing the phase transition line for the Ising model. Again this is straightforward to explain from the length coupling of the Ising model: As $\beta(j)$ falls of as $\frac{\beta_0}{j}$, larger spins $j$ get quickly suppressed. The spin that is suppressed the most is $j=3$, then $j=2$, etc. Thus close to the geometric phase transition this suppression favours a flow towards the $J=2$ background over the $J=3$ background and analogues for the $J=1$ and $J=2$ background. If the system is then moved to smaller $\alpha$, $j=3$ in turn becomes emphasized by the background such that the suppression from the Ising weights must be stronger for the $J=2$/$J=1$ phase to appear, which explains the advancing of both the $J=2$ and $J=1$ phase to smaller $\alpha$. At some point, the suppression is too strong, such that all spins $j>0$ are suppressed and the geometric transition `merges' into the transition towards geometries, for which only $j=0$ is allowed.

To conclude both the Ising and the background model react very sensitively to one another. From the effective coupling (and thus the position of the phase transition) seen by the Ising model, one can deduce which geometric configurations are preferred at a specific superposition of background intertwiner models. Here the superposition results in an emphasis towards edges carrying spin $j=0$, such that the effective coupling for neighbouring Ising spins is stronger. Conversely, the geometry is greatly affected by the matter coupling as well: if the Ising coupling is strong enough, a geometric transition to a different background model is induced. Admittedly this requires the background to be in a superposition, otherwise the deviation away from this fixed point caused by the Ising model is not sufficient. Also note that we did not find any phases beyond the `product phases' of the combined model. Whether more phases exist, e.g. close to both phase transitions, can probably only be answered by using a tensor network algorithm keeping more singular values in each iteration, yet close to the phase transition an infinite bond dimension is required.

In section \ref{sec:sup-area} we discuss the significantly different phase diagram for the area coupling.

\subsection{Area coupling} \label{sec:sup-area}

\begin{figure}
\includegraphics[width=0.45\textwidth]{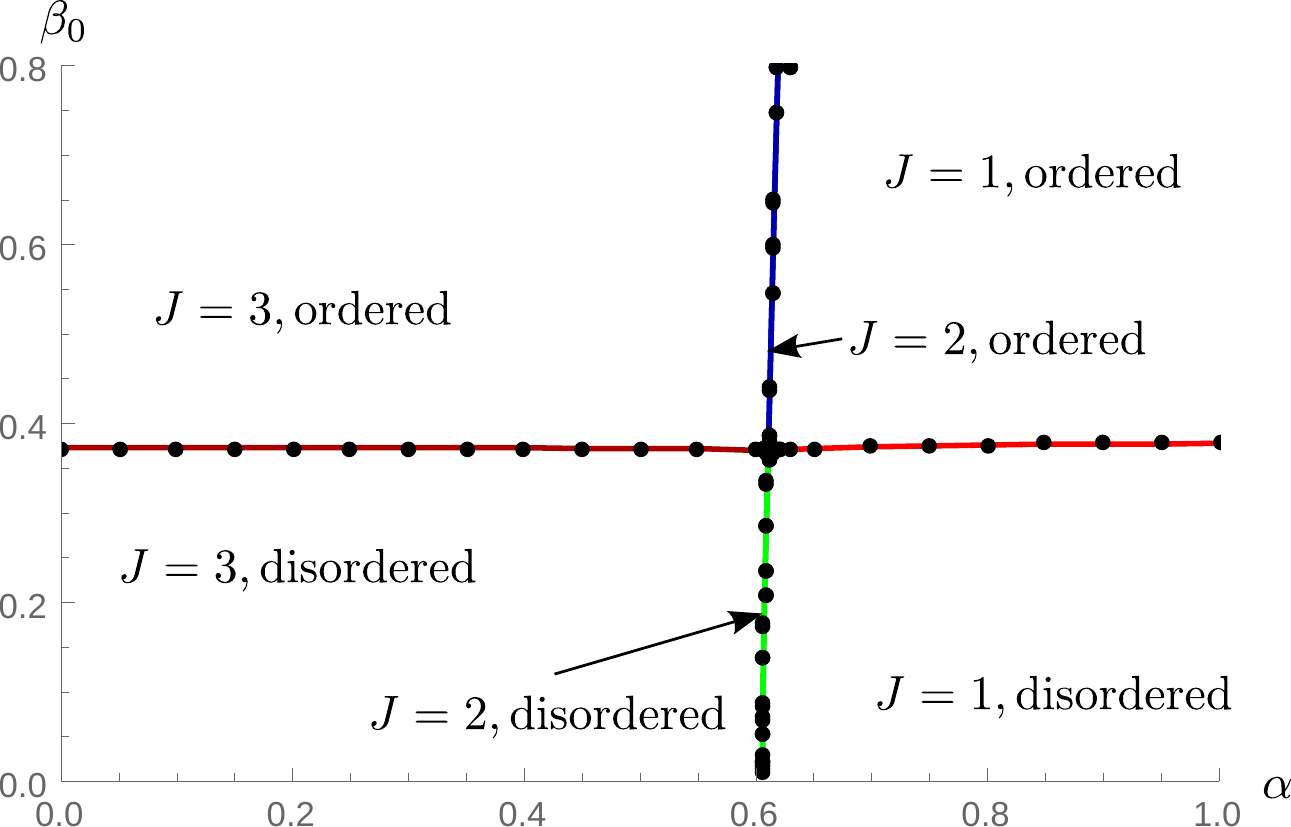}
\includegraphics[width=0.45\textwidth]{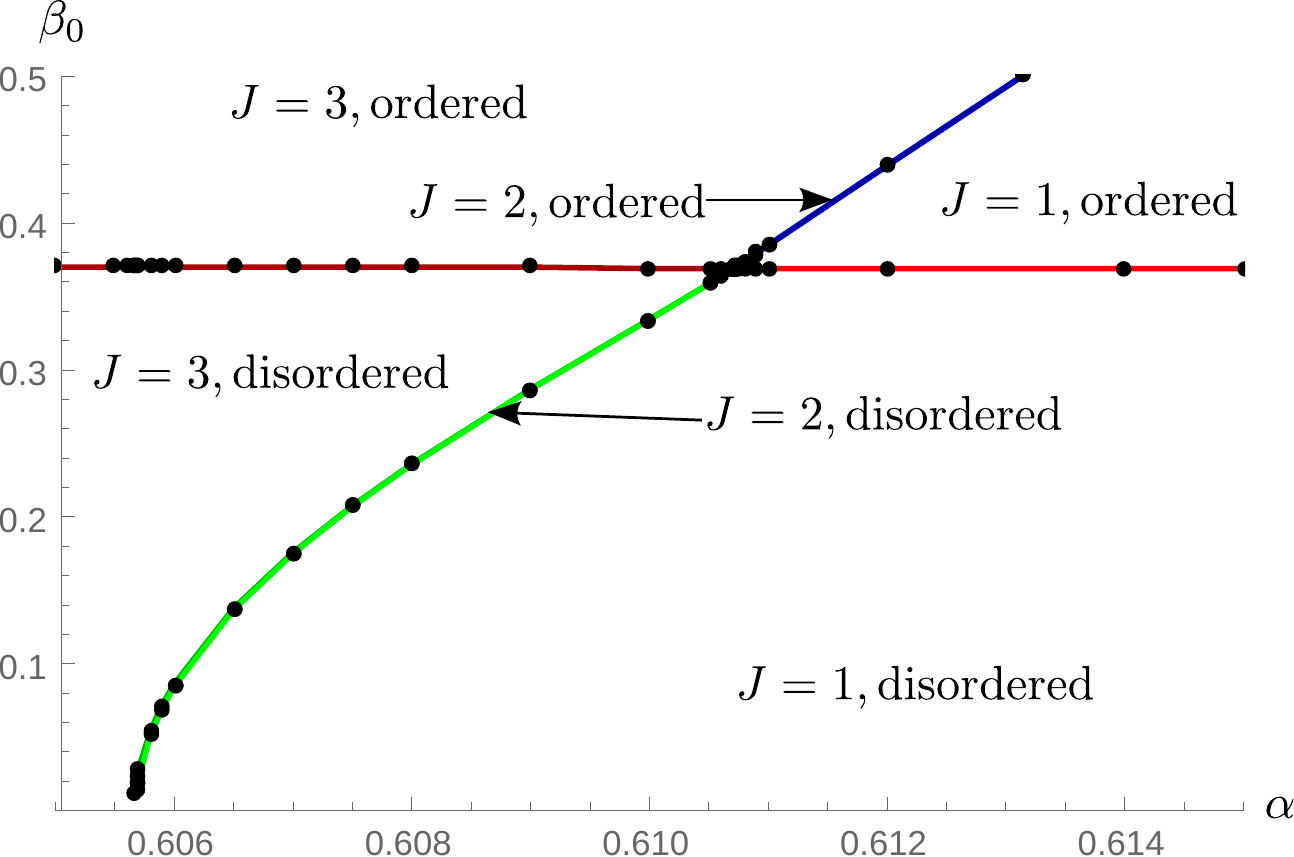}
\caption{Phase diagram for $k=6$, superimposing the models $J=1$ ($\alpha=1$) and $J=3$ ($\alpha=0$) in the area coupling. The plotstyle, in particular the colouring of the transitions, is similar to the one for the length coupling.
\label{fig:p=16-area}}
\end{figure}

The phase diagram for the area coupling can be found in fig. \ref{fig:p=16-area}.
As expected, we observe a very different phase diagram in comparison to the length coupling. First of all, the Ising model transitions for $\alpha=0$ ($J=3$) and $\alpha=1$ ($J=1$) are consistent with the results in section \ref{sec:top-background}. Again we observe two geometric phase transitions for $\beta_0 = 0$, in the same small interval as for the length coupling.

The horizontal line marking the phase transition of the Ising model from the disordered to the ordered phase (as $\beta_0$ is increased) is indeed almost horizontal, it only slightly decreases (from both sides) to roughly $\beta_0 \approx 0.371$ as the geometric phase transition is approached. Thus the Ising spins again see an effective coupling that is slighly stronger as the geometric transition is approached; as the phase transition occurs for larger $\beta_0$ for $\alpha=1$ (compared to $\alpha=0$), we can also conclude that this effect is stronger in the regime where the $J=1$ background dominates. Let us analyse this behaviour in steps: Recalling the discussion for the superposition of backgrounds we already know that the superposition of intertwiners affects the amplitudes such that the trivial representation $j=0$ is emphasized, while all others get suppressed, albeit only slightly in the case of $j=1$. As we have already discussed before the area coupling is less sensitive to the excited edge lengths alone, but rather how irregular the geometry is as seen from a single vertex. Indeed, the strongest coupling of the Ising spins (for a single vertex) occurs if the edges meeting at this vertex alternate between minimal ($j=0$) and maximal ($j=j_{\text{max}}$) length. Aware of this mechanism, the emphasis of $j=0$ close to the geometric transition explains the slightly stronger effective Ising coupling as follows: If $j=0$ is slightly favoured, while all also spins $j>0$ remain excited, this puts an emphasis on more irregular configurations. Additionally, from the perspective of the $J=1$ model, the superposition excites spins $j>1$ as well, such that more irregular configurations are possible, resulting in the slightly larger decline as the geometric transition is approached from $\alpha > 1$.

Similar to the transition of the Ising model, the two geometric transitions, i.e. from from $J=3$ to $J=2$ to $J=1$ as $\alpha$ is increased, is also less affected by the coupling to the Ising model than in the length coupling. The most crucial difference is the `bending' of the transition towards larger $\alpha$ instead of smaller $\alpha$. Again this is due to the different nature of the area coupling as it is rather sensitive to the (ir)regularity of geometries than just the excited edge lengths. As $\beta_0$ is increased, the irregular geometries are given more weight than the regular ones from which one can conclude a favouring of larger spins, since these allow for more irregular geometries, in particular with respect to $j=0$. Thus the transition to models with $J>1$ occurs at larger $\alpha$, yet this effect is much weaker than in the length coupling. Also, we do not observe a broadening of the $J=2$ phase similar to the situation in the length coupling.

Again, we would like to mention that additional geometric transitions occur for very large $\beta_0$ due to the divergence of the Ising weights, which occur for $J=3$ at roughly $\beta_0 > 3$ and for $J=1$ at roughly $\beta_0 > 10$. We will not cover these in detail in this work because of their complicated structure and since they occur in a regime far away from the interesting dynamics of the Ising model. To us this rather indicates that further modifications to the matter couplings must be done if these are to be avoided or $\beta_0$ should be restricted to smaller values.

Let us some up the results for the area coupling. Again we observe that both the Ising model and the geometric background are sensitive to each other, however to a much smaller degree than for the length coupling. The effect on the position of the Ising phase transition is comparatively small and can be well explained by the changes in the background geometry induced by the superposition of fixed point intertwiners, namely the slight preference of irregular geometries due to an emphasis of configurations containing edges with spin $j=0$. In contrast to the length coupling, the geometric transitions shift towards larger $\alpha$, i.e. towards the $J=1$ model, as $\beta_0$ grows. This is caused since for larger $\beta_0$ more weight is assigned to irregular configurations, which prefers the appearance of spins $j>1$, even though these are disfavoured by the background model. Again the coupling to the Ising model induces geometric transitions, here from $J=1$ to $J=2$ and then $J=3$. As already mentioned for the length coupling, a more thorough study might be worthwhile at the intersection of both transitions, yet this requires an algorithm capable of efficiently studying systems at criticality, which means obtaining good results, e.g. for critical exponents, at a finite bond dimension.

\section{Summary and discussion} \label{sec:discussion}

In this article we have studied a 2D toy model for coupling pure lattice Yang--Mills theory to a spin foam model, where the gravitational and the gauge theory are replaced by simpler models, which do preserve parts of the dynamical structure of the full theories. For the gravitational part, we have chosen topological intertwiner models defined in \cite{invariant_int}, which allow for a notion of intertwiner degrees of freedom similar to the spin foam case, where these occur due to the imposition of simplicity constraints. These intertwiner models are defined for the quantum group $\text{SU}(2)_k$, which comes with a natural cut--off on the representation labels \cite{biedenharn,yellowbook}. Moreover it is conjectured that spin foam models defined for quantum groups incorporate a non--vanishing cosmological constant \cite{turaev-viro,smolin-qg,smolin-qg2,smolin-qg3,karim-qg,q-spinfoam0,q-spinfoam,q-spinfoam2,lqg-lambda1,lqg-lambda2,lqg-lambda3,lqg-lambda4,carlo-francesca-curved}. For the matter system, we have considered spin systems, more precisely the $\mathbb{Z}_2$ Ising model, for which the local gauge symmetry of lattice Yang--Mills is replaced by a global symmetry. Again it is expected \cite{kogut-review} that these models share statistical properties with their 4D lattice gauge theory relatives.

Due to the structure of the two systems it is straightforward to coarse grain them via tensor network renormalization \cite{levin,guwen}. To do so the partition function of the system is rewritten as a contraction of a tensor network. By performing local manipulations of the tensors, in particular defining and identifying the most relevant coarse degrees of freedom via a singular value decomposition, the partition function is evaluated / approximated in steps until eventually the procedure reaches a fixed point. In context of this coarse graining procedure, we have briefly revisited the discussion in \cite{simone-3dYM} whether the matter degrees of freedom should be placed on the same lattice as the spin foam or rather the dual lattice as it is done e.g. in \cite{oriti-pfeiffer}. For the toy model under discussion, this choice has rather technical than physical consequences: First, the Ising model is self dual \cite{savit-review}, i.e. both choices are related by a duality transformation, which maps the strong coupling sector to the weak coupling sector and vice versa. However, in order to coarse grain these models, they have to be cast into tensors, in essence local amplitudes that are associated to the vertices of the lattice. As it turns out, if the Ising model is placed on the dual lattice, it is more suitable to keep the original Ising spin variables; in the process they have to be coarse grained like decorated tensor networks \cite{decorated-tensor}, where the Ising spins `decorate' the gravitational part. Conversely, if the Ising model `lives' on the same lattice, it is preferable to group Fourier transform the Ising variables. These insights for coarse graining coupled systems might be helpful, e.g. to examine the influence of the choice of discretisation, once one tackles more complicated theories, in particular if these theories are not self dual or the dual theories are not known. Nevertheless, we agree with \cite{simone-3dYM} that it is preferable to define both matter and gravitational degrees of freedom on the same discretisation, such that one can establish contact to the states and Hilbert spaces of the associated canonical theory.

Undoubtedly, the crucial part of this article is the matter coupling between the background system and the Ising model: The background intertwiner models endow a geometric interpretation onto the lattice, its $\text{SU}(2)_k$ representation labels $j_e$ on the edges assign a length $j_e + \frac{1}{2}$ to these edges. Intuitively, the interaction among Ising spins should be sensitive to the distance between them, spins further apart should have a weaker interaction than close spins. Therefore we have introduced a coupling between Ising spins and the background by choosing the Ising coupling constant $\beta$ to be a function of the background spins $j$, actually similar to \cite{oriti-pfeiffer,mikovic1}. In fact, it is also related to the idea expressed in \cite{oeckl-book} to assign local coupling constants (to parts of the discretisation) for lattice gauge theories defined on irregular lattices. Interestingly, the related construction of the Ising model on 2D causal dynamical triangulations \cite{cdt-ising1,cdt-ising2} is somewhat orthogonal to our idea: Instead of fixing the discretisation, in particular the combinatorics, and varying the edge lengths, one rather fixes the edge lengths and sums over all possible triangulations. Thus, the distance between neighbouring Ising spins is always the same, thus no modification of the coupling constant $\beta$, but the number of neighbouring Ising spins can vary.

In this article, we have constructed two modifications of the coupling constant, one for which $\beta (j) \sim \frac{1}{j}$ and one for which the distance is related to the (square root of the) area that is partially spanned by this edge $\beta (j) \sim \frac{\sum_{f \supset e} \sqrt{Ar(f)}}{j_e}$. While the first is a local interaction, assigning the highest weight to the shortest edge length, the latter is more non--local by relating the edge length to the overall geometry and is analogous to the modification introduced in \cite{oriti-pfeiffer} for 4D Yang--Mills coupled to the Barrett--Crane spin foam model \cite{Barrett-Crane}. There it was chosen such that the Wilson action scales as the volume of the 4--simplex it is defined on. Admittedly, these modifications are heuristic and appear ad hoc; one can certainly construct more realistic couplings, e.g. by considering Laplacians on discretisations, and it would be preferable to derive such a coupling from the classical theory in 4D. However, neither is this the main purpose of this article, nor does this avoid choosing a discretisation, which eventually results in lattice imperfections. It is rather to explore whether a simple modification in the matter part, which couples the two systems, can result in an interesting and reasonable dynamics for the coupled system or whether a different mechanism is necessary.

Indeed, the chosen couplings have a significant effect on the dynamics as they also emphasize different geometries of the background system. The straightforward length coupling simply considers edge lengths, the longer an edge the weaker the interaction between Ising spins is. As a result, the Ising model sees effectively a (much) weaker coupling the more edge lengths $j>0$ (and thus also configurations of the background with $j>0$) are allowed, such that the phase transition is clearly shifted towards larger $\beta_0$. However, this coupling has a direct effect on the background geometry as for $\beta_0 \gg 1$ spins $j>0$ get strongly suppressed causing a phase transition to geometries with just spins $j=0$ allowed. Moreover, the Ising model is very sensitive, in terms of the position of its phase transition, to changes in the background model. In particular the emphasis on the smallest possible edge length suggests two possibilities: It can imply that the coupling $\beta_0$ should be restricted to smaller values or one should introduce a modification to the coupling.

In contrast to that the area coupling is not very sensitive to edge lengths alone; it rather favours specific configurations due to its non--local nature. Interestingly, it assigns the weakest weights to regular, equilateral configurations (independent on the edge lengths as it is invariant under a global rescaling of all edge lengths) and the strongest to very irregular configurations, e.g. if around one vertex the edge lengths alternate between the minimal and maximal edge lengths. As a consequence, the Ising spins effectively see stronger couplings if spins $j>0$ are allowed, thus shifting the phase transition to smaller $\beta_0$, yet this shift is considerably smaller in comparison to the length coupling and far less sensitive to changes of the background model. E.g. when considering a superposition of background models in section \ref{sec:superposition}, the line marking the Ising phase transition is almost horizontal with only slight deviations consistent with the changes in the background geometry. In fact this is encouraging, since we do observe that this coupling allows for matter sensitive to the geometry it is defined on, yet without resulting in significant deviations away from the equilateral case.

Fortunately, also the gravitational background model is sensitive to the matter defined on it, albeit only significantly if it is not close to a topological fixed point. If the latter is the case, the deviations away from it caused by the Ising model are not strong enough, the background quickly flows back to the initial fixed point intertwiner and decouples from the Ising model, which remains sensitive to the background geometry. However, if we consider the superposition of two background fixed points, the interaction with the Ising model can trigger phase transitions in the geometry. In fact, depending on the geometries preferred by the coupling scheme, for growing $\beta_0$ the geometric transition is shifted away from its position in parameter space without the Ising model. Again this effect is far more profound in the length coupling than in the area coupling. Nevertheless, we have shown that by introducing a simple modification of the Ising model coupling constant, which couples it to a topological background, one can implement a dynamics, which exhibits regions (in parameter space) of strong coupling between the two systems, namely close to the phase transitions of both systems, and regions in which they are only slightly coupled, e.g. on the fixed points of the background model. Hence, it might be worthwhile to study a similar coupling mechanism for spin foams and lattice gauge theories in order to potentially identify similar regions as well. A region in which both systems are only weakly coupled might allow us to establish contact to lattice gauge theories on a fixed (flat) background, which could indirectly teach us more about the geometry arising from spin foam models.

As these coupling mechanisms of the Ising model to the background are the central piece of this article, let us also discuss some of the choices made and their possible consequences. In both cases, we have regularized the $j=0$ case by defining the $j+\frac{1}{2}$ as the length of an edge. We have motivated this choice with the asymptotics of (3D) spin foam models \cite{PR,pr-model}, but admittedly this is ad hoc and one can straightforwardly generalize this to $j + \epsilon$ instead, where $\epsilon \neq 0$.

For the length coupling, $\epsilon > 0$, i.e. essentially the case studied here, implies that $\beta(j)$ is always positive and the largest coupling constant is assigned to $j=0$, and thus also the largest Ising edge weight. The size of $\epsilon$ then determines how much $j=0$ is favoured, but the qualitative behaviour is more or less the same. However, if $\epsilon < 0$, $\beta(j)$ is negative for all $j < |\epsilon |$, such that for some edge lengths, the interaction for $j < |\epsilon|$ is antiferromagnetic and ferromagnetic for $j > |\epsilon |$. One can expect significant changes for the dynamics of such a model, if one tunes $\epsilon$ such that small edge lengths give antiferromagnetic and large edge lengths ferromagnetic interactions between the Ising spins. One can also choose $\epsilon$ such that the interaction is antiferromagnetic for all $j$, where $| \beta(j)|$ is then largest for $j_{\text{max}}$. Thus it might be possible to study other interesting phenomena, possibly a transition from antiferromagnetic to ferromagnetic or the modified interaction could cure the unintended emphasis on $j=0$ configurations. However one should not literally interpret $j + \epsilon$ as the length, but rather an effective description, in which the system rather avoids shortest distances, which one could associate with highly energetic configurations.

Concerning the area coupling, a modification to $\epsilon < 0$ can also have interesting effects, which are however not as straightforward to predict, as $\beta(j)$ can become complex (due to the squareroot). Another option can be to invert the modification, i.e. normalize the length by the square root of the area instead. However, it is not obvious whether this leads to a favouring of equilateral over irregular configurations, as both the product of weights around a single vertex remain unchanged for the equilateral and the alternating case. This has to be examined more thoroughly before one can draw a conclusion.

Before we conclude, we would like to make a few remarks on the coarse graining algorithm. Since we have used a symmetry protecting version similar to those introduced in \cite{abelian_tnw} for Abelian and in \cite{s3_tnw,q_spinnet} for non--Abelian and quantum group symmetries, we have been able to straightforwardly identify the phases of the model. The main advantage of the triangular tensor network algorithm, that we have used in this work, is its improved memory and computational cost, as we only save 3--valent tensors and contract one index less. This roughly reduces the computation time from $\chi^6$ to $\chi^5$ compared to the 4--valent algorithm in \cite{guwen}, where $\chi$ is the bond dimension (or index range) of the tensor. However, the algorithm has a caveat: Even though we have worked at rather low accuracy, the fixed point tensors have shown so--called CDL structure. This CDL structure means that part of the tensor is of a corner double line form, its edges being double indices, which are pairwise identified along the vertex. This structure is a fixed point of many tensor network renormalization schemes and for fixed initial parameters, the final fixed point tensor is cut--off dependent, see also the appendix of \cite{vidal-evenbly} for a nice explanation. Thus it obstructs a proper flow to the true fixed point of the system. In condensed matter physics, this structure is interpreted as short range entanglement, which is unintentionally promoted to a larger scale by the coarse graining procedure and can be cured by entanglement filtering \cite{vidal,guwen,mera,vidal-evenbly}. Due to the absence of a (background) scale, the interpretation in spin foam models is less clear. Fortunately in the system under discussion here, the different phases can still be clearly identified, such that the results (in the approximation scheme) are consistent.

Furthermore, we would like to briefly comment on the chosen cut--off scheme, namely one singular value per intertwiner channel $(j,k)$. Even though it appears to be a strong simplification, our results suggest that this approximation is good enough to capture many important aspects of the system. Therefore, if we would work with a higher accuracy, i.e. more singular values are taken into account in each iteration, we expect that the actual position of the phase transitions will change, however not the qualitative behaviour, e.g. which phases the model has or the qualitative reaction of the Ising spins interaction to the background. Close to both the Ising and geometric transition, i.e. where both systems are more strongly coupled, this approximation breaks down and one has to work with a larger bond dimension. There we also expect interesting dynamics to occur, possibly with additional phases beyond the product phases of the composite model. Furthermore, it would be interesting to compute critical exponents in this regime in order to learn more about the phase transitions. However many tensor network algorithms, including the one used here, require an infinite bond dimension to study systems at criticality and are not well suited for extracting critical exponents \cite{kadanoff2}. The algorithm recently developed by \cite{vidal-evenbly} is more promising to fulfil this task as it successfully deals with CDL structure, may allow for good approximations close to criticality for finite bond dimension and is claimed to give good results for critical exponents. Therefore, it is more promising to examine the toy model at criticality with a properly adapted algorithm.


Additionally, one may want to supplement the numerical analysis of the toy model by more analytical means, e.g. examining whether one can define the toy model to be discretisation independent and thus possibly fix the coupling of Ising spins and the background. Therefore one could attempt to find a model independent under Pachner moves, or at least perturb around the topological fixed points of the background and determine the Ising weights, such that the system is approximately discretisation independent. In fact we have examined these possibilities, however did not find a straightforward solution to either of them. One rather contrived option would be to construct the whole amplitude, that is Ising plus background, to be a topological fixed point from \cite{invariant_int}, which however requires the background to completely absorb the dependence on the Ising degrees of freedom. For the other option, the perturbation around a topological fixed point, the equations essentially force the Ising model to assign constant weights to all background configurations; the Ising model is forced to either be in the ordered or disordered phase. This is also reflected in the numerical simulations for the topological background in section \ref{sec:top-background}, where the system quickly flows back to these and only the Ising spins flow to either the ordered or disordered phase. In fact, this is already non--trivial to define the Ising model itself in a discretisation independent way (away from either the ordered or disordered phase), since one can expect it to be non--local as one can already deduce from a simple decimation scheme and as it is also well--known for other interacting systems, e.g. discrete gravity under Pachner moves \cite{bojowald-measure,regge_measure,non-local,sf-pachner}. A possible way to overcome this is to generalize the concept of discretisation independence by a means to relate theories defined on different discretisations. One way is to allow refining and coarse--graining of the boundary (data), in such a way that one can define embedding maps that relate between configurations / states defined on different boundaries. If these embedding maps are designed such that one can unambiguously relate different boundaries, one is able to identify states across them, i.e. recognize the same (physical) situation represented on different boundaries. This condition is known as cylindrical consistency, e.g. realized for the kinematical Hilbert space of loop quantum gravity \cite{thomasbook,ashtekar-lewan1,ashtekar-lewan2}. To realize this for interacting theories or the physical Hilbert space of loop quantum gravity, i.e. states annihilated by all constraints of the theory, is non--trivial, however see \cite{bianca-cyl,time-evo,bianca-solve} on how this problem can be tackled.

At last, we would like to raise a conceptual issue concerning the coupling of matter to spin foams. The Ising model, and also lattice Yang--Mills in the Wilson or heat kernel action \cite{kogut-review}, are essentially Wick--rotated theories that appear as $\exp(- S_{\text{matter}})$ in the partition function / path integral. However, spin foams, and also the discussed intertwiner models, are complex assigning $\sim \exp(i S)$ to the path integral (for one orientation), also for Riemannian signature. Moreover, it is well--known from spin foam asymptotics \cite{4dWilliams,baez10j,freidel-6j-10j,Conrady-Freidel,Frank3,bc-paper} that both orientations of a 4--simplex have to be considered, such that one rather finds $\sim \cos(S_{\text{Regge}})$, where $S_{\text{Regge}}$ is the Regge action \cite{regge} associated to the 4--simplex. To the author's best knowledge, it is not known whether it is consistent to couple a Wick--rotated matter theory to a non--rotated gravitational theory and, moreover, in which way matter should couple to different orientations. There exist attempts to remove the other orientation from the asymptotics \cite{orientation1,orientation2,orientation3,orientation4,orientation-zipfel}, whereas it has been argued in \cite{antispacetime} that fermionic degrees of freedom might be sensitive to changes in the orientation.

In this article we have introduced and tested a toy model to demonstrate that coupling matter degrees of freedom to a dynamical (discrete) background \`a la spin foams can be straightforwardly achieved and can additionally result in an interesting dynamics, where both systems are sensitive to one another. Moreover, we have identified regions in parameter space of strong and weak coupling between matter and background degrees of freedom. Therefore we suggest to generalize these ideas to the 4D theories, e.g. Yang--Mills theory and spin foams, and attempt to extract new insights for spin foam models. Of course these theories are very complex and a renormalization algorithm is (at the moment) not at hand. Therefore it will be necessary to introduce approximations and simplifications, which might still allow us to study interesting aspects, albeit with the caveat of lacking a continuum limit. Nevertheless, this may allow us to tentatively establish connections to other approaches based on discretisations, such as lattice gauge theories, or may give us hints on how to modify the matter coupling, the spin foam model or both. An application, which might be in reach, could be the identification of an effective (matter) dynamics consistent with lattice gauge theories on flat spacetime, which would massively support spin foam models as viable quantum gravity candidates.

\section*{Acknowledgements}
The author would like to thank Bianca Dittrich for encouraging the author to investigate this system further, for useful comments on an earlier draft and also for interesting discussions on the toy model, including the results, possible modifications and issues. Moreover, the author would like to thank Benjamin Bahr for discussions on coupling lattice gauge theories to spin foams. At last, the author would like to thank Erik Schnetter for discussions on the 3--valent tensor network algorithm.

This work was funded by the project BA 4966/1-1 of
the German Research Foundation (DFG).

\appendix

\section{Quantum group basics} \label{app:quantum-group}

In this appendix, we briefly introduce some basics on the quantum group $\text{SU}(2)_k$ and the diagrammatical calculus, which are necessary to understand the calculations, in particular in appendix \ref{app:tensor} and partially in section \ref{sec:renormalization}, however not essential to understand the main results of the paper. We use the notation and conventions introduced in \cite{biedenharn}, where one can also find a more detailed introduction to quantum groups, see also \cite{yellowbook}. Many of tools we discuss and use in this work have been originally developed in \cite{q_spinnet}.

If we refer to the quantum group $\text{SU}(2)_k$, we are actually referring to the q-deformation ${\cal U}_q(\mathfrak{su}(2))$ of the universal enveloping algebra $U(\mathfrak{su}(2))$ of the Lie algebra $\mathfrak{su}(2)$ as in \cite{biedenharn}. The algebra ${\cal U}_q(\mathfrak{su}(2))$ is generated by three operators $J_{\pm},J_z$ with commutation relations
\begin{eqnarray}
[J_z , J_{\pm} ]&=& \pm J_\pm       \nonumber \\
~ [J_+ , J_-] &=& \frac{q^{J_z}-q^{-J_z}}{q^{1/2}-q^{-1/2}} \; .
\end{eqnarray}
As mentioned in the main body of the paper, the finite dimensional representation of $\text{SU}(2)_k$ are labelled by $j \in \frac{\mathbb{N}}{2}$ and can be defined on $2(j+1)$ dimensional representation spaces $V_j$ as for $\text{SU}(2)$. The quantum dimension $d_j$ of representation $j$ is defined as the quantum number of the classical dimension:
\begin{equation}
d_j:=[2j+1]  \; ,
\end{equation}
where the brackets denote quantum numbers:
\begin{eqnarray}
[n]&=& \frac{q^{\frac{n}{2}}-q^{-\frac{n}{2}}  }{q^{\frac{1}{2}}-q^{-\frac{1}{2}}  } \; .   
\end{eqnarray}
In this paper, the deformation parameter $q$ is a root of unity, with $q=\exp(\frac{2\pi}{(k+2)}i)$. $k \in \mathbb{N}$ is called the level of the quantum group $\text{SU}(2)_k$. Quantum numbers are periodic 
\begin{eqnarray}
[n]&=&\frac{\sin(\frac{2\pi n}{2k+4})}{\sin(\frac{2\pi}{2k+4})}  \; ,
\end{eqnarray}
with zeros at $n=0$ and $n=k+2$. Thus $j=\frac{k}{2}$ with $d_{k/2}=1$ is the `last' representation with a strictly positive quantum dimension. Representations $j=0,\frac{1}{2},\ldots,\frac{k}{2}$ are called admissible, representations $j>\frac{k}{2}$ are of so--called quantum trace zero.

The tensor product of two representations $V_{j_1},V_{j_2}$ via the co--product $\Delta$. The action of the $\text{SU}(2)_k$ algebra on $V_{j_1}\otimes V_{j_2}$ is defined as
\begin{align}
\Delta(J_\pm) &=& q^{-J_z/2} \otimes J_{\pm} \,+\, J_\pm  \otimes q^{J_z/2} \nonumber \\
\Delta(J_z) &=& \mathbb{I} \otimes J_z \,+\, J_z\otimes \mathbb{I} \quad .
\end{align}
The tensor product $V_{j_1}\otimes V_{j_2}$ can be decomposed into a direct sum of irreducible representations plus a part consisting of trace zero representations (which are modded out). With an orthogonal basis $|j,m\rangle$ in the representation spaces, the decomposition is given by Clebsch-Gordan coefficients
\begin{eqnarray}
|j,m\rangle &=&\sum_{m_1,m_2} {}_qC^{j_1j_2j}_{m_1m_2 m} \,|j_1m_1\rangle \otimes |j_2,m_2\rangle \; .
\end{eqnarray}

If one couples three admissible representations $j_I$, $j_K$ and $j_L$ in this way, several conditions have to be satisfied for the Clebsch--Gordan coefficients to be non--vanishing:
\begin{eqnarray} \label{couplingcond}
j_I+j_K & \geq & j_L \; \text{for permutations} \; \{J,K,L\} \;\text{of}\; \{1,2,3\} \; ,\nonumber \\
j_1+j_2+j_3 &=& 0\!\mod 1 \; , \nonumber \\
j_1+j_2+j_3 &\leq & k \; .
\end{eqnarray}
The last condition in (\ref{couplingcond}) is special to the quantum deformed case at root of unity and indicates that  $V_{j_1}\otimes V_{j_2}$ can include trace zero parts, which can be modded out \cite{yellowbook}. However, some equations (for instance the definition of the $[6j]$ symbol) are only valid up to trace zero parts \cite{yellowbook}. 

In particular we have the completeness relation
\begin{eqnarray} \label{complete}
\sum_{m_3,\, j_3 \, \text{admiss.}} {}_qC^{j_1j_2j_3}_{m_1m_2m_3} \,\, {}_qC^{j_1j_2j_3}_{m'_1m'_2m_3}&=&\Pi^{j_1 j_2}_{m_1m_2\,,m'_1m'_2} \; ,
\end{eqnarray}
where $\Pi^{j_1 j_2}_{m_1m_2\,,m'_1m'_2}$ projects out the trace zero part in $V_{j_1}\otimes V_{j_2}$. The orthogonality relation for the Clebsch-Gordan coefficients is given as
\begin{equation} \label{ortho}
\sum_{m_1,m_2} {}_qC^{j_1j_2j}_{m_1m_2m} \,{}_qC^{j_1j_2j'}_{m_1m_2m'}\,\,=\,\,\delta_{jj'}\delta_{mm'} \theta_{j_1j_2j} \; ,
\end{equation}
where $\theta_{j_1j_2j}=1$ if the coupling conditions (\ref{couplingcond}) are satisfied and vanishing otherwise.

\subsection{Diagrammatic Calculus}

The introduction of a quantum group complicates some definitions known in the classical case, e.g. the notion of a dual, which is necessary to calculate the recoupling basis of the tensor in appendix \ref{app:tensor} and section \ref{sec:renormalization}. To do so, a convenient graphical representation has been introduced in \cite{q_spinnet}.
The quantum group requires to specify a special direction, which we will take as the vertical direction and can be interpreted as maps from a tensor product of representation spaces of $\text{SU}(2)_k$, represented by incoming lines from below, to a tensor product of representation spaces, drawn as outgoing lines on top. Each of these lines carries a representation label $j$ and a magnetic index $m$. One basic example of such a map are the Clebsch-Gordan coefficients, denoted by $_q\mathcal{C}_{m_1\,m_2\,m_3}^{j_1 \, j_2 \, j_3}$\footnote{This is not the standard Clebsch-Gordan coefficient defined in \cite{biedenharn}, but it is modified by the quantum dimension: ${}_q\mathcal{C}_{m_1\,m_2\,m_3}^{j_1 \, j_2 \, j_3} = {}_q C_{m_1\,m_2\,m_3}^{j_1 \, j_2 \, j_3 } \left(\sqrt{d_{j_3}}\right)^{-1}$.}. They are interpreted as a map $V_{j_1} \otimes V_{j_2} \rightarrow V_{j_3}$, symbolizing how the spins $j_1$ and $j_2$ (with their respective magnetic indices) couple to $j_3$. We have already introduced their graphical representation in equation \eqref{eq:cg1}:
\begin{equation}\label{eq:clebsch}
\begin{tikzpicture} [baseline,scale=0.75]
\draw (0,-0.75) node {$j_1$}
      (0,-0.5) -- (0.5,0) -- (1,-0.5)
      (1,-0.75) node {$j_2$}
      (0.5,0) -- (0.5,0.5) 
      (0.5,0.75) node {$j_3$};
\end{tikzpicture}
:= {}_q\mathcal{C}_{m_1\,m_2\,m_3}^{j_1 \, j_2 \, j_3} \quad .
\end{equation}
A particular version of this Clebsch-Gordan coefficient will be important later on: If we choose $j_1 = j_2 = j$ and take $j_3 = 0$, we define the `cap' as a map: $V_j\otimes V_j\rightarrow {\mathbb C}$, namely
\begin{equation} 
\begin{tikzpicture} [baseline,scale=0.75]
\draw (0,-0.75) node {$m$}
      (0,-0.5) -- (0,0) 
      (1,0) arc(-0:180:0.5)
      (0.5,0.75) node {$j$}
      (1,0) -- (1,-0.5)
      (1,-0.75) node {$m'$};
\end{tikzpicture}
:= {}_q \mathcal{C}^{j\,j\,0}_{m\,m'\,0} \sqrt{d_j} =(-1)^{j-m} q^{\frac{m}{2}} \delta_{m,-m'} \quad.
\end{equation}
From this `cap' we can similarly define a `cup' by requiring that they give the identity if we concatenate them:
\begin{equation}\label{identity}
\begin{tikzpicture}[baseline,scale=0.75]
\draw (0,-0.75) node {$m$}
      (0,-0.5) -- (0,0) 
      (1,0) arc(-0:180:0.5)
      (1,0) -- (1,-0.1)
      (1,-0.1) arc(-180:0:0.5)
      (2,-0.1) -- (2,0.6)
      (2,0.85) node {$m''$};
\end{tikzpicture}
= 
\begin{tikzpicture} [baseline,scale=0.75]
\draw (0,-0.75) node {$m$}
      (0,-0.5) -- (0,0.6)
      (0,0.85) node {$m''$};
\end{tikzpicture}
= \delta_m^{m''} \quad ,
\end{equation}
which gives:
\begin{equation}
\begin{tikzpicture}[baseline,scale=0.75]
\draw (0,0.75) node {$m$}
      (0,0.5) -- (0,0)
      (0,0) arc(-180:0:0.5)
      (1,0) -- (1,0.5)
        (0.5,-0.75) node {$j$}
      (1,0.75) node {$m'$};
\end{tikzpicture}
= (-1)^{j+m} q^{\frac{m}{2}} \delta_{m,-m'} \quad .
\end{equation}
Using these `cups' and `caps', we can construct the Clebsch-Gordan coefficients for the quantum group with inverse (here: complex conjugate) deformation parameter $\bar{q}$, already given in \eqref{eq:cg2}, by `bending up' one of the lower legs of the Clebsch-Gordan in \eqref{eq:clebsch}.
\begin{equation}
{}_{\bar{q}} \mathcal{C}_{m_1\, m_2 \, m_3}^{j_1 \, j_2 \, j_3} =
\begin{tikzpicture}[baseline,scale=0.75]
\draw (0,-0.55) -- (0,0) -- (-0.5,0.5)
      (0,0) -- (0.5,0.5)
      (0,-0.8) node {$j_3$}
      (-0.5,0.75) node {$j_2$}
      (0.5,0.75) node {$j_1$};
\end{tikzpicture}
\; = \;
\begin{tikzpicture}[baseline,scale=0.75]
\draw (-0.5,-1) -- (-0.5,-0.5) -- (0,0)
      (0,0) -- (0,0.7)
      (0,0) -- (0.5,-0.5)
      (0.5,-0.5) arc (-180:0:0.5)
      (1.5,-0.5) -- (1.5,0.7)
      (-0.5,-1.25) node {$j_3$}
      (0,0.95) node {$j_2$}
      (1.5,0.95) node {$j_1$};
\end{tikzpicture}
\; = \;
\begin{tikzpicture} [baseline,scale=0.75]
\draw (-1.5,0.7) -- (-1.5,-0.5)
      (-1.5,-0.5) arc (-180:0:0.5)
      (-0.5,-0.5) -- (0,0) -- (0,0.7)
      (0,0) -- (0.5,-0.5) -- (0.5,-1)
      (-1.5,0.95) node {$j_2$}
      (0,0.95) node {$j_1$}
      (0.5,-1.25) node {$j_3$};
\end{tikzpicture} \quad .
\end{equation}
This map can hence be interpreted as mapping $V_{j_3} \rightarrow V_{j_1} \otimes V_{j_2}$, thus it is  dual to \eqref{eq:clebsch}. With a `cap' we can `pull down' one of the legs again and arrive back at \eqref{eq:clebsch}:
\begin{equation}
\begin{tikzpicture} [baseline,scale=0.75]
\draw (-0.5,0.5) -- (0,0) -- (0,-0.7)
      (0,0) -- (0.5,0.5) 
      (1.5,0.5) arc(0:180:0.5)
      (1.5,0.5) -- (1.5,-0.7)
      (-0.5,0.75) node {$j_5$}
      (0,-0.95) node {$j_3$}
      (1.5,-0.95) node {$j_4$};
\end{tikzpicture}
\; = \;
\begin{tikzpicture} [baseline,scale=0.75]
\draw (-0.5,-0.5) -- (0,0) -- (0,0.7)
      (0,0) -- (0.5,-0.5)
      (-0.5,-0.75) node {$j_3$}
      (0.5,-0.75) node {$j_4$}
      (0,0.95) node {$j_5$};
\end{tikzpicture}
\; = \;
\begin{tikzpicture} [baseline,scale=0.75]
\draw (-1.5,-0.7) -- (-1.5,0.5) 
      (-0.5,0.5) arc(0:180:0.5)
      (-0.5,0.5) -- (0,0) -- (0,-0.7)
      (0,0) -- (0.5,0.5)
      (-1.5,-0.95) node {$j_3$}
      (0,-0.95) node {$j_4$}
      (0.5,0.75) node {$j_5$};
\end{tikzpicture} \quad .
\end{equation}
Concatenating these two maps, we obtain a map $V_{j_3} \rightarrow V_{j_3}$ proportional to the identity.
\begin{equation}
\begin{tikzpicture} [baseline,scale=0.75]
\draw (0,-1) -- (0,-0.5) -- (-0.5,0) -- (0,0.5) -- (0,1)
      (0,-0.5) -- (0.5,0) -- (0,0.5)
      (0,-1.25) node {$j_3$}
      (-0.75,0) node {$j_1$}
      (0.75,0) node {$j_2$}
      (0,1.25) node {$j_3$};
\end{tikzpicture}
\; = \;
\begin{tikzpicture}[baseline,scale=0.75]
\draw (-1,-1) -- (-0.5,-0.5) -- (-0.5,0) -- (0,0.5) -- (0,1)
      (-0.5,-0.5) -- (0,-1) arc(-180:0:0.25) -- (0.5,0) -- (0,0.5)
      (-1,-1.25) node {$j_3$}
      (-0.75,0) node {$j_1$}
      (0.75,0) node {$j_2$}
      (0,1.25) node {$j_3$};
\end{tikzpicture}
\; = \; (-1)^{j_1 + j_2 - j_3} d_{j_3}^{-1} \delta_{m_3\,m'_3} \quad.
\end{equation}

Given these graphical ingredients, several important identities can be derived, which we append here:

We start with the expression giving the dependence of the 4--valent intertwiner $\mathcal{P}^4_v$ \eqref{eq:4-valent-int} on the magnetic indices:
\begin{equation} \label{eq:4-valent-basis}
\begin{tikzpicture}[baseline,scale=0.75]
\draw (-0.5,-1) -- (0,-0.5) -- (0,0.5) -- (-0.5,1)
      (0.5,-1) -- (0,-0.5)
      (0,0.5) -- (0.5,1)
      (-0.5,-1.25) node {$j_3$}
      (0.5,-1.25) node {$j_4$}
      (0.5,1.25) node {$j_1$}
      (-0.5,1.25) node {$j_2$}
      (0.25,0) node {$j_5$};
\end{tikzpicture}
\; = \; \sum_{m_5} {}_{\bar{q}} \mathcal{C}^{j_1\,j_2\,j_5}_{m_1\,m_2\,m_5} \; {}_q \mathcal{C}^{j_3\,j_4\,j_5}_{m_3\,m_4\,m_5} \quad .
\end{equation}
Its dual is defined by placing `cups' on its bottom legs and `caps' on its top ones (such that these do not cross). In terms of Clebsch--Gordan coefficients this reads:
\begin{align}\label{dualize}
\begin{tikzpicture}[baseline,scale=0.75]
\draw (-0.25,-0.75) -- (0,-0.5) -- (0,0.5) -- (-0.25,0.75) arc(0:180:0.25)
      (0.25,-0.75) -- (0,-0.5)
      (0,0.5) -- (0.25,0.75) arc(0:180:0.75)
      (-0.25,-0.75) arc(0:-180:0.25)
      (0.25,-0.75) arc(0:-180:0.75)
      (-0.75,0.5) node {$j_1$}
      (-1.25,0.5) node {$j_2$}
      (-0.75,-0.5) node {$j_4$}
      (-1.25,-0.5) node {$j_3$}
      (0.25,0) node {$j_5$};
\end{tikzpicture}
\; = & \;
\begin{tikzpicture}[baseline,scale=0.75]
\draw (-0.5,0.75) -- (-0.25,1) arc(180:0:0.5) -- (0.75,-1) arc(0:-180:0.5) -- (-0.5,-0.75)
      (-0.25,1) -- (0,0.75)
      (-0.25,-1) -- (0,-0.75)
      (-0,0.5) node {$j_1$}
      (-0.5,0.5) node {$j_2$}
      (-0,-0.5) node {$j_4$}
      (-0.5,-0.5) node {$j_3$}
      (1,0) node {$j_5$};
\end{tikzpicture}
\; = \; \nonumber \\
= & \;(-1)^{2 j_5} \sum_{m_5}\; q^{m_5} \; {}_{\bar{q}} \mathcal{C}^{j_1 \, j_2 \, j_5}_{m_1 \, m_2 \, m_5} \; {}_q \mathcal{C}^{j_3 \, j_4 \, j_5}_{m_3 \, m_4 \, m_5}
\quad ,
\end{align}
In fact, if we connect the diagrams \eqref{eq:4-valent-basis} and \eqref{dualize}, we obtain the following diagram
\begin{equation}
\begin{tikzpicture}[baseline,scale=0.8]
\draw (-0.25,-0.75) -- (0,-0.5) -- (0,0.5) -- (-0.25,0.75) arc(0:180:0.25) -- (-1,0.5) -- (-1,-0.5)
      (0.25,-0.75) -- (0,-0.5)
      (0,0.5) -- (0.25,0.75) arc(0:180:0.75) -- (-1,0.5)
      (-0.25,-0.75) arc(0:-180:0.25) -- (-1,-0.5)
      (0.25,-0.75) arc(0:-180:0.75) -- (-1,-0.5)
      (-1.25,0) node {$j_5$}
      (0.25,0) node {$j'_5$}
      (-0.5,0.75) node {$j_1$}
      (-0.5,-0.75) node {$j_3$}
      (-0.5,1.75) node {$j_2$}
      (-0.5,-1.75) node {$j_4$};
\end{tikzpicture}
\; = \; (-1)^{j_1+j_2+j_3+j_4} \left(d_{j_5}\right)^{-1} \delta_{j_5 j'_5}\; ,
\end{equation}
from which we can deduce that the map
\begin{align}
& \mathbf{P}{\left(\{m\},\{m'\}\right)}(j_1,j_2,j_3,j_4) := \nonumber \\
= & \sum_{j_5} (-1)^{j_1+j_2+j_3+j_4}d_{j_5}
\begin{tikzpicture}[baseline,scale=0.75]
\draw (-0.5,0.75) -- (-0.25,1) arc(180:0:0.5) -- (0.75,-1) arc(0:-180:0.5) -- (-0.5,-0.75)
      (-0.25,1) -- (0,0.75)
      (-0.25,-1) -- (0,-0.75)
      (-0,0.5) node {$j_1$}
      (-0.5,0.5) node {$j_2$}
      (-0,-0.5) node {$j_4$}
      (-0.5,-0.5) node {$j_3$}
      (1,0) node {$j_5$};
\end{tikzpicture}
\; \otimes \;
\begin{tikzpicture}[baseline,scale=0.75]
\draw (-0.5,-1) -- (0,-0.5) -- (0,0.5) -- (-0.5,1)
      (0.5,-1) -- (0,-0.5)
      (0,0.5) -- (0.5,1)
      (-0.5,-1.25) node {$j_3$}
      (0.5,-1.25) node {$j_4$}
      (0.5,1.25) node {$j_1$}
      (-0.5,1.25) node {$j_2$}
      (0.25,0) node {$j_5$};
\end{tikzpicture}
\quad ,
\end{align}
can be used to project onto the basis \eqref{eq:4-valent-basis}, from which also follows that the components of a 4--valent tensor in that basis, e.g. $\hat{T}$ in appendix \ref{app:tensor} / section \ref{sec:renormalization}, can be computed by contracting that tensor with \eqref{dualize}. Moreover, we can also compute the change of basis related to a different 4--valent recoupling scheme:

We are looking for the relation
\begin{equation}
\begin{tikzpicture}[baseline,scale=0.75]
\draw (-1,0.5) -- (-0.5,0) -- (0,0) -- (0.5,0.5)
      (-0.5,0) -- (-1,-0.5)
      (0,0) -- (0.5,-0.5)
      (-1,-0.75) node {$j_3$}
      (-1,0.75) node {$j_2$}
      (0.5,0.75) node {$j_1$}
      (0.5,-0.75) node {$j_4$}
      (-0.25,0.25) node {$j_6$};
\end{tikzpicture}
\; = \; \sum_{j_5} c(j_5,j_6) 
\begin{tikzpicture}[baseline,scale=0.75]
\draw (-0.5,-0.75) -- (0,-0.25) -- (0,0.25) -- (-0.5,0.75)
      (0.5,-0.75) -- (0,-0.25)
      (0,0.25) -- (0.5,0.75)
      (-0.75,-0.75) node {$j_3$}
      (0.75,-0.75) node {$j_4$}
      (0.75,0.75) node {$j_1$}
      (-0.75,0.75) node {$j_2$}
      (0.25,0) node {$j_5$};
\end{tikzpicture}
\quad .
\end{equation}
In order to find the coefficient $ c(j_5,j_6) $ we contract the above expression with the diagram \eqref{dualize}:
\begin{align} \label{eq:change-of-basis}
\begin{tikzpicture}[baseline,scale=0.8]
\draw (-1,0.5) -- (-0.5,0) -- (0,0) -- (0.5,0.5) -- (-0.25,1) arc(180:0:0.75) -- (1.25,-1) arc(0:-180:0.75) -- (0.5,-0.5)
      (-0.5,0) -- (-1,-0.5) -- (-0.25,-1)
      (0,0) -- (0.5,-0.5)
      (-1,0.5) -- (-0.25,1)
      (-1,-0.75) node {$j_3$}
      (-1,0.75) node {$j_2$}
      (0.5,0.75) node {$j_1$}
      (0.5,-0.75) node {$j_4$}
      (-0.25,0.25) node {$j_6$}
      (1,0) node {$j'_5$};
\end{tikzpicture}
\; & = \; \sum_{j_5} c(j_5,j_6) 
\begin{tikzpicture}[baseline,scale=0.75]
\draw (-0.5,-0.75) -- (0,-0.25) -- (0,0.25) -- (-0.5,0.75) -- (0,1.25) arc(180:0:0.75) -- (1.5,-1.25) arc(0:-180:0.75) -- (0.5,-0.75)
      (0.5,-0.75) -- (0,-0.25)
      (0.5,0.75) -- (0,1.25)
      (-0.5,-0.75) -- (0,-1.25)
      (0,0.25) -- (0.5,0.75)
      (-0.75,-0.75) node {$j_3$}
      (0.75,-0.75) node {$j_4$}
      (0.75,0.75) node {$j_1$}
      (-0.75,0.75) node {$j_2$}
      (0.25,0) node {$j_5$}
      (1.25,0) node {$j'_5$};
\end{tikzpicture}
\end{align}
The diagram on the left hand side is the $6j$ symbol of $\text{SU}(2)_k$, given as a particular contraction of four Clebsch--Gordan coefficients:
\begin{align} \label{6j-def}
&
\begin{tikzpicture}[baseline,scale=0.6]
\draw (0,-2) -- (0,-1.5) -- (-0.5,-1) -- (-0.5,-0.5) -- (0.5,0.5) -- (0.5,1) -- (0,1.5) -- (0,2) arc(180:0:1) -- (2,-2) arc(0:-180:1)
      (0.,-1.5) -- (1,-0.5) -- (1,0) -- (0.5,0.5)
      (-0.5,-0.5) -- (-1,0) -- (-1,0.5) -- (0,1.5)
      (-0.75,-1) node {$j_1$}
      (1.3,-0.5) node {$j_2$}
      (-1.25,0) node {$j_3$}
      (0.8,1) node {$j_4$}
      (2.3,0) node{$j_5$}
      (-0.2,0.25) node {$j_6$};
\end{tikzpicture}
\; = \;
\begin{tikzpicture}[baseline,scale=0.6]
\draw (0,-2) -- (0,-1.5) -- (-1,-0.5) -- (-1,0) -- (-0.5,0.5) -- (-0.5,1) -- (0,1.5) -- (0,2) arc(180:0:1) -- (2,-2) arc(0:-180:1)
      (0.,-1.5) -- (0.5,-1) -- (0.5,-0.5) -- (1,0) -- (1,0.5) -- (0,1.5)
      (0.5,-0.5) -- (-0.5,0.5)
      (-1.25,-0.5) node {$j_1$}
      (0.8,-1) node {$j_2$}
      (1.3,0) node {$j_4$}
      (-0.75,1) node {$j_3$}
      (2.3,0) node {$j_5$}
      (0.2,0.25) node {$j_6$};
\end{tikzpicture}
\; = \; \nonumber \\
= & \;
\left \{
\begin{matrix}
\,j_1\, & \,j_2\, & \,j_5\, \\
\,j_4\, & \,j_3\, & \,j_6\,
\end{matrix}
\right \} =: \frac{(-1)^{j_1+j_2+j_3+j_4}}{\sqrt{d_{j_5}d_{j_6}}}\left[
\begin{matrix}
\, j_1 \, & \, j_2 \,  & \, j_5 \, \\
\, j_4 \, & \, j_3 \, & \, j_6 \,
\end{matrix}
\right] \quad .
\end{align}
Calculating both sides of \eqref{eq:change-of-basis}, we find
\begin{align}
& (-1)^{j_1+j_2+j_3+j_4} \left( d_{j_5}d_{j_6} \right)^{-\frac{1}{2}}
\left[
\begin{matrix}
\, j_1 \, & \, j_2 \, & \, j'_5 \, \\
\, j_3 \, & \, j_4 \, & \, j_6 \,
\end{matrix}
\right]
= \; \nonumber \\
 = & (-1)^{j_1 + j_2 + j_3 + j_4} \sum_{j_5} c(j_5,j_6) \left(d_{j_5} \right)^{-1} \delta_{j_5,j'_5} \quad,
\end{align}
from which we deduce
\begin{align}
\begin{tikzpicture}[baseline,scale=0.75]
\draw (-1,0.5) -- (-0.5,0) -- (0,0) -- (0.5,0.5)
      (-0.5,0) -- (-1,-0.5)
      (0,0) -- (0.5,-0.5)
      (-1,-0.75) node {$j_3$}
      (-1,0.75) node {$j_2$}
      (0.5,0.75) node {$j_1$}
      (0.5,-0.75) node {$j_4$}
      (-0.25,0.25) node {$j_6$};
\end{tikzpicture}
\; & = \; \sum_{j_5}  \sqrt{\frac{d_{j_5}}{d_{j_6}}} 
\left[
\begin{matrix}
\, j_1 \, & \, j_2 \, & \, j'_5 \, \\
\, j_3 \, & \, j_4 \, & \, j_6 \,
\end{matrix}
\right] 
\begin{tikzpicture}[baseline,scale=0.75]
\draw (-0.5,-0.75) -- (0,-0.25) -- (0,0.25) -- (-0.5,0.75)
      (0.5,-0.75) -- (0,-0.25)
      (0,0.25) -- (0.5,0.75)
      (-0.75,-0.75) node {$j_3$}
      (0.75,-0.75) node {$j_4$}
      (0.75,0.75) node {$j_1$}
      (-0.75,0.75) node {$j_2$}
      (0.25,0) node {$j_5$};
\end{tikzpicture}
\quad .
\end{align}
As a last step, let us briefly explain how the diagram \eqref{eq:9j-symbol} enters into \eqref{eq:new-effective-tensor} and \eqref{eq:new-effective-tensor2} and also why \eqref{eq:9j-symbol} can be split into two $6j$ symbols for the 3--valent algorithm in section \ref{sec:renormalization}.

If we consider just the magnetic indices of the intermediate 3--valent tensors $S_i$ (equations \eqref{eq:S1} to \eqref{eq:S4}) and connect the 3--valent vertices according to \eqref{eq:new-effective-tensor} or figure fig. \ref{fig:algo1}, we find a diagram with two bottom legs and two top legs connected by a square. To compute the components of \eqref{eq:new-effective-tensor} in the basis \eqref{eq:4-valent-basis}, it has to be contracted with \eqref{dualize} resulting in \eqref{eq:new-effective-tensor2}. The diagram that we have to compute is thus:
\begin{equation} \label{eq:9j-graph}
\begin{tikzpicture}[baseline,scale=0.8]
\draw (-1,-1) -- (-0.5,-0.5) -- (-0.5,0.5) -- (-1,1) -- (0,1.5) arc(180:0:0.75) -- (1.5,-1.5) arc(0:-180:0.75) -- (-1,-1)
      (0,-1.5) -- (1,-1) -- (0.5,-0.5) -- (0.5,0.5) -- (1,1) -- (0,1.5)
      (-0.5,-0.5) -- (0.5,-0.5)
      (-0.5,0.5) -- (0.5,0.5)
      (0,-0.75) node {$j_3$}
      (-0.75,0) node {$j_2$}
      (0.75,0) node {$j_4$}
      (0,0.75) node {$j_1$}
      (-1.25,-1.) node {$j_7$}
      (1.25,-1.) node {$j_8$}
      (1.25,1.) node {$j_5$}
      (-1.25,1.) node {$j_6$}
      (1.75,0) node {$j_9$};
\end{tikzpicture} \quad .
\end{equation}
Instead of doing this directly we use the following two identities to simplify it first:
\begin{equation} \label{eq:identity1}
\begin{tikzpicture}[baseline,scale=0.75]
\draw (0,-1) -- (0,1) 
(1,-1) -- (1,1) 
         (-0.25,0) node {$j_2$}
      (1.25,0) node {$j_4$};
\end{tikzpicture}
\; = \;
\begin{tikzpicture}[baseline,scale=0.75]
\draw (-0.5,-1)-- (0,-0.5) -- (0,0.5) --(-0.5,1)
	(0,0.5) --(0.5,1)
	(0.5,-1)-- (0,-0.5)
      (0.3,0) node {$j_9$}
      (-0.75,1) node {$j_2$}
      (0.75,1) node {$j_4$}
       (-0.75,-1) node {$j_2$}
      (0.75,-1) node {$j_4$};
\end{tikzpicture}
\; (-1)^{j_2+j_4-j_9}{d_{j_9}} \; ,
\end{equation}
\begin{equation} \label{eq:identity2}
\begin{tikzpicture}[baseline,scale=0.75]
\draw [pattern=north east lines] (0,0) circle[radius = 0.5];
\draw (0,0.5) -- (0,1.5) 
      (0,-0.5) -- (0,-1.5)
      (0.25,-1) node {$j_9$}
      (0.25,1) node {$j_9$};
\end{tikzpicture}
\; = \;
\begin{tikzpicture}[baseline,scale=0.75]
\draw [pattern=north east lines] (0,0) circle[radius = 0.5];
\draw (0,0.5) -- (0,1) arc(180:0:0.5) -- (1,-1) arc(0:-180:0.5)
      (0,-0.5) -- (0,-1)
      (1.25,0) node {$j_9$};
\end{tikzpicture}
\; \frac{(-1)^{2j_9}}{d_{j_9}} \;
\begin{tikzpicture}[baseline,scale=0.75]
\draw (0,-1) -- (0,1)
      (0.3,0) node {$j_9$};
\end{tikzpicture} \quad .
\end{equation}
Thus we can manipulate \eqref{eq:9j-graph} in the following way:
\begin{align}
\begin{tikzpicture}[baseline,scale=0.8]
\draw (-1,-1) -- (-0.5,-0.5) -- (-0.5,0.5) -- (-1,1) -- (0,1.5) arc(180:0:0.75) -- (1.5,-1.5) arc(0:-180:0.75) -- (-1,-1)
      (0,-1.5) -- (1,-1) -- (0.5,-0.5) -- (0.5,0.5) -- (1,1) -- (0,1.5)
      (-0.5,-0.5) -- (0.5,-0.5)
      (-0.5,0.5) -- (0.5,0.5)
      (0,-0.75) node {$j_3$}
      (-0.75,0) node {$j_2$}
      (0.75,0) node {$j_4$}
      (0,0.75) node {$j_1$}
      (-1.25,-1.) node {$j_7$}
      (1.25,-1.) node {$j_8$}
      (1.25,1.) node {$j_5$}
      (-1.25,1.) node {$j_6$}
      (1.75,0) node {$j_9$};
\end{tikzpicture}
\; = & \;
(-1)^{j_2 + j_4 - j_9} d_{j_9}
\begin{tikzpicture}[baseline,scale=0.9]
\draw (-0.5,-0.75) -- (0,-0.25) -- (0,0.25) -- (-0.5,0.75) -- (0,1.25) arc(180:0:0.75) -- (1.5,-1.25) arc(0:-180:0.75) -- (0.5,-0.75)
      (0.5,-0.75) -- (0,-0.25)
      (0.5,0.75) -- (0,1.25)
      (-0.5,-0.75) -- (0,-1.25)
      (0,0.25) -- (0.5,0.75)
      (-0.5,0.75) -- (0.5,0.75)
      (-0.5,-0.75) -- (0.5,-0.75)
       (0,-0.5) node {$j_3$}
      (0,0.55) node {$j_1$}
      (-0.6,1) node {$j_6$}
      (0.6,1) node {$j_5$}
      (0.6,0.5) node {$j_4$}
      (-0.6,0.5) node {$j_2$}
      (0.6,-0.4) node {$j_4$}
      (-0.6,-0.4) node {$j_2$}
      (-0.6,-1) node {$j_7$}
      (0.6,-1) node {$j_8$}
      (0.25,0) node {$j_9$}
      (1.75,0) node {$j_9$};
\end{tikzpicture}  \nonumber \\
= & \;
(-1)^{j_2+j_4-j_9}
\begin{tikzpicture}[baseline,scale=0.9]
\draw (0,1.75) arc(180:0:0.5) -- (1.,0.75) arc(0:-180:0.5)
      (0,-0.75)  arc(180:0:0.5) -- (1,-1.75) arc(0:-180:0.5)
      (0,0.75)--(0.5,1.25)
        (0,1.75)--(0.5,1.25)
        (0,1.75)--(-0.5,1.25)
           (0,0.75)--(-0.5,1.25)
           (-0.5,1.25)--(0.5,1.25)
            (0,-0.75)--(0.5,-1.25)
        (0,-1.75)--(0.5,-1.25)
        (0,-1.75)--(-0.5,-1.25)
           (0,-0.75)--(-0.5,-1.25)
           (-0.5,-1.25)--(0.5,-1.25)
      (0.,-1.) node {$j_3$}
      (0.,1.1) node {$j_1$}
      (-0.6,1.6) node {$j_6$}
      (0.6,1.6) node {$j_5$}
      (0.6,0.75) node {$j_4$}
      (-0.6,0.75) node {$j_2$}
      (0.6,-0.75) node {$j_4$}
      (-0.6,-0.75) node {$j_2$}
      (-0.6,-1.6) node {$j_7$}
      (0.6,-1.6) node {$j_8$}
      (1.25,-1.25) node {$j_9$}
      (1.25,1.25) node {$j_9$};
\end{tikzpicture}
\quad .
\end{align}
The diagrams on the right hand side are again $6j$-symbols defined in \eqref{6j-def}. Eventually, we find
\begin{align}
\begin{tikzpicture}[baseline,scale=0.8]
\draw (-1,-1) -- (-0.5,-0.5) -- (-0.5,0.5) -- (-1,1) -- (0,1.5) arc(180:0:0.75) -- (1.5,-1.5) arc(0:-180:0.75) -- (-1,-1)
      (0,-1.5) -- (1,-1) -- (0.5,-0.5) -- (0.5,0.5) -- (1,1) -- (0,1.5)
      (-0.5,-0.5) -- (0.5,-0.5)
      (-0.5,0.5) -- (0.5,0.5)
      (0,-0.75) node {$j_3$}
      (-0.75,0) node {$j_2$}
      (0.75,0) node {$j_4$}
      (0,0.75) node {$j_1$}
      (-1.25,-1.) node {$j_7$}
      (1.25,-1.) node {$j_8$}
      (1.25,1.) node {$j_5$}
      (-1.25,1.) node {$j_6$}
      (1.75,0) node {$j_9$};
\end{tikzpicture}
= &
\frac{(-1)^{j_2 + j_4 + j_9} (-1)^{j_5+j_6+j_7+j_8} }{d_{j_9} \sqrt{d_{j_1} d_{j_3}}} \; \times \nonumber \\
& \times \;
 \left[
\begin{matrix}
\, j_2 \, & \, j_4 \,  & \, j_9 \, \\
\, j_5 \, & \, j_6 \, & \, j_1 \,
\end{matrix}
\right] 
\left[
\begin{matrix}
\, j_2 \, & \, j_4 \,  & \, j_9 \, \\
\, j_8 \, & \, j_7 \, & \, j_3 \,
\end{matrix}
\right] \; .
\end{align}

\section{4--valent tensor network renormalization: general idea and derivation of symmetry protecting algorithm} \label{app:tensor}

In this appendix we will explain the 4--valent tensor network algorithm in more detail. As mentioned already in section \ref{sec:renormalization}, the general idea is to encode the entire dynamics of a (discrete) system, in particular the partition function, into a tensor network, i.e. a local contraction of multidimensional arrays, and evaluate it in steps via coarse graining. The latter is done via a local manipulation of the network, in which one defines new coarse degrees of freedom from the finer ones and introduces a truncation, which allows for a control on the error being made. Therefore, it is necessary to know the relevance of the coarse degrees of freedom (with respect to the other ones).

\begin{figure}
\includegraphics[width=0.45\textwidth]{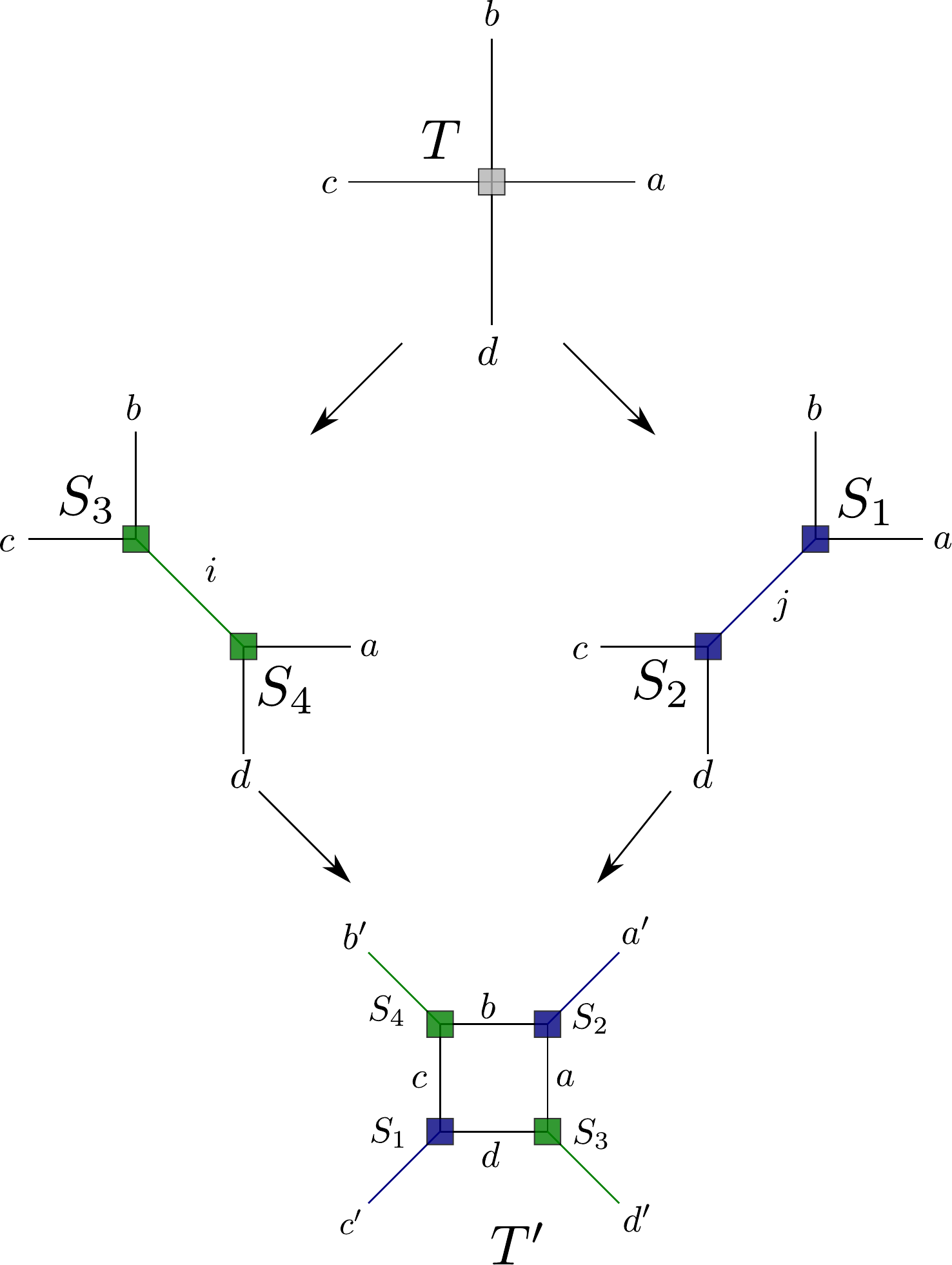}
\caption{Tensor network renormalization for a regular square lattice. The tensor $T$ is split into 3--valent tensors in two different ways by grouping two exterior legs together and performing an SVD. A truncation is performed on the new indices $i$ by only taking the largest $\chi$ singular values into account. The new truncated tensors $S_i$ are combined into a new 4--valent tensors by summing over the old indices.
\label{fig:algo1}}
\end{figure}

Such variable transformations are computed via a singular value decomposition (SVD). To explain this let us discuss the concrete algorithm developed in \cite{levin,guwen}. Consider a tensor network on a 2D square lattice with identical 4--valent tensors $T_{abcd}$ on all vertices. Divide the tensors into even and odd ones, and split them according to fig. \ref{fig:algo1}, where the odd ones are green and the even ones blue. This splitting is performed via a singular value decomposition acting on the following matrices:
\begin{align}
T_{(ab);(cd)} =: & M^{(1)}_{(ab),(cd)} \nonumber \\
= & \sum_i U^{(1)}_{(ab),i} \, \lambda^{(1)}_i \, (V^{(1)})^\dagger_{(cd),i} \quad , \\
T_{(da);(bc)} =: & M^{(2)}_{(da),(bc)} \nonumber \\
= & \sum_i U^{(2)}_{(da),i} \, \lambda^{(2)}_i \, (V^{(2)})^\dagger_{(bc),i} \quad ,
\end{align}
where $U^{(l)}_{(ab),i}$ and $V^{(l)}_{(cd),i}$ are the singular vectors, $\lambda^{(l)}_i$ the singular values of the matrix $M^{(l)}_{(ab),(cd)}$\footnote{$(ab)$ denotes that the indices $a,b$ have been combined into one index, such that the entries of the tensor are arranged as the indices of a matrix.}. $U$ and $V$ are unitary matrices, i.e. the singular vectors are orthonormal, while the singular values are non--negative and ordered in size, $\lambda_1 \geq \lambda_2 \geq ...\geq \lambda_N \geq 0$.

Such a singular value decomposition can be applied to any matrix $M$, say of rank $N$, where the size of the $N$ non--vanishing singular values $\lambda_i$ signifies the significance of a singular vector. Conversely, given the decomposition of the matrix, it can be approximated by a matrix of rank $\chi$ by only keeping the $\chi$ largest singular values. Actually this matrix is the best approximation of $M$ by a matrix of rank $\chi$ with respect to the least square error. Whether this is a good approximation can be directly estimated from the size of the discarded singular values with respect to the kept ones, in particular the largest.

In the context of tensor network renormalization, the matrices $U$ and $V$ are variable transformations, in fact isometries. They map the fine degrees of freedom, encoded in the indices $a$,$b$, etc., into coarse degrees of freedom that are (for the moment) labelled by the number of the singular value. They also give the intermediate 3--valent tensors:
\begin{eqnarray}
(S_{1,3})_{(ab),i} :=& U^{(1,2)}_{(ab),i} \, \sqrt{\lambda_i} \quad , \\
(S_{2,4})_{(ab),i} :=& V^{(1,2)}_{(ab),i} \, \sqrt{\lambda_i} \quad .
\end{eqnarray}
Truncating the number of singular values results in a truncation on the index range on the new edges of the intermediate tensor network. As a final step, the fine edges, labelled by the original indices $a,b,...$, are contracted and one defines a new tensor $T'$:
\begin{equation}
T'_{a'b'c'd'} = \sum_{a,b,c,d} (S_1)_{dc,c'} (S_2)_{ba,a'} (S_3)_{ad,c'} (S_4)_{cb,a'} \quad .
\end{equation}
The new tensor network consisting of effective tensors $T'$ is coarser and tilted by 45 degrees. From here on the procedure is iterated; the system has reached a fixed point if the tensor does not change under consecutive iterations, described by a fixed point tensor $T^*$.

While this truncation solves the practical problem of exponentially growing index ranges, the interpretation of the new degrees of freedom is unclear. In each iteration of the algorithm, the fine variables get redefined and lose their meaning (in terms of the original variables) if one forgets these variable redefinitions, yet even keeping them is not very feasible after many iterations. Moreover, if a system possesses symmetries, e.g. from an underlying (quantum) group, these symmetries will also be present in the tensor, e.g. the $\mathbb{Z}_2$ delta--function present on each vertex in the Ising model. This symmetry will also survive the coarse graining procedure, but may not be obvious to identify after several variable redefinitions. Hence it is preferable to use an algorithm that explicitly preserves the symmetries.

On the level of the matrix that is to be split one can achieve this by considering representation theory of the underlying group(s). Since we are considering intertwiners, i.e. elements of the invariant subspace of a tensor product of representation vector spaces, on the vertices, the irreducible representations on the adjacent edges must couple to the trivial representation, e.g. $k=0$ for $\mathbb{Z}_2$. 
This 4--valent intertwiner space on the vertices can be split into 3--valent intertwiner spaces, analogous to the splitting of the tensor network. The new edge connecting the 3--valent vertices carries a new representation, which for $\mathbb{Z}_2$ is uniquely determined. If the 4--valent intertwiner space is not one dimensional, as for $\text{SU}(2)_k$, the intermediate label can take multiple values. Conversely, a specific intermediate label can allow several pairs of representations on the fine edges. Crucially, the splitted intertwiner admits the same configurations of exterior edges as the original one, but allows them to be arranged according to the intermediate label.

The same holds for the matrices $M$ subject to the SVD. Their entries can be rearranged according to the intermediate label, which turns them into a block diagonal form. It is thus sufficient to only consider the blocks for each intermediate label and neglect the forbidden configurations. This not only has computational advantages, since we can perform more SVDs on smaller matrices, but also interpretative ones. We call these blocks intertwiner channels, see also \cite{s3_tnw,q_spinnet}: Their label will be endowed onto the new edge of the coarser tensor network, which is the same label as the original model. Thus we also explicitly preserve the intertwiner structure on the vertices. The algorithm also slightly changes because we perform a single SVD for each block. In order to obtain a correct truncation of the $\chi$ largest singular values, it is necessary to compare all values from all blocks and take the $\chi$ largest of them. As a result, some intertwiner labels can appear with a multiplicity larger than one and lead to a generalized class of models with respect to the original one.

In the system at hand this block diagonal form depends on two parameters, a $\text{SU}(2)_k$ representation $j_i$ and a $\mathbb{Z}_2$ representation $k_i$. For $\text{SU}(2)_k$ this form is computed by expanding it in a particular recoupling basis, essentially of Clebsch--Gordan coefficients, which capture the dependence on the magnetic indices which is unchanged during the coarse graining procedure. This idea has been developed for (finite) non--Abelian groups in \cite{s3_tnw} and has been derived in great detail for quantum groups in \cite{q_spinnet} and will not be repeated here. We add a small section in the appendix \ref{app:quantum-group} to provide a short tour for the interested reader. For the Ising model, it is straightforward and can be found in detail in \cite{abelian_tnw,abelian_tnw2}, which deals with finite Abelian groups $\mathbb{Z}_q$ with $q\geq 2$. Here we will briefly present the idea:

As discussed above, the $\mathbb{Z}_2$ representations $k_e$ meeting at an edge have to sum to zero (modulo 2), which is encoded in the $\mathbb{Z}_2$ delta--function on the vertex. If we split the 4--valent vertex into two 3--valent ones, we replace one delta--function by two and introduce an additional variable $k_i$. The delta--functions enforce that $k_1 + k_2 = k_i = k_3 + k_4$, such that solving one constraint for $k_i$ restores the original delta--function. Conversely, if both $\delta$--functions are solved for e.g. $k_2$ and $k_4$, the tensor is in block diagonal form with parameter $k_i$.

To sum up this discussion, the tensors $T$ that are meant to be split, can be written in the following form:
\begin{widetext}
\begin{eqnarray} \label{eq:form-of-tensor}
T (\{j_i\},\{m_i\},\{k_i\}) = & \sum_{j_5} {\hat{T}_1}^{(j_5,k_5)}(\{j_i\},\{k_i\}) \, \delta^{(2)}(k_1 + k_2 - k_5) \, \delta^{(2)}(k_3 + k_4 - k_5) \,
\begin{tikzpicture}[baseline,scale=0.75]
\draw (-0.5,-1) -- (0,-0.5) -- (0,0.5) -- (-0.5,1)
      (0.5,-1) -- (0,-0.5)
      (0,0.5) -- (0.5,1)
      (-0.5,-1.25) node {$j_3$}
      (0.5,-1.25) node {$j_4$}
      (0.5,1.25) node {$j_1$}
      (-0.5,1.25) node {$j_2$}
      (0.25,0) node {$j_5$};
\end{tikzpicture}
 \quad , \\
T (\{j_i\},\{m_i\},\{k_i\}) = & \sum_{j_6} {\hat{T}_2}^{(j_6,k_6)}(\{j_i\},\{k_i\}) \, \delta^{(2)}(k_2 + k_3 - k_6) \, \delta^{(2)}(k_4 + k_1 - k_6) \,
\begin{tikzpicture}[baseline,scale=0.75]
\draw (-1,0.5) -- (-0.5,0) -- (0,0) -- (0.5,0.5)
      (-0.5,0) -- (-1,-0.5)
      (0,0) -- (0.5,-0.5)
      (-1,-0.75) node {$j_3$}
      (-1,0.75) node {$j_2$}
      (0.5,0.75) node {$j_1$}
      (0.5,-0.75) node {$j_4$}
      (-0.25,0.25) node {$j_6$};
\end{tikzpicture}
 \quad . 
\end{eqnarray}
\end{widetext}
The two different basis correspond to the two different splittings of the tensors, where $j_5$, $k_5$ and $j_6$, $k_6$ label the intermediate labels respectively, the graphs pictorially encode the dependence on magnetic indices. In principle, the delta--functions encoding the constraints of the Ising model can be implicitly included in the $\hat{T}_i$, yet we write them out to underline the fact that this symmetry is preserved by the algorithm. In the next step, the SVD is applied to $\hat{T}$ for each choice of intermediate labels. One computes the following 3--valent tensors:
\begin{widetext}
\begin{align} \label{eq:S1}
(S_1)^{j_5,k_5}_{m_5}(\{I\}_{\{1,2\}},i)= &\sqrt{ d_{j_5} (\lambda_1)^{(j_5,k_5)}_{ii}}
(U_1)^{(j_5,k_5)}_{\{j_1,k_1,j_2,k_2\},i} \, \delta^{(2)}(k_1 + k_2 - k_5)
\begin{tikzpicture}[baseline,scale=0.75]
\draw  (0,-0.75) -- (0,0) -- (-0.5,0.5)
      (0,0) -- (0.5,0.5)
         (0.5,0.75) node {$j_1$}
      (-0.5,0.75) node {$j_2$}
      (0.25,-0.5) node {$j_5$};
\end{tikzpicture}
\quad ,\\ \label{eq:S2}
(S_2)^{j_5,k_5}_{m_5}(\{I\}_{\{3,4\}},i)= &\sqrt{ d_{j_5}(\lambda_1)^{(j_5,k_5)}_{ii}}
(V_1)^{(j_5,k_5)}_{i,\{j_3,k_3,j_4,k_4\}} \, \delta^{(2)}(k_3 + k_4 - k_5)
\begin{tikzpicture}[baseline,scale=0.75]
\draw (-0.5,-0.5) -- (0,0) -- (0,0.75)
      (0.5,-0.5) -- (0,0)
      (-0.5,-0.75) node {$j_3$}
      (0.5,-0.75) node {$j_4$}
      (0.25,0.5) node {$j_5$};
\end{tikzpicture}
\quad ,\\ \label{eq:S3}
(S_3)^{j_6,k_6}_{m_6}(\{I\}_{\{2,3\}},i)= &\sqrt{ d_{j_6} (\lambda_2)^{(j_6,k_6)}_{ii}}
(U_2)^{(j_6,k_6)}_{\{j_2,k_2,j_3,k_3\},i} \, \delta^{(2)}(k_2 + k_3 - k_6)
\begin{tikzpicture}[baseline,scale=0.75]
\draw (-1.25,0.5) -- (-0.75,0) --(0,-0.3)
      (-0.75,0) -- (-1.25,-0.5)
      (-1,-0.75) node {$j_3$}
      (-1,0.75) node {$j_2$}
      (-0.25,0.25) node {$j_6$};
\end{tikzpicture} 
\quad ,\\ \label{eq:S4}
(S_4)^{j_6,k_6}_{m_6}(\{I\}_{\{4,1\}},i)= &\sqrt{ d_{j_6}(\lambda_2)^{(j_6,k_6)}_{ii}}
(V_2)^{(j_6,k_6)}_{i,\{j_4,k_4,j_1,k_1\}}\; \delta^{(2)}(k_4 + k_1 - k_6)
\begin{tikzpicture}[baseline,scale=0.75]
\draw (-0.75,0.25) -- (0,0) -- (0.5,0.5)
      (0,0) -- (0.5,-0.5)
      (0.5,0.75) node {$j_1$}
      (0.5,-0.75) node {$j_4$}
      (-0.5,-0.1) node {$j_6$};
\end{tikzpicture}
\quad .
\end{align}
The $I$ in the argument of $S_i$, $i=1,...,4$, summarizes both $\text{SU}(2)_k$ and $\mathbb{Z}_2$ representations. Note that both the Clebsch--Gordan coefficients and the $\mathbb{Z}_2$ delta--functions split trivially due to our choice of basis. The single non--trivial step is the SVD on the $\hat{T}_i$. As the final step, these four 3--valent tensor are combined according to the fig. \ref{fig:algo1} to form the new tensor $T'$. To obtain the recurrence relation for $\hat{T}'$, we have to contract the magnetic indices and obtain (see also appendix \ref{app:quantum-group}):
\begin{align} \label{eq:new-effective-tensor}
\hat{T}_1^{(j_5,k_5)}(I_1,I_2;I_3,I_4)& = \sum_{a,b,c,d}\sum_{\{m\}}(-1)^{j_1+j_2,+j_3,+j_4}
\begin{tikzpicture}[baseline,scale=0.75]
\draw (-0.5,0.75) -- (-0.25,1) arc(180:0:0.5) -- (0.75,-1) arc(0:-180:0.5) -- (-0.5,-0.75)
      (-0.25,1) -- (0,0.75)
      (-0.25,-1) -- (0,-0.75)
      (-0,0.5) node {$j_1$}
      (-0.5,0.5) node {$j_2$}
      (-0,-0.5) node {$j_4$}
      (-0.5,-0.5) node {$j_3$}
      (1,0) node {$j_5$};
\end{tikzpicture}
\nonumber\\
&\times
(S_2)^{j_1,k_1}_{m_1}(\{I\}_{\{b,a\}},i_1) \,
(S_4)^{j_2,k_2}_{m_2}(\{I\}_{\{b,c\}},i_2)
\nonumber\\
&\times
(S_1)^{j_3,k_3}_{m_3}(\{I\}_{\{d,c\}},i_3) \,
(S_3)^{j_4,k_4}_{m_4}(\{I\}_{\{a,d\}},i_4)\quad .
\end{align}
After substituting the equations for all $S_i$ and contracting all magnetic indices, we obtain the final recurrence relation:
\begin{align} \label{eq:new-effective-tensor2}
\hat{T}_1^{(j_5,k_5)}(I_1,I_2;I_3,I_4)&=\sum_{a,b,c,d}\sum_{\{m\}}
\frac{(-1)^{j_c + j_a + j_5}} {d_{j_5}\sqrt{d_{j_b} d_{j_d}}}
\sqrt{ d_{j_1}d_{j_2}d_{j_3}d_{j_4}}\nonumber\\
&\times
\sqrt{(\lambda_1)^{(j_1,k_1)}_{i_1i_1}\,
(\lambda_2)^{(j_2,k_2)}_{i_2i_2}\, (\lambda_1)^{(j_3,k_3)}_{i_3i_3}\,(\lambda_2)^{(j_4,k_4)}_{i_4i_4}}
\nonumber\\
&\times
(V_1)^{(j_1,k_1)}_{i_1,\{j_b,k_b,j_a,k_a\}} \,
(V_2)^{(j_2,k_2)}_{i_2,\{j_b,k_b,j_c,k_c\}} \,
(U_1)^{(j_3,k_3)}_{\{j_d,k_d,j_c,k_c\},i_3} \,
(U_2)^{(j_4,k_4)}_{\{j_a,k_a,j_d,k_d\},i_4}\nonumber\\
&\times
\left[
\begin{matrix}
\, j_c \, & \, j_a \,  & \, j_5 \, \\
\, j_1 \, & \, j_2 \, & \, j_b \,
\end{matrix}
\right] 
\left[
\begin{matrix}
\, j_c \, & \, j_a \,  & \, j_5 \, \\
\, j_4 \, & \, j_3 \, & \, j_d \,
\end{matrix}
\right] 
\quad ,
\end{align}
where we have used the following identity of Clebsch--Gordan coefficients that is proven in \cite{q_spinnet} and appendix \ref{app:quantum-group}:
\begin{align} \label{eq:9j-symbol}
\begin{tikzpicture}[baseline,scale=0.75]
\draw (-1,-1) -- (-0.5,-0.5) -- (-0.5,0.5) -- (-1,1) -- (0,1.5) arc(180:0:0.75) -- (1.5,-1.5) arc(0:-180:0.75) -- (-1,-1)
      (0,-1.5) -- (1,-1) -- (0.5,-0.5) -- (0.5,0.5) -- (1,1) -- (0,1.5)
      (-0.5,-0.5) -- (0.5,-0.5)
      (-0.5,0.5) -- (0.5,0.5)
      (0,-0.75) node {$j_d$}
      (-0.75,0) node {$j_c$}
      (0.75,0) node {$j_a$}
      (0,0.75) node {$j_b$}
      (-1.25,-1.) node {$j_3$}
      (1.25,-1.) node {$j_4$}
      (1.25,1.) node {$j_1$}
      (-1.25,1.) node {$j_2$}
      (1.75,0) node {$j_5$};
\end{tikzpicture}& 
=\frac{(-1)^{j_c + j_a + j_5} (-1)^{j_1+j_2+j_3+j_4} }{d_{j_5}\sqrt{d_{j_b} d_{j_d}}}
\left[
\begin{matrix}
\, j_c \, & \, j_a \,  & \, j_5 \, \\
\, j_1 \, & \, j_2 \, & \, j_b \,
\end{matrix}
\right] 
\left[
\begin{matrix}
\, j_c \, & \, j_a \,  & \, j_5 \, \\
\, j_4 \, & \, j_3 \, & \, j_d \,
\end{matrix}
\right] \quad .
\end{align}
\end{widetext}

This concludes the discussion of this particular tensor network algorithm. Interestingly, it can be simplified further by `cutting it in half' and using instead an algorithm based on 3--valent vertices, which is straightforward to derive from the 4--valent one and the subject of section \ref{sec:renormalization}.

\bibliographystyle{utphys}
\bibliography{toy-model-ref}

\end{document}